\documentclass[twocolumn,twocolappendix,appendixfloats]{aastex631}
\usepackage{apjfonts}
\usepackage{amsmath}
\usepackage{scalerel}

\renewcommand*{\thefootnote}{\fnsymbol{footnote}}

\DeclareSymbolFont{matha}{OML}{txmi}{m}{it}
\DeclareMathSymbol{\varv}{\mathord}{matha}{29}

\newcommand\msol{$M_{\odot}$}%
\newcommand{\hei}{\ion{He}{1}}%
\newcommand{\feii}{\ion{Fe}{2}}%
\newcommand{\feiii}{\ion{Fe}{3}}%

\newcommand{\fex}{\ion{Fe}{2}\protect\scaleto{$/III$}{1.4ex}}%

\newcommand{\coiii}{\ion{Co}{3}}%
\newcommand{\caii}{\ion{Ca}{2}}%
\newcommand{\niii}{\ion{Ni}{2}}%
\newcommand{\nai}{\ion{Na}{1}}%

\newcommand{\cii}{\ion{C}{2}}%
\newcommand{\oi}{\ion{O}{1}}%
\newcommand{\siii}{\ion{Si}{2}}%
\newcommand{\sii}{\ion{S}{2}}%
\def\simgt{\lower.5ex\hbox{$\; \buildrel > \over \sim \;$}}%
\def\simlt{\lower.5ex\hbox{$\; \buildrel < \over \sim \;$}}%
\def\sni{SN Ia}%
\def\he1{${\rm{^{1}He}}$}%
\def\ni56{${\rm{^{56}Ni}}$}%
\def\sni58{${\rm{^{58}Ni}}$}%
\def\co56{${\rm{^{56}Co}}$}%
\def\fer56{${\rm{^{56}Fe}}$}%
\def\fe52{${\rm{^{52}Fe}}$}%
\def\chr48{${\rm{^{48}Cr}}$}%
\def\dm15{$\Delta M_{15}(B)$}%
\def\tas{Type Ia SN}%
\def\tase{Type Ia SNe}%
\def\sn1a{SNe Ia}%

\def\uname{SN~2021aefx}
\def\sbv{$s_{BV}$}%
\def\tp{$t_{\rm p}$}%
\def\t0{$t_{\rm 0}$}%
\def\chisq{$\chi^2$}%
\def\chisqr{$\chi^2_{\rm R}$}%
\def\d6s{D$^{\wedge}$6}%
\def\vi{\mbox{$V\!-\!i$}}%

\def\scl{1.2}%

\shorttitle{Split-Velocity Infant Type Ia with Excess Emission \& Redward Color Evolution}
\shortauthors{Ni et al.}

\graphicspath{{./}{figures/}}

\begin{document}

\title{Origin of high-velocity ejecta, excess emission and redward color evolution in the infant Type Ia Supernova 2021aefx}

\author[0000-0003-3656-5268]{Yuan Qi Ni}
\affiliation{David A. Dunlap Department of Astronomy and Astrophysics, University of Toronto, 50 St. George Street, Toronto, ON M5S 3H4, Canada}

\author[0000-0003-4200-5064]{Dae-Sik Moon}
\affiliation{David A. Dunlap Department of Astronomy and Astrophysics, University of Toronto, 50 St. George Street, Toronto, ON M5S 3H4, Canada}

\author[0000-0001-7081-0082]{Maria R. Drout}
\affiliation{David A. Dunlap Department of Astronomy and Astrophysics, University of Toronto, 50 St. George Street, Toronto, ON M5S 3H4, Canada}

\author[0000-0001-9732-2281]{Christopher D. Matzner}
\affiliation{David A. Dunlap Department of Astronomy and Astrophysics, University of Toronto, 50 St. George Street, Toronto, ON M5S 3H4, Canada}

\author[0009-0003-4229-1184]{Kelvin C. C. Leong}
\affiliation{David A. Dunlap Department of Astronomy and Astrophysics, University of Toronto, 50 St. George Street, Toronto, ON M5S 3H4, Canada}

\author[0000-0001-9670-1546]{Sang Chul Kim}
\affiliation{Korea Astronomy and Space Science Institute, 776, Daedeokdae-ro, Yuseong-gu, Daejeon 34055, Republic of Korea}
\affiliation{Korea University of Science and Technology (UST), 217 Gajeong-ro, Yuseong-gu, Daejeon 34113, Republic of Korea}

\author[0000-0002-3505-3036]{Hong Soo Park}
\affiliation{Korea Astronomy and Space Science Institute, 776, Daedeokdae-ro, Yuseong-gu, Daejeon 34055, Republic of Korea}
\affiliation{Korea University of Science and Technology (UST), 217 Gajeong-ro, Yuseong-gu, Daejeon 34113, Republic of Korea}

\author[0000-0002-6261-1531]{Youngdae Lee}
\affiliation{Department of Astronomy and Space Science, Chungnam National University, Daejeon 34134, Republic of Korea}

\correspondingauthor{Yuan Qi Ni}
\email{chris.ni@mail.utoronto.ca}

\begin{abstract}
\object{SN 2021aefx} is a normal Type Ia Supernova (SN) showing excess emission and redward color evolution over the first $\sim$~2 days. We present analyses of this SN using our high-cadence KMTNet multi-band photometry, spectroscopy, and publicly available data, including first measurements of its explosion epoch (MJD~59529.32~$\pm$~0.16) and onset of power-law rise ($t_{\rm PL}$~=~MJD~59529.85~$\pm$~0.55; often called ``first light'') associated with the main ejecta ${\rm{^{56}Ni}}$ distribution. The first KMTNet detection of SN~2021aefx precedes $t_{\rm PL}$ by $\sim$~0.5~hours, indicating presence of additional power sources. Our peak-spectrum confirms its intermediate Type Ia sub-classification between Core-Normal and Broad-Line, and we estimate an ejecta mass of $\sim$~1.34~$M_{\odot}$. The spectral evolution identifies material reaching $>$~40,000~km~s$^{-1}$ (fastest ever observed in Type Ia SNe) and at least two split-velocity ejecta components expanding homologously: (1) a normal-velocity ($\sim$~12,400~km~s$^{-1}$) component consistent with typical photospheric evolution of near-Chandrasekhar-mass ejecta; and (2) a high-velocity ($\sim$~23,500~km~s$^{-1}$) secondary component visible during the first $\sim$~3.6~days post-explosion, which locates the component within the outer $<$~16\% of the ejecta mass. Asymmetric subsonic explosion processes producing a non-spherical secondary photosphere provide an explanation for the simultaneous appearance of the two components, and may also explain the excess emission via a slight ${\rm{^{56}Ni}}$ enrichment in the outer $\sim$~0.5\% of the ejecta mass. Our 300~days post-peak nebular-phase spectrum advances constraints against non-degenerate companions and further supports a near-Chandrasekhar-mass explosion origin. Off-center ignited delayed-detonations are likely responsible for the observed features of SN~2021aefx in some normal Type Ia SNe.
\end{abstract}


\section{Introduction} \label{sec:intro}

Type Ia supernovae (SNe) are thought to be from white dwarf (WD) stars that explode as a result of mass transfer in binary systems. 
They are the main producers of iron-peak elements in the Universe \citep{Matteucci2012book} and a crucial tool for measuring cosmological distances, leading to the discovery of the accelerated cosmological expansion and dark energy \citep{Riess1998aj, perlmutter1999apj}.
Despite the extensive effort to understand their origins, the details of the \tase\ explosion process remain poorly understood \citep{Maoz2014araa}, especially for the majority class of ``normal'' \tase\ comprising $\sim$ 70\% of the entire \tas\ population \citep{Blondin2012aj} that is most widely used for cosmological distance measurements \citep{Wang2009apj, Zhang2021mnras}.
This includes the nature of the companion star \citep{Whelan&Iben1973apj, Iben&Tutukov1984apjs, Aznar2015mnras}, the mass transfer process \citep{Guillochon2010apj, Pakmor2012apj, Kushnir2013apj}, the method of core carbon ignition \citep{Hoeflich1995apj, Polin2019apj, Townsley2019apj, Shen2021apjl}, non-spherical symmetry \citep{Maeda2010natur, Boos2021apj}, and the distribution of \ni56\ in the outer layers \citep{Piro&Nakar2014apj, Magee2020aa, Ni2022natas}, just to name a few.
Furthermore, how these properties and processes relate to
the differences between the two main subtypes of normal \tase\ \citep{Parrent2014apss}---``Core-Normal/Normal-Velocity'' (CN/NV) and ``Broad-Line/High-Velocity'' (BL/HV)---and the explosion mass of the WD progenitor---near-Chandrasekhar-mass \citep[$\sim$ 1.38~\msol][]{Mazzali2007sci} or sub/super-Chandrasekhar-mass \citep{Scalzo2019mnras, Brown2014apj}---cast another challenge towards our understanding of \tas\ origins.

Of particular interest are the origins of excess emission and high-velocity silicon and calcium lines seen in the early phases,
and their relation to the \tas\ explosion mechanisms and progenitor systems.
Excess emission has been identified within $\sim$ 5 days post-explosion in a few normal events \citep{Marion2016apj, Hosseinzadeh2017apj, Jiang2018apj, Jiang2021apj, Sai2022mnras, Deckers2022mnras, Dimitriadis2019apj, Dimitriadis2023mnras}, 
and high-velocity \siii\ and \caii\ features,
which are faster than the normal photospheric velocities by $\gtrsim$ 7,000~km~s$^{-1}$ 
\citep{Wang2003apj, Gerardy2004apj, Quimby2006apj, Marion2013apj, Silverman2015mnras},
have been observed in a number of normal events before $B$-band maximum.
In the case of the earliest \tas\ detected to date, SN~2018aoz, infant-phase excess emission during 1--12 hours since first light showed suppressed $B$-band flux attributed to line-blanket absorption by surface iron-peak elements \citep{Ni2022natas, Ni2023apj}.

\uname, which we study in detail in this paper 
based on new high-cadence, multi-band photometry,
new spectra from critical epochs, as well as the data from previous studies, 
provides a rare opportunity for 
answering some of these questions, being a normal \tas\ with a prominent early excess emission in the first $\sim$ 2 days, possible high-velocity Si lines in the excess emission phase, but uncertain classification between the CN/NV and BL/HV subtypes \citep{Ashall2022apj, Hosseinzadeh2022apj}.
Radio observations from 2--8 days and a late-phase ($+$118 days since peak) spectrum rule out wind from a symbiotic binary progenitor and disfavours the presence of swept-up H, He, and O from the companion \citep{Hosseinzadeh2022apj}.
In addition, nebular-phase (255 and 323 days post-peak) spectra from the James Webb Space Telescope (JWST) reveal offset ($\sim$ 1000~km~s$^{-1}$) core velocities and stable \sni58, indicating that \uname\ may originate from an off-center ignited explosion \citep{Kwok2023apj} of a $\sim$ 1.38~\msol WD \citep{DerKacy2023apj}, though this is argued to be inconclusive \citep{Blondin2023arxiv}.

Although these previous observations have provided important 
constraints on the origin and evolution of \uname, 
some of the key parameters---such as 
the exact explosion epoch, reliable subtype classification at $B$-band maximum, ejecta mass, and the 
nebular-phase ($>$ 200 days since peak) constraint on swept-up 
H and He---remain to be estimated.
The nature of its early optical excess emission 
and the presence of high-velocity Si also require further examination. 
In this paper, we provide estimations of these key parameters and 
conduct detailed investigations into the early excess emission
and velocity evolution of \uname.
We describe our observations and data analysis in Section~\ref{sec:obs}.
Section~\ref{sec:early} shows the characterization of the photospheric-phase evolution of the SN, including measurements of its epochs of ``first light'' and explosion and establishment of the presence of high-velocity Si lines before $B$-band maximum, followed by Section~\ref{sec:peak} showing the peak-spectrum classification of \uname\ and estimation of its ejecta mass using pseudo-bolometric luminosity.
In Section~\ref{sec:excess}, we investigate the nature of the observed early excess emission by modelling the predicted emission from \ni56\ heating processes and ejecta-companion/CSM interactions.
In Section 6, we examine possible origins for the observed high-velocity Si lines, including ejecta with two distinct components (``split-velocity ejecta''), ejecta affected by a reverse-shock interaction, as well as ejecta with a local opacity enhancement,
and find that the former is the most likely origin.
In Section~\ref{sec:neb}, based on our spectrum from the nebular phase,
we diagnose the strength of \caii\ emission to be incompatible with sub-Chandrasekhar-mass explosions 
and provide advanced upper limits on the presence of H, He, and O from the companion/CSM.
Finally, we discuss the origin of the early excess emission and high-velocity lines observed in \uname\ 
and their implications for the explosion mechanisms and progenitor systems 
of \tase\ in Section~\ref{sec:conc}.

\section{Observations and Data Analysis} \label{sec:obs}

The discovery of \uname\ was made by 
the Distance Less than 40 Mpc Survey \citep[DLT40;][]{Hosseinzadeh2022apj} 
at an unfiltered brightness of 17.24 mag
at 12h14m Universal Time (UT) on November 11, 2021 (MJD 59529.343).
A subsequent spectrum obtained by Southern African Large Telescope \citep[SALT;][]{Brown2013pasp}
at MJD 59529.872 was H-poor with broad \siii\ absorption features \citep{Bostroem2021tnscr},
identifying the source to be a \tas. 

\begin{figure*}[t!]
\epsscale{\scl}
\begin{center}
\includegraphics[width=0.9\textwidth]{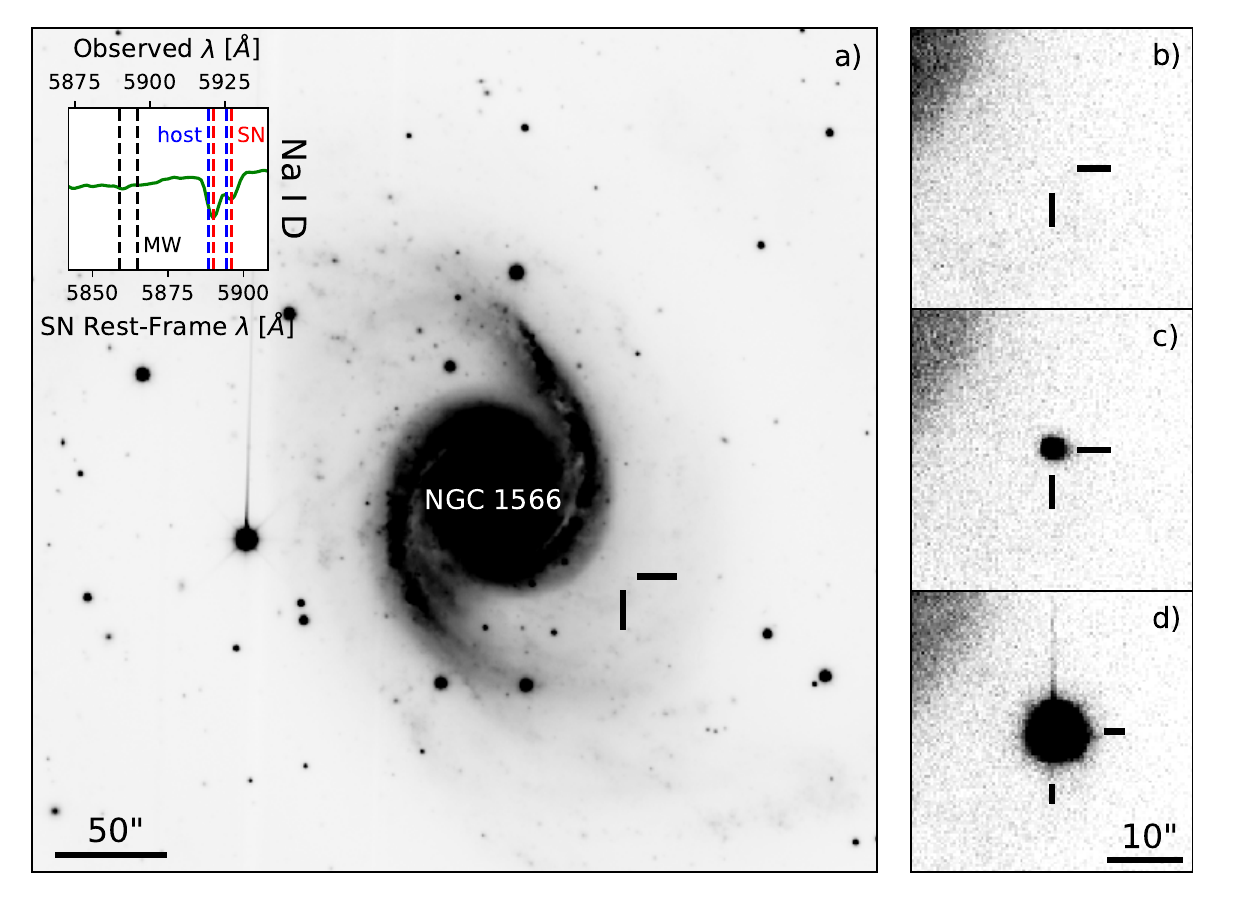}
\end{center}
\caption{(a) Deep KMTNet $I$-band stacked image containing the locations of \uname\ and the host galaxy, NGC~1566, obtained by stacking 837 individual 60~s exposures taken prior to November 6, 2021, well before the SN explosion. The location of \uname\ is indicated with the crosshair. North is up, East is to the left.
The inset shows our near-peak Gemini spectrum of the source from $-$0.5 days since $B$-band maximum zoomed in on the detected \nai~D feature at 5890 and 5896~\AA\ from the SN rest-frame (red vertical dashed lines) and non-detections in the frame of the Milky Way (MW; black vertical dashed lines).
Panels (b)–(d) are individual 60~s $I$-band exposures centered on the location of \uname: (b) the image obtained 2 days prior to the first detection of the source, (c) the $I$-band first detection image obtained at MJD 59529.83008, and (d) the image with the maximum $I$-band apparent brightness obtained 13.86 days from the first detection. A color version of the field of view of NGC 1566 as shown in panel (a) is available as an animation in the online journal, including the images shown in panels (b-d) and a selection of other images acquired by the KSP. The animation shows the evolution of \uname\ beginning on Oct. 1st, 2021 and ending on Feb. 25th, 2022. The real-time duration of the animation is 6 seconds.}
\label{fig:det}
\end{figure*}

\begin{deluxetable*}{cccccc}
\tabletypesize{\footnotesize}
\tablecolumns{6} 
\tablewidth{0.99\textwidth}
 \tablecaption{Sample KSP photometry of \uname.}
 \tablehead{
 \colhead{Date [MJD]} & \colhead{Band} & \colhead{Magnitude$\rm ^a$} & \colhead{1-$\sigma$ (detection)} & \colhead{1-$\sigma$ (total)$\rm ^b$} & \colhead{S/N}
 } 
\startdata 
  59529.82709& $B$ &      16.678&         0.016&          0.019&           69.3\\
  59529.82859& $V$ &      16.506&         0.014&          0.017&           79.3\\
  59529.83008& $i$ &      16.842&         0.012&          0.017&           91.7\\
  59530.82063& $B$ &      16.261&         0.018&          0.021&           61.4\\
  59530.82198& $V$ &      15.682&         0.015&          0.018&           74.1\\
  59530.82333& $i$ &      16.191&         0.011&          0.017&           95.6\\
  59530.86225& $B$ &      16.256&         0.017&          0.020&           65.2\\
  59530.86361& $V$ &      15.644&         0.014&          0.018&           76.0\\
  59530.86512& $i$ &      16.156&         0.011&          0.016&          101.2\\
\enddata
\tablenotetext{{\rm a}}{The $BV$-band magnitudes are in the Vega system,
while the $i$-band magnitudes are in the AB system (see Section~\ref{sec:phot} text).}
\tablenotetext{{\rm b}}{Total 1-$\sigma$ error includes the detection S/N, the photometric calibration error, and the S--correction error (see Section~\ref{sec:phot} text).}
\tablecomments{The entire observed magnitudes are available in the electronic edition.} 
\end{deluxetable*} 
\label{tab:lc}

\begin{deluxetable*}{lccccc}
\tabletypesize{\footnotesize}
\tablecolumns{5} 
\tablewidth{0.99\textwidth}
 \tablecaption{GMOS spectroscopy of \uname.}
 \tablehead{
 \colhead{Date (UT)} & \colhead{Phase$^{\rm a}$} & \colhead{Telescope} & \colhead{Instrument} & \colhead{R} & \colhead{Wavelength [\AA]}
 } 
\startdata 
2021 November 28.12 & $-$0.5 days & Gemini S & GMOS & 1690 & 4000--10000\\
2022 January 20.20 & $+$52.3 days & Gemini S & GMOS & 1690 & 4000--10000\\
2022 September 27.37 & $+$302.1 days & Gemini S & GMOS & 1690 & 4000--10000\\
\enddata
\tablenotetext{{\rm a}}{Phase is rest-frame days since $B$-band maximum ($t_{\rm max}$ = MJD 59546.67; Table~\ref{tab:param}).}
\end{deluxetable*} 
\label{tab:spec}

\subsection{Photometry} \label{sec:phot}

We have conducted high-cadence, multi-color ($BVI$) monitoring observations of
the field containing the spiral galaxy NGC~1566
using Korea Microlensing Telescope Network \citep[KMTNet;][]{Kim2016jkas} 
as part of the KMTNet Supernova Program \citep{Moon2016spie}
optimized for detecting early SNe and transients
\citep[e.g.,][]{Afsariardchi2019apj, Moon2021apj, Lee2022apj, Ni2022natas, Ni2023apj}.
The KMTNet is composed of three 1.6-m telescopes all equipped with a wide-field CCD camera 
with 4 square degree field-of-view in Chile, South Africa, and Australia,
providing 24-hour continuous sky coverage.
Between November 1, 2021 and February 25, 2022, we obtained $\sim$ 1200 total images 
of the field with 60~s exposure time for individual frames reaching an average 3-$\sigma$ detection 
limit of 21.3 mag. 
On average, one set of $BVI$-band images were obtained every $\sim$ 7 hours.
Compared to previously reported photometric observations of \uname, our cadence is $\sim$ 3.4 and 1.7~$\times$ higher than the multi-band cadences of Las Cumbres Observatory (LCO) and the Precision Observations of Infant Supernova Project (POISE), respectively, and each of our images reach a $\sim$ 2 mag deeper detection limit than unfiltered DLT40 photometry \citep{Hosseinzadeh2022apj, Ashall2022apj}.

The first KSP detection of \uname\ with signal-to-noise (S/N) $>$ 3 was made at 19h51m UT on November 11, 2021 (= MJD 59529.82709) in a $B$-band image, followed by $V$- and $I$-band detections 2 and 4 minutes later, respectively. 
The initial $BVi$-band\footnote{Note that the $BV$-band observations are calibrated to the Vega system,
while the $I$-band observations are calibrated to the AB system. We, therefore, use ``$BVI$'' to refer to the KMTNet filter system, while we use ``$BVi$'' to refer to their corresponding photometric magnitudes.} apparent brightness of the source was 16.68 $\pm$ 0.02, 16.51 $\pm$ 0.02, and 16.84 $\pm$ 0.02 mag, respectively.
The source was not detected in any bands observed by the KSP during the last two epochs prior to this.
At 0.98 days prior (= MJD 59528.85) to our first detection, we measured 3-$\sigma$ detection limits of 16.88 mag in $B$ band and 18.20 mag in $i$ band, and at 1.99 days prior (= MJD 59527.84), we measured $BVi$-band limits of 21.82, 21.46, and 21.59 mag, respectively.
These substantially tighten the constraints on the last non-detection of the source compared to previous observations: 19.35 mag (unfiltered) at MJD 59524.33 and 17.70 mag ($g$-band) at MJD 59527.36 from DLT40 \citep{Valenti2021tnstr} and the All-Sky Automated Survey for Supernovae \citep[ASAS-SN;][]{Ashall2022apj}, respectively.
The source remained detected until February 25, 2022, after which it was too close to the Sun for continuous monitoring.
The source attained peak apparent brightnesses of 11.98 in $B$ band, 11.90 in $V$ band, and 12.66 mag in $i$ band, on MJD 59546.94, 59549.33, and 59544.69, respectively.
The average J2000 coordinate of the source
is ($\alpha$, $\delta$) = ($\rm 04^h19^m53^s.38, -54\degr56\arcmin53\farcs24$), consistent with the coordinates reported on the Transient Name Server\footnote{\url{https://www.wis-tns.org/object/2021aefx/}}.
Figure~\ref{fig:det} presents KMTNet images of the field where the measured position of the source is indicated with crosshairs.
In the figure, panel (a) is an $I$-band stacked image composed of images obtained before the SN explosion.
Panels (b), (c), and (d) are the $I$-band images from two days prior to first detection, first detection, and the maximum brightness, respectively.

Point-spread function (PSF) photometry of \uname\ was performed using the SuperNova Analysis Package (SNAP)\footnote{\url{https://github.com/niyuanqi/SNAP}}, a custom python-based pipeline for supernova photometry and analysis.
A local PSF was obtained by fitting a Moffat function \citep{Moffat1969aap, Trujillo2001mnras} to nearby reference stars and simultaneously fitting sky background emission with a first-order polynomial.
We fit the flux of \uname\ using this PSF, and the detection S/N of the source reported in this paper is equal to the best-fit flux divided by its 1-$\sigma$ uncertainty.
Photometric calibration was performed against more than 10 reference stars within 15\arcmin\ distance from the source in the AAVSO Photometric All-Sky Survey\footnote{\url{https://www.aavso.org/apass}} database whose apparent magnitudes are in the range of 15--16 mag.
The observations in the $BVI$ KMTNet filters were calibrated against reference stars in the nearest AAVSO filters (Johnson~$BV$, and Sloan $i'$; or $BVi$).

The KSP instrumental magnitudes for the AAVSO reference stars were transformed to standard $BVi$ magnitudes using the equations from \citet{Park2017apj}.
However, the spectra of SNe, particularly after the peak, are significantly different from the AAVSO standard stars used to derive the transformation equations. In order to account for this, we applied spectrophotometric (S)--corrections, which are magnitude corrections between instrument and standard filters derived by performing synthetic photometry on spectra obtained at the same epoch \citep{Stritzinger2002aj}. For this, we used the optical spectra compiled in \citet{Hosseinzadeh2022apj} covering the epochs from 20h55m UT on November 11, 2021 (MJD 59529.872; 1.1 hours since the first KSP detection) until 11h30m UT on February 20, 2022 (MJD 59630.480; 4.6 days before the last KSP detection). The calibrated and S-corrected photometry is presented in Table~\ref{tab:lc}, and the early light curves are shown in Figure~\ref{fig:lc}.

\subsection{Spectroscopy} \label{sec:spec}

We obtained spectra of \uname\ from three different epochs, including peak and nebular phase, using the
Gemini Multi-Object Spectrograph \citep[GMOS;][]{Hook2004} on the 8m Gemini-South telescope. The observations are summarized in Table~\ref{tab:spec}.
The spectra were reduced using the custom \texttt{gmos} suite of IRAF tasks. Bias and flat-field corrections were performed on the two-dimensional frames, one-dimensional spectra were extracted, and wavelength calibration was performed using calibration lamps taken immediately after target exposures.
Flux calibration was performed using the IRAF tasks \texttt{standard} and \texttt{calibrate} using photometric standards observed in the same filters.
In our analysis below we supplement this data with publicly available spectroscopy from \citet{Hosseinzadeh2022apj}.

\subsection{Host Galaxy, Distance and Extinction} \label{subsec:host}

The host galaxy of \uname\ is the nearby spiral galaxy NGC~1566 \citep[the ``Spanish Dancer'';][]{Skrutskie2003ycat} centered at ($\alpha$, $\delta$) = ($\rm 04^h20^m00^s.42, -54\degr56\arcmin16\farcs1$), $\sim$ 1\farcm2 (= 6.0 kpc) northeast of the position of \uname\ (see Fig~\ref{fig:det}).
The host galaxy redshift of $z$ = 0.005017 $\pm$ 0.000007 \citep{Allison2014mnras} and the corresponding Hubble flow distance modulus (DM) of 31.40 $\pm$ 0.15 mag in the cosmology of \citet{Riess2016apj} with corrections for peculiar velocities due to the Virgo Supercluster, Great Attractor, and Shapley Supercluster \citep{Mould2000apj}, 
as well as the \citet{Tully1977aa} and tip of the red giant branch (TRGB) distances of DM = 31.28 $\pm$ 0.23 mag and 31.27 $\pm$ 0.49, respectively, to NGC~1566 have been used in previous studies of \uname\ \citep{Hosseinzadeh2022apj, Ashall2022apj, Kwok2023apj}.
Given that all of those studies identify \uname\ to be a normal \tas, 
we adopt DM = 31.16 $\pm$ 0.08 mag based on normal \tas\ template fitting (see Section~\ref{subsec:templ}), corresponding to the luminosity distance of 17.06 $\pm$ 0.63~Mpc.
This is consistent with the Hubble flow, \citet{Tully1977aa}, and TRGB distances within their uncertainties, but slightly more accurate.

\uname\ suffers from relatively little Galactic extinction since it is located near the Galactic south pole with $E(B-V)_{\rm gal}$ = 0.0079 $\pm$ 0.0002 mag in its direction according to the extinction model of \citet[][S\&F 2011]{Schlafly&Finkbeiner2011apj}. This corresponds to Galactic extinction corrections of 
0.033, 0.025, and 0.016 mag in the $B$, $V$, and $i$ bands, respectively, using an $R_V$ = 3.1 \citet{Fitzpatrick1999pasp} reddening law.
The inset of Figure~\ref{fig:det} (a) shows our near-peak Gemini spectrum zoomed into the vicinity of the \nai~D absorption lines at 5890 and 5896~\AA\ associated with dust extinction. 
The lines are absent in the frame of the Milky Way (black vertical dashed lines), confirming the low value of Galactic extinction towards the direction of the SN, while they are prominently detected near the expected redshift of the host galaxy ($z$ = 0.005017; blue vertical dashed lines).

We measure the host galaxy extinction in the vicinity of \uname\ by fitting the equivalent widths of the \nai~D lines with a Voigt doublet profile, assuming a Milky-Way-like correlation between \nai~D and dust extinction \citep{Poznanski2012mnras}.
The mean of the measurements gives $E(B-V)_{\rm host}$ = 0.124 $\pm$ 0.033 mag, consistent with the reported host galaxy extinction of $E(B-V)_{\rm host}$ = 0.097 based on a similar measurement from $\sim$ 16 days post-peak \citep{Hosseinzadeh2022apj}.
We adopt the latter because it is from a much higher resolution (R $\sim$ 40000) spectrum and therefore more accurate, though we note that the scatter associated with $E(B-V)_{\rm host}$ estimated using the method of \citet{Poznanski2012mnras} can be as high as 68\% \citep{Phillips2013apj}.
The mean redshift of the fitted Na I D lines is $z$ = 0.005284 $\pm$ 0.000005, which implies a SN peculiar velocity of (80.0 $\pm$ 3.6)~km~s$^{-1}$ in the frame of the host galaxy, consistent with the measured host galaxy rotation velocity of (65 $\pm$ 60)~km~s$^{-1}$ at the SN location \citep{Elagali2019mnras} that was adopted in previous studies \citep{DerKacy2023apj}.
Hereafter, we adopt $z$ = 0.005284 for the SN redshift since it is measured from the source position and more accurate.

\begin{deluxetable*}{lr}[ht!]
\tabletypesize{\footnotesize}
\tablecolumns{2} 
\tablewidth{0.99\textwidth}
 \tablecaption{\uname\ basic properties.}
 \tablehead{
 \colhead{Parameter} & \colhead{Value}
 } 
\startdata 
    ($\alpha$, $\delta$) (J2000) & ($\rm 04^h19^m53^s.38, -54\degr56\arcmin53\farcs24$) \\
    Redshift ($z$)                             & 0.005284 $\pm$ 0.000005 \\
    KSP first detection: UT \& MJD                 & 19$h$51$m$ on November 11, 2021 \& MJD  59529.82709 \\
    Peak apparent magnitude                    & 11.98 ($B$), 11.90 ($V$), 12.66 ($i$) mag  \\ 
    Apparent peak epochs                                & MJD 59546.94 ($B$), 59549.33 ($V$), 59544.69 ($i$) \\
    Distance Modulus (DM)                       & 31.16 $\pm$ 0.08 mag \\
    Peak absolute magnitude                    & --19.56 ($B$), --19.54 ($V$), --18.66 ($i$) mag \\
    Post-peak decline rate (\dm15)             & 0.89 $\pm$ 0.08 mag \\
    Color stretch parameter (\sbv)             & 1.00 $\pm$ 0.04 \\
    Rest-frame peak epochs
     & MJD 59546.67 ($B$)$^{\rm a}$, 59549.22 ($V$), 59544.70 ($i$)\\
    Onset of power-law rise ($t_{\rm PL}$)$^{\rm b}$               & MJD 59529.85 $\pm$ 0.55    \\
    Early light curve power-law index  ($\alpha$)    &  2.15 $\pm$ 0.26 ($B$), 1.58 $\pm$ 0.19 ($V$), 1.65 $\pm$ 0.19 ($i$) \\ 
    Explosion epoch ($t_{\rm exp}$)               & MJD 59529.32 $\pm$ 0.16    \\
    $\varv_{\rm peak}$ (\siii~$\lambda$6355~\AA)             &  (12.44 $\pm$ 0.02) $\times$ 10$^{3}$ km s$^{-1}$ \\
    pEW$_{\rm peak}$ (\siii~$\lambda$6355~\AA)     &  100.3 $\pm$ 1.0 \AA\ \\
    pEW$_{\rm peak}$ (\siii~$\lambda$5972~\AA)     &  12.7 $\pm$ 0.5 \AA\ \\
    Peak bolometric luminosity ($L_{\rm peak}$)  & (1.54 $\pm$ 0.02) $\times$ 10$^{43}$ erg s$^{-1}$   \\
    Bolometric peak epoch                      & MJD 59545.86 $\pm$ 0.21 days \\
    Bolometric post-peak decline rate ($\Delta M_{15, \rm bol}$) & 0.75 $\pm$ 0.04 mag \\
    \ni56\ mass ($M_{\rm Ni}$)               & 0.631 $\pm$ 0.009~\msol\ \\
    $\tau_m$ parameter                        &
    11.80 $\pm$ 0.31~days \\
    Ejecta mass ($M_{\rm ej}$)                & 1.34 $\pm$ 0.07~\msol\  \\
    Ejecta kinetic energy ($E_{\rm ej}$)       & (1.24 $\pm$ 0.07) $\times$ 10$^{51}$ erg \\
\enddata
\tablenotetext{{\rm a}}{$t_{\rm max}$ = MJD 59546.67 is the epoch that is referred to as $B$-band maximum throughout the text.}
\tablenotetext{{\rm b}}{The epoch of ``first light'' in \tase\ is often estimated by fitting the observed power-law rise (see Section~\ref{sec:early}).}
\tablecomments{All time/date values are given in the observer frame, except for $\tau_m$ which is in rest frame.}
\end{deluxetable*} 
\label{tab:param}

\subsection{Template Fitting Distance and K--correction} \label{subsec:templ}

The light curves of normal \tase\ in the photospheric phase from $-$10 to 15 days since $B$-band maximum form a one-parameter family of functions, often parameterized by the Phillips parameter, \dm15, defined as the post-peak decline in $B$-band magnitude between 0 and 15 days \citep{Phillips1999aj}.
To include peculiar \tase\ such as the rapidly evolving 91bg-like subtype, the stretch parameter defined by $s_{BV} = t_{BV}/($30 days$)$, where $t_{BV}$ is the time between $B$-band peak to the maximum post-peak $B-V$ color \citep{Burns2014apj}, has been used as an alternative to \dm15.
We measure the distance to \uname\ by fitting its $BgVri$-band light curves from LCO \citep{Hosseinzadeh2022apj} with a normal \tas\ template based on that of \citet{Hsiao2007apj} using SNooPy \citep[\texttt{EBV\_model2};][]{Burns2011aj}, noting that its light curves were determined to be normal using multiple other independent distance measures (Section~\ref{subsec:host}).
The fitted parameters of \texttt{EBV\_model2} are \sbv, time of $B$-band maximum (\tp), and DM.
In the fitting process, we applied $K$--corrections between the observer- and rest-frame filter response functions and Galactic extinction corrections to the template.
The best-fit is obtained with \tp, DM, and \sbv\ of MJD 59547.2 $\pm$ 0.3, 31.16 $\pm$ 0.08 mag, and 1.05 $\pm$ 0.03 respectively, providing a reduced \chisq\ (= \chisq\ normalized by the number of degrees of freedom; \chisqr) of 5.5.

We measure $K$--corrections between the observer and rest-frame filter response functions for the near-peak KSP light curves of \uname\
by performing a separate SNooPy fit, replacing the $BVi$-band light curves with those from the KSP.
In this fit, we fix the distance modulus to be be 31.16 mag,
obtaining the best-fit template with \tp\ and \sbv\ of MJD 59547.5 $\pm$ 0.3 and 1.14 $\pm$ 0.03 respectively, and \chisqr\ = 7.71.
We derive rest-frame magnitudes for the near-peak light curves of \uname\ by applying the $K$--corrections from this best-fit template, DM = 31.16 mag, and Galactic extinction correction.
Note that the $K$--corrections are insubstantial ($<$ 0.025 mag) given the relatively small redshift of the SN. 
With these rest-frame near-peak light curves, we used polynomial fitting to measure the peak absolute magnitudes of each $BVi$-band light curve to be $-$19.56 $\pm$ 0.02, $-$19.54 $\pm$ 0.03, and $-$18.66 $\pm$ 0.03 mag, respectively.
The rest-frame $BVi$-band light curves each attained their peaks at MJD 59546.67 $\pm$ 0.67, 59549.22 $\pm$ 0.68, and 59544.70 $\pm$ 0.89 days (in the observer frame), respectively.
We also measured the Phillips parameter and color stretch parameter to be \dm15\ = 0.89 $\pm$ 0.08 mag and \sbv\ = 1.00 $\pm$ 0.04, respectively.
These parameters obtained from the KSP light curves are consistent with the range of what has been reported in other studies of \uname\ \citep{Ashall2022apj, Hosseinzadeh2022apj}.
The parameters of \uname\ measured in this section and others are summarized in Table~\ref{tab:param}.

\begin{figure*}[t!]
\epsscale{\scl}
\begin{center}
\includegraphics[width=0.7\textwidth]{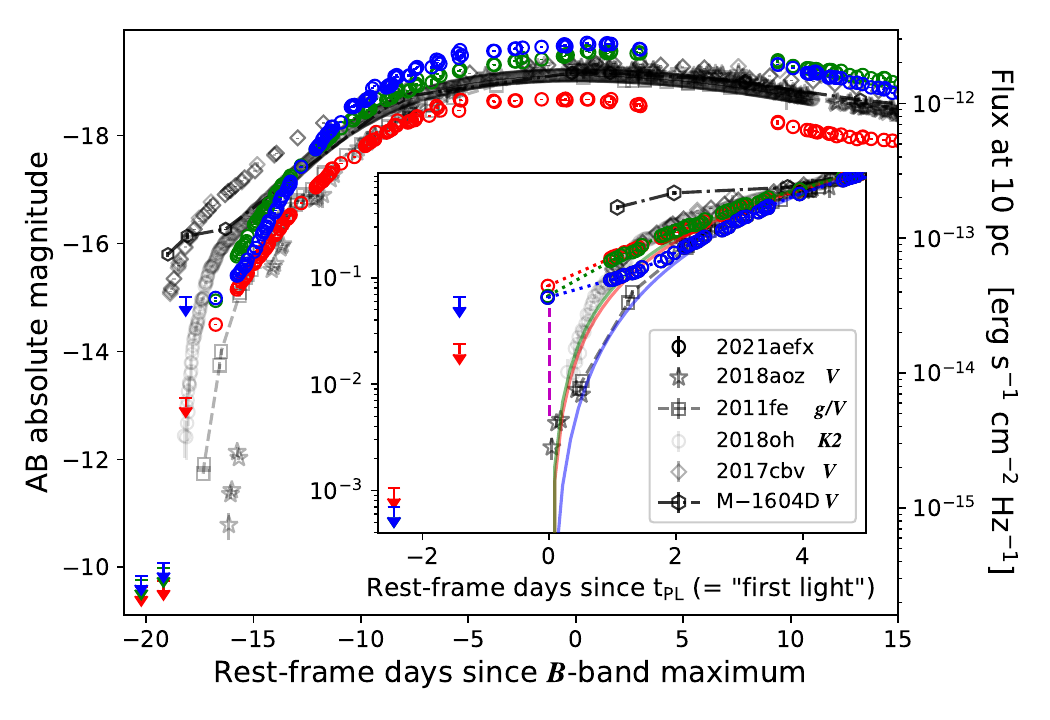}
\end{center}
\caption{The $BVi$ (blue, green, and red data points, respectively) light curves of \uname\ are compared to those of normal
\tase~2018aoz \citep{Ni2022natas, Ni2023apj}, 2011fe \citep{Guillochon2017apj}, 
2018oh \citep{Dimitriadis2019apj}, and 2017cbv \citep{Hosseinzadeh2017apj} as well as the peculiar \tas\ MUSSES1604D \citep{Jiang2017nat} in their nearest-to-$V$ bands centered at maximum light. The errorbars represent the 1-$\sigma$ uncertainty level and inverted arrows are 3-$\sigma$ detection limits in this figure and all of the following.
The inset shows the early light curves within 5 days of their $t_{\rm PL}$ (often called epoch of ``first light''), where excess emission has previously been identified in some \tase\ \citep{Dimitriadis2019apj, Hosseinzadeh2017apj, Stritzinger2018apj}, normalized to their fluxes at 5 days.
SN~2018oh, SN~2017cbv, and MUSSES1604D each show excess emission between 1 and 5 days, whereas SNe~2011fe and 2018aoz show a typical power-law-like rise.
The power-law rise of \uname\ based on fitting a power-law $+$ Gaussian model (see Fig~\ref{fig:gaus}) is shown with blue, green, and red curves. 
The early excess emission in \uname, indicated with the magenta vertical dashed line, is clearly different from those of other SNe.
\label{fig:lc}}
\end{figure*}

\begin{figure}[t!]
\epsscale{\scl}
\plotone{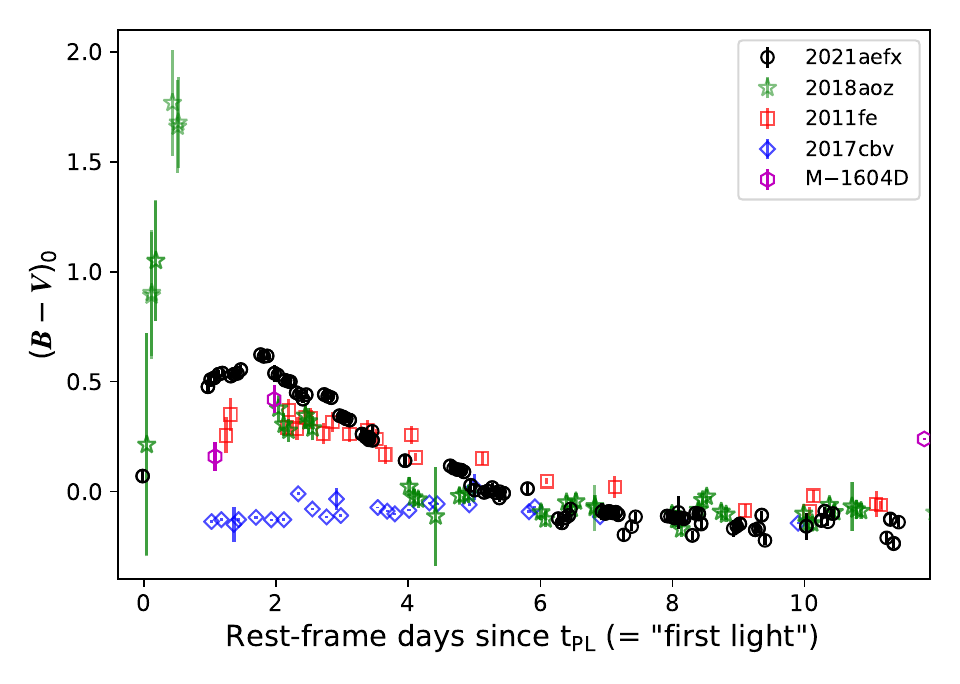}
\caption{The early $B-V$ color of \uname\ is compared to those of other \tase\ showing that \uname\ undergoes a unique redward color evolution over the first $\sim$ 2 days that is faster than that of MUSSES1604D and slower than that of SN~2018aoz before evolving bluewards, falling in line with early-red events including SNe~2018aoz and 2011fe \citep{Stritzinger2018apj}.
\label{fig:col}}
\end{figure}

\section{Early Evolution} \label{sec:early}

Here, we describe the early photometric and spectroscopic evolution of \uname, beginning with a summary of the early light curve and color evolution of the SN in Figures~\ref{fig:lc} and \ref{fig:col}.
Figure~\ref{fig:lc} compares the early KSP light curves of \uname\ in the rest frame with those of other \tase\ in their nearest-to-$V$ bands.
\uname\ appears to have a higher peak luminosity and faster pre-peak rise compared to the other events.
The inset shows the light curves of \uname\ and the other SNe centered on their respective power-law fitted epochs of ``first light'' and normalized to their observed fluxes 5 days after.
The blue, green, and red solid curves show the fitted power-laws to the $B$-, $V$-, and $i$-band rising light curves of \uname, respectively.

Typically, the onset of the rising early light curve in \tase\ is estimated using power-law fitting. For \uname\ we find a date of MJD 59529.85 $\pm$ 0.55 using this technique (see Section~\ref{subsec:gaus} below). We note that while this epoch ($t_{\rm PL}$) has often been referred to as the epoch of ``first light'', since the power-law rise of \tase\ is thought to be driven by radioactive emission from the main distribution of centrally-concentrated \ni56\ in the ejecta \citep{Piro&Nakar2013apj, Piro&Nakar2014apj}, the precise quantity estimated by $t_{\rm PL}$ is the onset of the central \ni56-driven power-law rise.
In the case of \uname, there is apparent excess emission to the power-law over the first $\sim$ 2 days whose onset likely precedes $t_{\rm PL}$ (Section~\ref{subsec:gaus}).
Notably, the excess emission
rises to a different flux level and on a different timescale than what is seen in the other events in the figure---SNe~2018aoz and 2018oh, and MUSSES1604D---that also show early excess emissions, making \uname\ a unique \tase.

Figure~\ref{fig:col} shows that the \bv\ color evolution of \uname\ undergoes a redward color evolution during the same period of $\sim$ 0--2 days where the early excess emission is observed, a behavior that is also reported in LCO and POISE photometry \citet{Ashall2022apj, Hosseinzadeh2022apj}.
By fitting a polynomial, we find that the \bv\ color of \uname\ rises to a peak of 0.568 $\pm$ 0.004 mag over the first 1.58 $\pm$ 0.02 days.
This is redder than any other \tas\ observed at these epochs other than SN~2018aoz.
Compared with the two other events in the figure that also show an early redward \bv\ color evolution, the color of \uname\ peaks later and bluer than what is seen in SN~2018aoz, while it does earlier and redder than in MUSSES1604D.

\subsection{Early Excess Emission and Onset of Power-law Rise}\label{subsec:gaus}

\begin{figure*}[t!]
\epsscale{\scl}
\begin{center}
\includegraphics[width=0.88\textwidth]{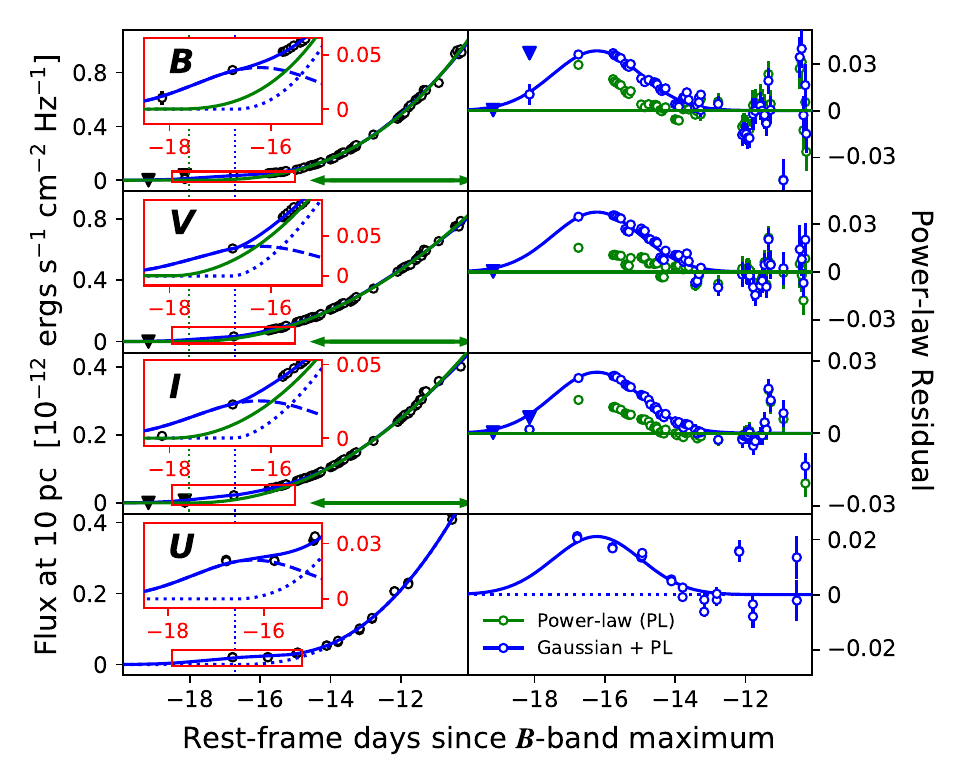}
\end{center}
\caption{(Left) The early ($<-$ 10 days) $UBVi$-band light curves of \uname\ are compared to the following models: (1) pure power-law (green) obtained by fitting the optical ($BVi$-band) light curves over an extensive set of fit intervals between $-$ 20 and $-$ 10 days; and (2) power-law $+$ Gaussian fitted to the light curves over the entire interval.
The vertical green and blue dotted lines indicate the onsets of the power-law components for the correspondingly colored models.
The inset zooms in on the excess emission in the first $\sim$ 3 days.
(Right) The early light curves compared to the same models, with the power-laws of the models subtracted. The green data points represent the result of subtracting the pure power-law, which leaves residuals between the data points and the model (green solid curve).
The blue data points represent that of  subtracting the power-law component of the power-law $+$ Gaussian fit, which leaves residuals that are captured by the Gaussian component. The onset of the power-law component of the power-law $+$ Gaussian fit (vertical blue dotted line) is used to estimate $t_{\rm PL}$ (Table~\ref{tab:param}).
\label{fig:gaus}}
\end{figure*}

\tase\ early light curves are mainly powered by a smooth, centrally-concentrated \ni56\ distribution with a power-law-like tail towards the ejecta surface.
The light curve luminosity is expected to rise as $t^{\alpha}$, where $\alpha$ depends on the steepness of the tail \citep{Piro&Nakar2014apj}.
The observed early optical light curves of most \tase\ have been found to be consistent with such a power-law rise up to $\lesssim$ 40\% of maximum brightness in \citep{Olling2015nat}, with a distribution of $\alpha$ centered on $\sim$ 2 \citep{Miller2020apj}.
Thus, the epoch of ``first light'' for \tase\ is usually inferred by determining the onset of this power-law rise.
However, as shown in Figure~\ref{fig:lc} (inset), 
the earliest emission of \uname\ shows excess luminosity 
in all three bands over the first 2 days before power-law rise dominates.
This is similar to the handful of \tase\ that have been found with ``early excess emission'' (see Figure~\ref{fig:lc}), including SN~2018aoz \citep{Ni2023apj, Ni2022natas}, SN~2018oh \citep{Dimitriadis2019apj}, SN~2017cbv \citep{Hosseinzadeh2017apj}, and MUSSES1604D \citep{Jiang2017nat}. 
To determine the onset of power-law dominated rise in \uname, 
we applied a power-law fit to its rest-frame KSP light curves over an extensive set of fit intervals of different lengths \citep[e.g., see][]{Olling2015nat, Dimitriadis2019apj}, starting from as early as $-$20 days and ending at $-$10 days since $B$-band maximum, during which the brightness is $\lesssim$ 40\% of maximum brightness.
(Note that the KSP bands span the optical spectral range and provide higher cadence and S/N than other multi-band datasets; Section~\ref{sec:phot}).
The best-fit starting epoch is $-$14.5 days, after which the light curves are fit by the power-law, $f_{\nu} \propto (t - t_{\rm PL})^{\alpha_{\nu}}$, with \chisqr\ = 2.1 and $\alpha_{(B, V, i)}$ = (2.75 $\pm$ 0.07, 2.041 $\pm$ 0.05, 2.12 $\pm$ 0.05).
The onset of this power-law is $t_{\rm PL}$ = $-$18.01 $\pm$ 0.13 days (or MJD 59528.56 $\pm$ 0.13). 
Figure~\ref{fig:gaus} (rows 1--3) shows the early light curves of \uname\ compared to this fit (green solid curves; left panels),
in which we can identify that the fit does not generalize well before $-$14.5 days where the residuals increase (right panels).

The discrepancy between the fitted power-law
and observed light curves before $-$14.5 days in Figure~\ref{fig:gaus} (rows 1--3) is due to the presence of the early excess emission during the period.
In order to isolate the power-law and excess emission components of the rising light curves of \uname\ in this phase, we fit the light curves between $-$20 and $-$10 days with a simple analytic model consisting of a power-law $+$ Gaussian in each band.
The three $BVi$-band light curves share the same Gaussian central epoch, $\mu$, and width $\sigma$, but each Gaussian is scaled independently, while we use the same power-law above.
The best-fit power-law $+$ Gaussian with \chisqr\ = 2.4 (Figure~\ref{fig:gaus}, blue solid curves) is obtained with $\mu$ = $-$16.23 $\pm$ 0.14 days (or MJD 59530.35 $\pm$ 0.14), $\sigma$ = 1.25 $\pm$ 0.14 days, $t_{\rm PL}$ = $-$16.73 $\pm$ 0.55 days (or MJD 59529.85 $\pm$ 0.55), and
$\alpha_{(B, V, i)}$ = (2.15 $\pm$ 0.26, 1.58 $\pm$ 0.19, 1.65 $\pm$ 0.19).
This fit provides excellent
agreement to the observed light curves of \uname,
showing that the power-law from this model fit better represents the underlying rise of the SN.
We note that other conceivable models similarly combining an early excess with the underlying SN rise can also provide a good fit, such as a two-component power-law model \citep{Ashall2022apj} or the combination of a non-Gaussian physical model for the early excess emission with a power-law (see Section~\ref{sec:excess}) or \tas\ template \citep{Hosseinzadeh2022apj}.
However, since the underlying SN rise is generally expected to follow the analytic power-law, and physical models of early excess emission in \tase\ mostly predict peaked Gaussian-like light curves, the power-law and Gaussian components generically represent the emission from the underlying SN rise and excess emission, respectively.

The fit parameters of the power-law $+$ Gaussian model estimate 
the onset of the power-law rise
to be $t_{\rm PL}$ = MJD 59529.85 $\pm$ 0.55, following 0.5 hours after our first S/N $>$ 3 $BVI$ detections of \uname\ at MJD 59529.83 (Table~\ref{tab:lc}), near the peak of the observed early excess emission (see Figure~\ref{fig:gaus}, left panels).
This indicates that the onset of emission from \uname\ likely \emph{precedes} power-law rise, with an estimated probability of $\gtrsim$ 82\% based on the first S/N $>$ 3 detection by DLT40 at MJD 59529.343.
Note that, to our knowledge, this is the first reported detection of \tas\ emission prior to the onset of power-law rise from the main distribution of centrally-concentrated \ni56\ in the ejecta, which indicates the presence of additional power sources outside of this centrally-concentrated distribution.
The 16.73-day rise time of \uname\ from $t_{\rm PL}$ to $B$-band maximum in rest frame is relatively short within the observed rise time distribution of the normal \tas\ population \citep[ranging in 15--22 days;][]{Miller2020apj}.
The Gaussian-like excess emission component of the early light curve peaks between $\sim -$0.75 and 1.75 days since $t_{\rm PL}$ (called ``excess emission phase'' hereafter).
During this phase, the light curve of \uname\ is dominated by the excess emission component, which emitted a total of $\Phi_{\rm ex} \sim$ 1.35 $\times$ 10$^{-6}$ ergs~cm$^2$ into the $BVi$ bands along the line of sight (or 4.70 $\times$ 10$^{46}$ ergs, assuming spherically symmetric emission).
Extending the power-law $+$ Gaussian fit to include the LCO $U$-band light curve \citep{Hosseinzadeh2022apj} results in \chisqr\ = 3.1 and estimates the $U$-band power-law index $\alpha_U$ = 2.49 (Figure~\ref{fig:gaus}; row 4), where the $t_{\rm PL}$, $\mu$, and $\sigma$ obtained from the $BVi$ fit is applied to the $U$-band.
The addition of $U$-band increases $\Phi_{\rm ex}$ by 17.9\%, indicating that the excess emission mainly emerges in the optical bands from $B$ to $i$ (also see Figure~\ref{fig:uvsup}).
The nature of this excess emission is explored in Section~\ref{sec:excess}.
Following the excess emission phase, the power-law component dominates the light curve rise, with power-law indices consistent with what has been inferred from the light curves of other normal \tase\ \citep[$\sim$ 1--3;][]{Miller2020apj}.

\subsection{Evolution of Early Photospheric Velocities} \label{subsec:sifit}

\begin{figure}[t!]
\epsscale{\scl}
\plotone{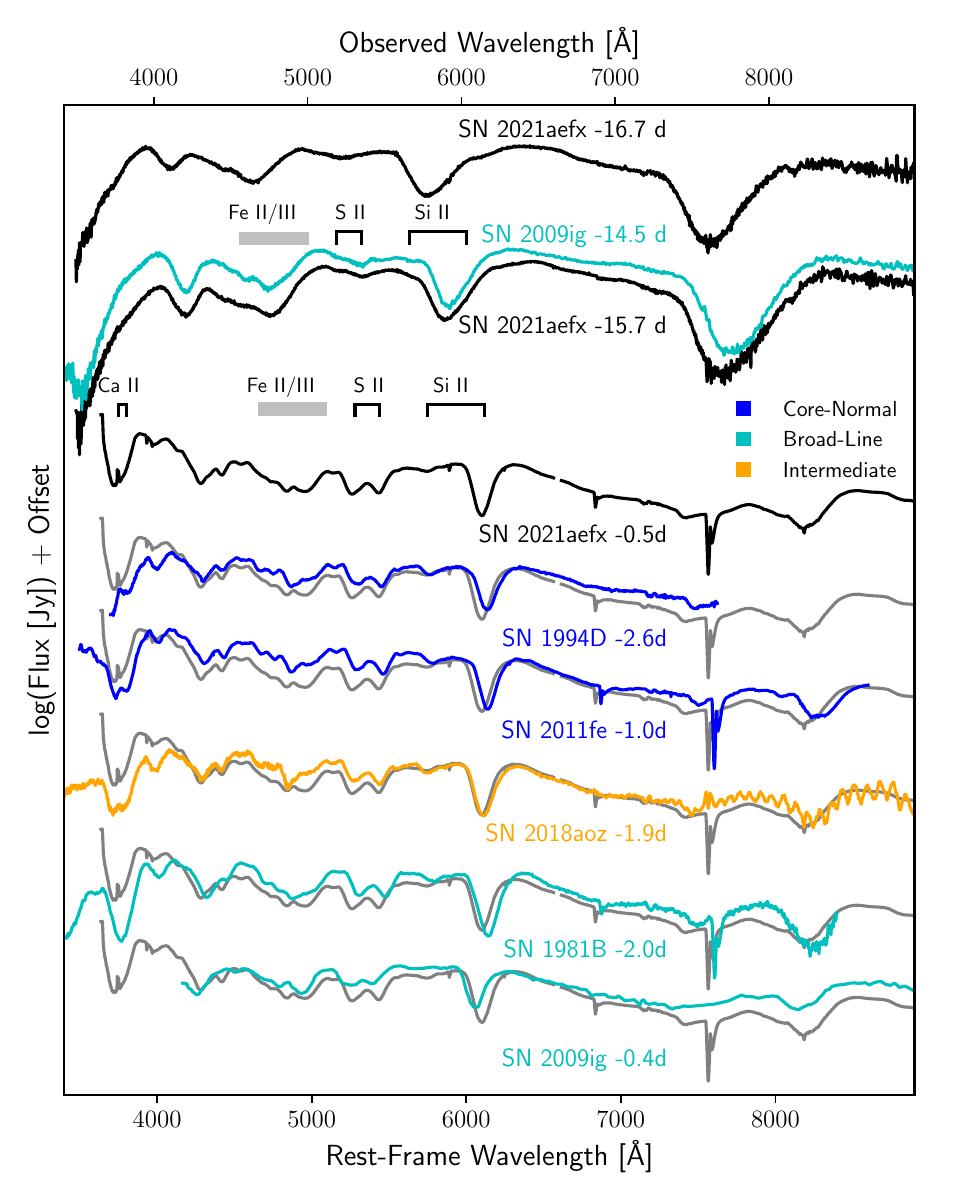}
\caption{Spectra of \uname\ (black) from $-$16.7, $-$15.7, and $-$0.5 days (as labelled) since $B$-band maximum in rest frame are compared to those of other normal \tase. The first two spectra were obtained by SALT during the excess emission phase---i.e., prior to 1.75 days since $t_{\rm PL}$ (see Section~\ref{subsec:gaus})---while the third is our spectrum from near $B$-band maximum.
The near-peak spectrum is translated downwards (grey) by subtracting a constant value to compare with those of five other normal \tase: Core-Normal events (blue) SNe~1994D \citep{Meikle1996mnras} and 2011fe \citep{Parrent2012apj}, Broad-Line events (cyan) SNe~1981B \citep{Branch1983apj} and 2009ig \citep{Foley2012apj}, and intermediate event SN~2018aoz \citep[orange;][]{Ni2023apj}.
Notable spectral features from the excess emission phase and near-peak phase are marked above the spectra from each phase, respectively.
The spectrum of \uname\ appears to resemble that of SN~2009ig during the excess emission phase prior to $\sim -$15 days, though it becomes more similar to SN~2018aoz near $B$-band maximum.
\label{fig:specevol}}
\end{figure}

\begin{figure*}[t!]
\epsscale{\scl}
\begin{center}
\includegraphics[width=0.9\textwidth]{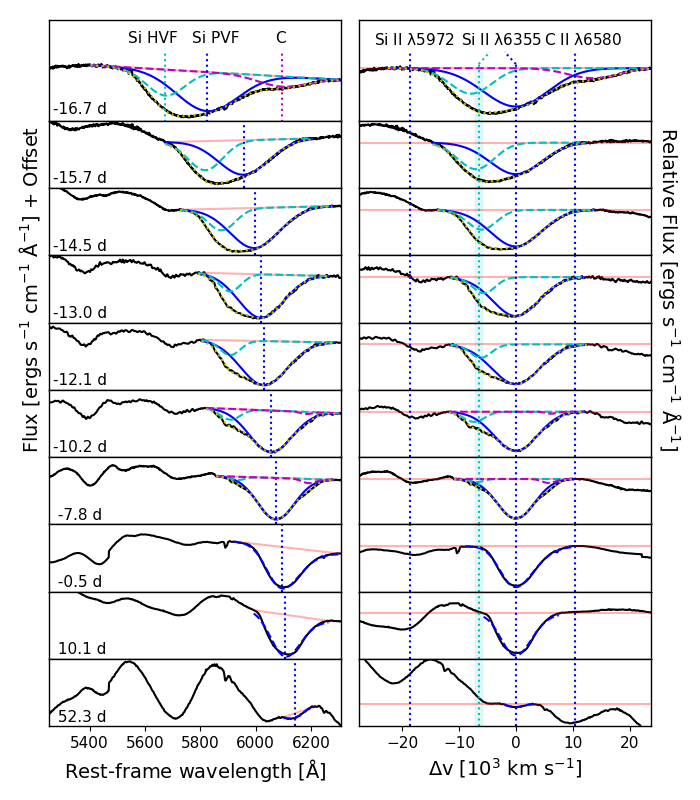}
\end{center}
\caption{(Left) Spectra of \uname\ (black) in the vicinity of the \siii~$\lambda$6355~\AA\ feature compared to multi-component Gaussian fits (yellow dotted curve) with PVF (blue solid curves), HVF (cyan dashed curves) and \cii\ (magenta dashed curves) components at ten evolutionary phases, as labelled on the bottom-left in rest-frame days since $B$-band maximum. Rows (1--3) are during the early PVF and HVF decline, (4--6) span the HVF plateau (see Figure~\ref{fig:siexp}), (7) is the last epoch when the HVF is detected, (8) is at $B$-band maximum, (9) has the largest \chisqr\ (see Table~\ref{tab:sifit}), and (10) is the last phase when the \siii~$\lambda$6355~\AA\ feature is detected. The red shaded line is the estimated continuum of the feature, while the vertical dotted lines indicate the fitted minima of the correspondingly colored Gaussian components. (Right) The same spectra and fits, but the wavelength axis is converted to velocity in the rest frame of the fitted PVF minima at each phase (middle blue vertical dotted line at 0~km~s$^{-1}$). 
The cyan vertical dotted line with shaded region represents the mean and 1-$\sigma$ velocity of the fitted HVF minima in this frame.
The blue vertical dotted lines flanking the PVF and HVF are the expected positions of the photospheric-velocity \siii~$\lambda$5972~\AA\ and \cii~$\lambda$6580~\AA\ features, as labelled at the top of the first panel. 
\cii~$\lambda$6580~\AA\ is detected in only a few spectra (see Table~\ref{tab:sifit}) near this photospheric velocity.
\label{fig:sifit}}
\end{figure*}

The Doppler shifts of the \siii\ $\lambda$6355~\AA\ absorption feature \citep[or ``\siii\ velocity'';][]{Parrent2014apss, tanaka2008ap, Wang2009apj, Ni2022natas}, which persists until $\sim$ 30 days after $B$-band maximum, 
have been used for studying the velocity evolution of \tas\ ejecta in the photosphere.
Some \tase\ have shown deep and broad \siii\ features in the first few days post-explosion
whose absorption profiles are best-fit with two velocity components \citep[e.g., SNe~2009ig and 2012fr;][]{Marion2013apj, Contreras2018apj}, suggesting the presence of a short-lived ``high-velocity feature'' (HVF) in addition to the typical photospheric-velocity feature (PVF) that is seen in all \tase.

Figure~\ref{fig:specevol} shows three spectra of \uname---including two of the earliest obtained by SALT \citep{Bostroem2021tnscr, Ashall2022apj, Hosseinzadeh2022apj} and our spectrum from nearest the epoch of $B$-band maximum---compared to those of other normal \tase. 
\citet{Hosseinzadeh2022apj} reported that the earliest spectrum of \uname\ from $-$16.8 days had an unusually deep and broad absorption feature around 570~nm in the rest frame, suggesting possible contributions from HVF and PVF features of \siii\ $\lambda$6355~\AA\ and/or blending with another \siii\ line at $\lambda$5972~\AA.
By $-$15.8 days, the 570~nm feature of \uname\ appears nearly identical to that of the BL/HV \tas~2009ig at a similar phase ($\sim -$15 days since $B$-band maximum).
SN~2009ig exhibits clear HVFs and PVFs of \siii\ $\lambda$6355~\AA\ \citep{Marion2013apj}, indicating that the shape of the 570~nm feature in \uname\ may also be attributable to HVF and PVF components of the same line.
At $B$-band maximum, the SN shows relatively blended \fex\ absorption features resembling BL SNe~1981B and 2009ig, while the \siii\ features match those of SN~2018aoz, which is an intermediate event between CN and BL (see Section~\ref{subsec:class} below for the precise subtype classification).

We fit the observed \siii\ $\lambda$6355~\AA\ feature in the spectra of \uname\ with multiple Gaussian functions
in order to characterize the velocity evolution of the HVF and PVF components,
following the method in \citet{Silverman2015mnras}.
First, the spectral continuum is measured by fitting the relatively broad local maxima on the red and blue edges of the feature 
with concave-downward quadratic functions,
where the peaks of the quadratics representing the local maxima  
are considered to be the red and blue endpoints of the feature.
The spectral continuum is adopted to be the line that connected the red and blue endpoints (see Figure~\ref{fig:sifit}, red shaded lines).
When the quadratic peaks appear to measure the feature endpoints incorrectly by either being too far inside or outside of the feature, the endpoints are selected manually.
Second, a sum of one, two, or three Gaussians is fitted to the residual of the spectrum after subtracting the spectral continuum.
For one-Gaussian fitting, a skewed Gaussian profile is used in order to better approximate the expected shape (P cygni) of isolated SN absorption features,
whereas a symmetric one is used for each component of multiple (2 or 3) Gaussian fitting in order to mitigate over-fitting. 
The mitigation of over-fitting is seen in the \chisqr\ values of the fits (Table~\ref{tab:sifit}), where \chisqr\ is similar between the one-component and multi-component fits.
Finally, we obtain the velocities of each component by applying the relativistic Doppler effect formula to the fitted wavelengths of the Gaussian minima.

Figure~\ref{fig:sifit} (left panels) shows the multi-component Gaussian fits to the \siii\ $\lambda$6355~\AA\ feature for a sample of the spectra of \uname. 
The bottom-most panel shows the last detection of the feature at 3-$\sigma$ confidence, where $\sigma$ is estimated by the rms noise of the residual after smoothing the spectrum using a second-order Savitsky-Golay filter with a width of 100~\AA.
(Note that the feature width is $>$ 100~\AA\ in all epochs).
For each fit, the subtracted continuum level (transparent red lines), as well as the PVF (blue solid curves) and HVF (cyan dashed curves) components of the fit are shown.
For the fourteen earliest spectra $\leq -$7.8 days, we find that a single Gaussian inadequately fits the shape of the feature while a multi-component Gaussian fit results in a (2.5--12)$\times$ improvement in \chisqr.
In particular, five of those early spectra require a third Gaussian component (magenta dashed curves) to accommodate the presence of an additional absorption feature at the red edge of the feature, resulting in (1.3--7.6)$\times$ improvements in \chisqr\ over the two-component fits.
This third Gaussian component is usually attributed to \cii\ $\lambda$6580~\AA\ in \tase\ \citep[e.g., see][]{Silverman2015mnras}.
Of the two remaining components, the one closer to the red edge of the feature is identifiable as the \siii\ PVF based on its velocity evolution (see Section~\ref{subsec:expl} and Figure~\ref{fig:siexp}), associating the one closest to the blue edge of the feature to be the \siii\ HVF. The absence of other likely contributing lines in the vicinity (Section~\ref{subsec:hvfid}) and the resulting \siii\ velocities further support this association.
Our measured \siii\ PVF and HVF velocities for \uname\ have a mean separation of $\Delta\varv$ = (6500 $\pm$ 600) km~s$^{-1}$, consistent with separations found in other normal \tase\ with HVFs \citep[$\Delta\varv\sim-$~6000~km~s$^{-1}$;][]{Silverman2015mnras}. The PVF attains a velocity of (12,440 $\pm$ 20)~km~s$^{-1}$ at $B$-band maximum, which is a typical normal \tas\ ejecta velocity (Section~\ref{subsec:class}), while the HVF velocities exceed 20,000~km~s$^{-1}$ in all epochs.

For the sixteen near- and post-peak spectra following $-$7.8 days, we are unable to identify the HVF.
Using two Gaussians in those epochs results in unreliable fits, with PVF and HVF separations ranging in (1100--3100)~km~s$^{-1}$ and HVF velocities $<$ 15,000~km~s$^{-1}$ in all epochs.
Note that these differ substantially from the PVF and HVF properties seen in prior epochs (see above),
and that the fitting method of \citet{Silverman2015mnras} is unreliable for distinguishing velocity components that are so closely separated.
Thus, following $-$7.8 days, we adopt a single Gaussian which appears to adequately fit the feature and gives acceptable \chisqr\ values, even for the spectrum from 10.1 days (second last row in the figure) which has the highest \chisqr\ of 14.2 among the sixteen spectra.
The measured PVF, HVF, and \cii\ velocities for all of the spectra are presented in Table~\ref{tab:sifit}. Figure~\ref{fig:siexp} shows the velocity evolution (see Section~\ref{subsec:expl}).

\begin{deluxetable*}{lccccccccc}
\tabletypesize{\footnotesize}
\tablecolumns{10} 
\tablewidth{0.99\textwidth}
 \tablecaption{Multi-component line velocity measurements in \uname.}
 \tablehead{
 \colhead{Date (MJD)} & \colhead{Phase$^{\rm a}$} & \colhead{Telescope} & \colhead{\chisqr$_{\rm Si}$} & \colhead{\siii\ PVF} & \colhead{\siii\ HVF} & \colhead{\cii\ $\lambda$6580~\AA} & \colhead{\chisqr$_{\rm Ca}$} & \colhead{\caii\ PVF} & \colhead{\caii\ HVF}
 }
\startdata 
59529.86 & $-$16.71 & SALT & 2.1 & $-$26.04 $\pm$ 0.29 & $-$33.86 $\pm$ 0.11 & $-$22.90 $\pm$ 0.24 & 1.6 & $-$31.97 $\pm$ 0.21 & $-$40.77 $\pm$ 0.10 \\
59530.87 & $-$15.70 & SALT & 5.5 & $-$19.42 $\pm$ 0.13 & $-$26.39 $\pm$ 0.07 & & 1.9 & $-$26.66 $\pm$ 0.18 & $-$37.62 $\pm$ 0.06 \\
59531.12 & $-$15.47 & SOAR & 2.3 & $-$18.86 $\pm$ 0.25 & $-$25.63 $\pm$ 0.12 & & 1.8 & $-$25.56 $\pm$ 0.29 & $-$36.76 $\pm$ 0.09 \\
59531.70 & $-$14.89 & FTS & 2.4 & $-$17.67 $\pm$ 0.15 & $-$23.83 $\pm$ 0.08 & & 1.3 & $-$23.79 $\pm$ 0.34 & $-$35.39 $\pm$ 0.17 \\
59532.07 & $-$14.52 & SALT & 5.9 & $-$17.43 $\pm$ 0.06 & $-$23.40 $\pm$ 0.03 & &  &  &  \\
59532.84 & $-$13.75 & SALT & 5.2 & $-$16.74 $\pm$ 0.06 & $-$22.47 $\pm$ 0.03 & &  &  &  \\
59533.63 & $-$12.98 & FTS & 2.4 & $-$16.28 $\pm$ 0.07 & $-$21.88 $\pm$ 0.04 & & 1.4 & $-$22.53 $\pm$ 0.16 & $-$32.37 $\pm$ 0.06 \\
59533.86 & $-$12.74 & SALT & 3.8 & $-$16.08 $\pm$ 0.05 & $-$21.98 $\pm$ 0.03 & &  &  &  \\
59534.50 & $-$12.11 & FTS & 3.1 & $-$15.72 $\pm$ 0.05 & $-$21.88 $\pm$ 0.04 & & 1.4 & $-$22.93 $\pm$ 0.32 & $-$31.09 $\pm$ 0.11 \\
59535.07 & $-$11.53 & SALT & 7.5 & $-$15.35 $\pm$ 0.02 & $-$21.73 $\pm$ 0.03 & &  &  &  \\
59536.45 & $-$10.17 & FTS & 1.5 & $-$14.53 $\pm$ 0.02 & $-$21.86 $\pm$ 0.05 & $-$18.68 $\pm$ 0.13 & 1.3 & $-$13.37 $\pm$ 0.14 & $-$26.00 $\pm$ 0.06 \\
59536.86 & $-$9.76 & SALT & 2.0 & $-$14.30 $\pm$ 0.01 & $-$21.21 $\pm$ 0.04 & $-$18.79 $\pm$ 0.03 &  &  &  \\
59538.07 & $-$8.55 & SALT & 1.5 & $-$13.85 $\pm$ 0.01 & $-$20.64 $\pm$ 0.03 & $-$18.46 $\pm$ 0.03 &  &  &  \\
59538.85 & $-$7.77 & SALT & 0.8 & $-$13.60 $\pm$ 0.01 & $-$20.53 $\pm$ 0.03 & $-$18.25 $\pm$ 0.02 &  &  &  \\
59546.12 & $-$0.54 & Gemini S & 7.0 & $-$12.44 $\pm$ 0.02 & & & 8.0 & $-$11.95 $\pm$ 0.04 & $-$19.31 $\pm$ 0.09 \\
59550.60 & $+$3.91 & FTS & 13.8 & $-$11.95 $\pm$ 0.05 & & & 1.1 & $-$12.76 $\pm$ 0.08 &  \\
59556.81 & $+$10.09 & SALT & 14.2 & $-$11.92 $\pm$ 0.09 & & &  &  &  \\
59558.48 & $+$11.75 & FTS & 10.7 & $-$11.73 $\pm$ 0.08 & & & 1.1 & $-$11.51 $\pm$ 0.05 &  \\
59559.09 & $+$12.35 & SOAR & 9.3 & $-$11.81 $\pm$ 0.07 & & & 40.9 & $-$11.67 $\pm$ 0.03 &  \\
59560.99 & $+$14.25 & SALT & 10.8 & $-$11.58 $\pm$ 0.05 & & &  &  &  \\
59561.48 & $+$14.73 & FTS & 7.6 & $-$11.52 $\pm$ 0.07 & & & 1.3 & $-$11.08 $\pm$ 0.04 &  \\
59564.50 & $+$17.73 & FTS & 7.6 & $-$11.18 $\pm$ 0.07 & & & 1.2 & $-$10.98 $\pm$ 0.05 &  \\
59567.67 & $+$20.89 & FTS & 4.8 & $-$10.67 $\pm$ 0.33 & & & 1.0 & $-$11.17 $\pm$ 0.10 &  \\
59568.01 & $+$21.23 & SALT & 5.5 & $-$10.81 $\pm$ 0.14 & & &  &  &  \\
59572.54 & $+$25.74 & FTS & 0.5 & $-$10.69 $\pm$ 0.06 & & & 2.1 & $-$10.68 $\pm$ 0.06 &  \\
59577.48 & $+$30.65 & FTS & 1.0 & $-$10.44 $\pm$ 0.16 & & & 2.2 & $-$10.40 $\pm$ 0.06 &  \\
59582.54 & $+$35.68 & FTS & 1.1 & $-$10.66 $\pm$ 0.38 & & & 3.9 & $-$10.75 $\pm$ 0.06 &  \\
59589.59 & $+$42.70 & FTS & 1.3 & $-$10.44 $\pm$ 0.36 & & & 1.9 & $-$10.56 $\pm$ 0.06 &  \\
59594.59 & $+$47.67 & FTS & 1.0 & $-$10.17 $\pm$ 0.86 & & & 1.4 & $-$10.35 $\pm$ 0.09 &  \\
59599.20 & $+$52.29 & Gemini S & 1.9 & $-$10.22 $\pm$ 0.67 & & & 59.1 & $-$10.38 $\pm$ 0.04 &  \\
\enddata
\tablenotetext{{\rm a}}{Phase is rest-frame days since $B$-band maximum ($t_{\rm max}$ = MJD 59546.67; Table~\ref{tab:param}).}
\tablecomments{Two of these spectra are from Gemini S (see Table~\ref{tab:spec}), while the remainder are publicly available spectra from SALT, SOAR, and FTS \citep{Hosseinzadeh2022apj}. Columns 4--7 provide the best-fit \chisqr\ and measured velocities of the PVF, HVF, and \cii\ $\lambda$6580~\AA\ components of the \siii\ $\lambda$6355~\AA\ feature, respectively, from multi-component Gaussian fitting as detailed in Section~\ref{subsec:sifit}, while columns 8--10 provide those from fitting the \caii\ NIR triplet with PVF and HVF components (Appendix~\ref{sec:cafit}). All velocities are presented in 10$^3$~km~s$^{-1}$.} 
\end{deluxetable*} 
\label{tab:sifit}

\subsection{Existence of a High-Velocity Secondary Photosphere}\label{subsec:hvfid}

We determine that the observed HVF of the \siii\ $\lambda$6355~\AA\ feature in \uname\ is formed by a high-velocity ``secondary photosphere''---distinct from the ``primary photosphere'' of the SN ejecta that produces the \siii\ PVF---as follows.
First, the observed \siii\ HVF is unlikely to be produced by any other line than \siii~$\lambda$6355~\AA.
The right panels of Figure~\ref{fig:sifit} show the same spectra as the ones in the left panels after subtracting the spectral continuum and shifting the wavelength axes to velocities in the rest frame of the SN photosphere. 
This rest frame is typically adopted to be that of the \siii\ $\lambda$6355~\AA\ PVF (blue solid curves) minimum \citep{tanaka2008ap, Piro&Nakar2014apj}, represented by the middle vertical blue line at 0~km~s$^{-1}$ in the figure.
The observed \siii\ HVF (cyan dashed curves) minima occupy a narrow range of velocities to the left of the PVFs.
The vertical cyan dotted line with a shaded region represents the measured mean separation and 1-$\sigma$ range of the observed HVFs from the PVFs ($\Delta\varv$ = $-$6500 $\pm$ 600 km~s$^{-1}$; Section~\ref{subsec:sifit}).
The vertical blue dotted lines flanking the HVFs and PVFs on the left and right sides represent the expected locations of the nearest other lines, \siii\ $\lambda$5972~\AA\ and \cii\ $\lambda$6580~\AA, respectively, in the photosphere (i.e., at PVF velocity) that can contribute to the observed \siii\ $\lambda$6355~\AA\ feature.
As seen in the figure, in order for those lines to be responsible for the observed HVF, \siii\ and \cii\ would need to be expanding $\sim$ 12,000~km~s$^{-1}$ slower and $\sim$ 17,000~km~s$^{-1}$ faster than the photosphere, respectively, making it much more likely for the HVF to be a $\sim$ 6,500~km~s$^{-1}$ faster component of \siii~$\lambda$6355~\AA.

Second, we identify a component of the \caii\ NIR triplet that traces the velocity evolution of the \siii\ HVF in the early phase ($\lesssim -$~12 days since $B$-band maximum; see Appendix~\ref{sec:cafit} and \caii\ PVF in Table~\ref{tab:sifit} and Figure~\ref{fig:siexp}).
The co-evolution of this \caii\ component with the \siii\ HVF (together called ``co-evolving HV Ca and Si components'' hereafter) indicates that both components were likely formed in a high-velocity secondary photosphere, which produces a broadband SN spectrum at their common velocity.
The simultaneous appearance of both PVFs and HVFs of \siii\ in the earliest spectra indicates that the overall photosphere of \uname\ is non-spherically symmetric in the outer layers of the ejecta, resulting in primary and secondary photospheric velocities integrated along different sightlines.
In Section~\ref{sec:velocity}, we investigate the origin of the \siii\ HVF in detail.

\subsection{\siii~Velocity Evolution and Epoch of Explosion}\label{subsec:expl}

\begin{figure}[t!]
\epsscale{\scl}
\plotone{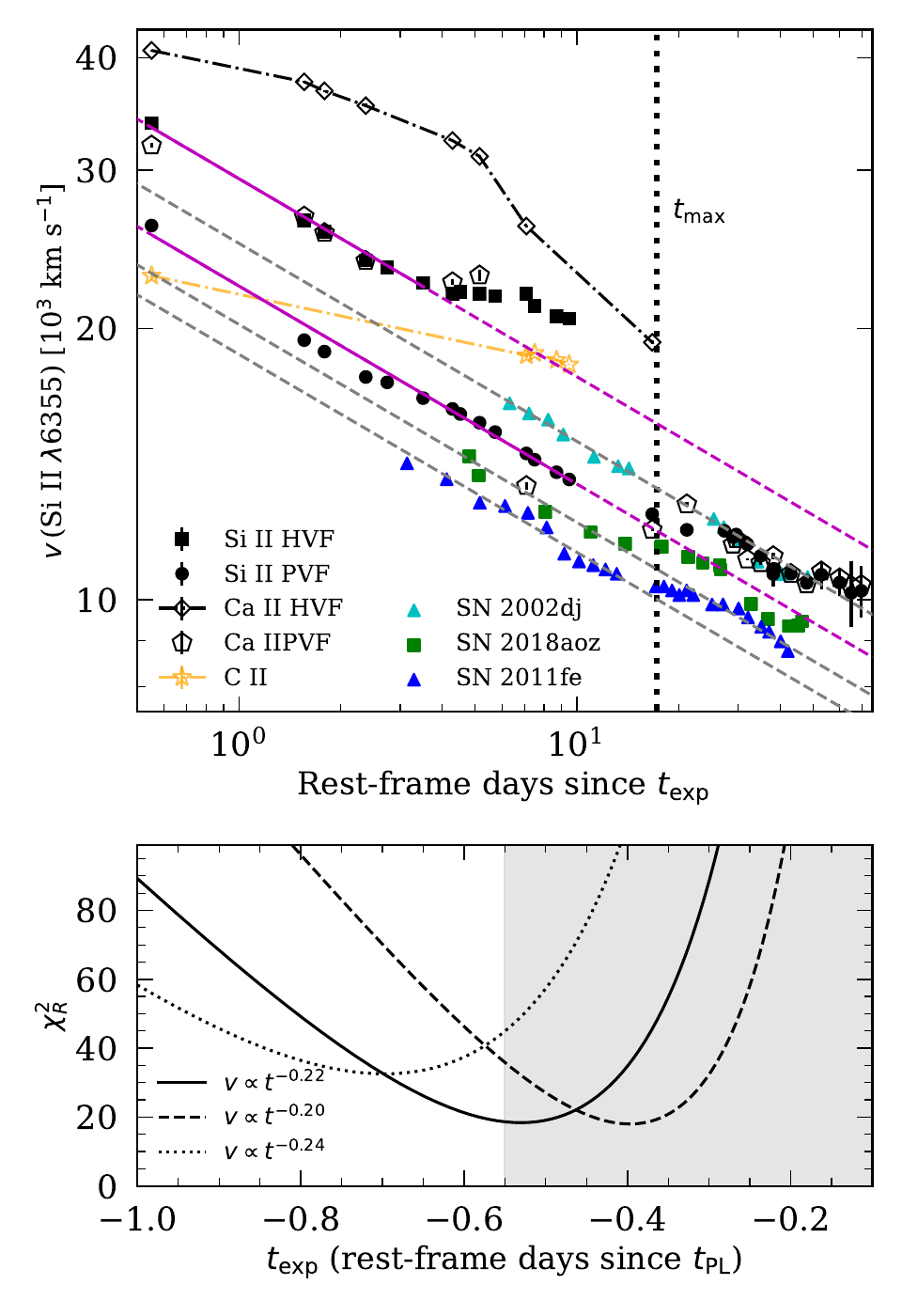}
\caption{(Top) The observed velocity evolution of the \siii\ PVF and HVF of \uname\ is compared to those of other normal \tase, those of the \caii\ NIR triplet PVF and HVF (Appendix~\ref{sec:cafit}) and the \cii\ feature of \uname, and the best-fit $\beta$ = $-0.22$ power-law (solid magenta lines with dashed extrapolations).
The x-axis represents time in days since the fitted explosion epoch (= $t_{\rm exp}$) obtained using this power-law.
The grey dashed lines show the $t^{-0.22}$ evolution of the pre-peak \siii\ velocities in the other SNe. (Bottom) \chisqr\ of fitting $t^{-\beta}$ power-laws to the \siii\ PVF velocity evolution of \uname\ before $B$-band maximum and the \siii\ HVF velocity evolution before the onset of its HVF plateau ($\sim$ 3.6 rest-frame days post-explosion).
Note that $\beta$ = $-0.22$ is the theoretically expected photospheric evolution for \tas\ ejecta.
\label{fig:siexp}}
\end{figure}

We measure the explosion epoch of \uname\ by fitting the photospheric velocity evolution traced by the \siii\ HVFs and PVFs (or ``\siii\ velocity''; Section~\ref{subsec:hvfid}) with power-law models of the form $\varv_{\rm Si} \propto (t - t_{\rm exp})^{-\beta}$, where $\varv_{\rm Si}$ is the \siii\ velocity and $(t - t_{\rm exp})$ is time (days) since the epoch of explosion.
Such a power-law model with index $\beta$ = 0.22 is theoretically expected for a homologously expanding \tas\ with polytropic (n = 3) ejecta structure \citep{Piro&Nakar2013apj}, and has been found to be a good fit to the pre-peak $\varv_{\rm Si}$ evolution in other \tase\ \citep{Piro&Nakar2014apj, Ni2022natas}.
As seen in Figure~\ref{fig:siexp} (top panel), the \siii\ HVF (black filled squares) and PVF (black filled circles) velocities both follow $\beta$ = 0.22 power-law declines (magenta dashed lines) in the earliest phase, consistent with them both initially originating from photospheres of homologously expanding polytropic material in the SN ejecta.
The features eventually plateau, the HVF after $\sim$ 3.6 days and the PVF after $B$-band maximum,
indicating that the expansion of \uname\ has revealed the bulk of this polytropically-distributed material by those phases.
In particular, since the PVF originates from the primary SN ejecta that is co-distributed with centrally-concentrated \ni56\ responsible for the \tas\ optical emission near peak, 
the onset of the PVF plateau is understood to coincide with $B$-band maximum.

The bottom panel of Figure~\ref{fig:siexp} shows the \chisq\ distribution (black solid curve) obtained by fitting only the \siii\ HVF and PVF velocity measurements from before their respective plateau times, which results in the best-fit $t_{\rm exp}$ of $-$0.53 $\pm$ 0.04 days since $t_{\rm PL}$ (or MJD 59529.32 $\pm$ 0.04 days).
This time difference between the epochs of explosion and the onset of power-law rise in \tase\ has been referred to as the ``dark phase'', and our measurement is similar to those from other normal \tase\ 2011fe \citep[$\sim$ 0.5 days;][]{Piro&Nakar2014apj} and 2018aoz \citep[$\sim$ 0.4 days;][]{Ni2022natas}.
The top panel shows that the $\beta$ = $-0.22$ power-law (magenta solid lines) adequately fits the HVF and PVF declines with a best-fit HVF/PVF velocity ratio of 1.32 and \chisqr\ = 18.
Following \citet{Piro&Nakar2014apj}, we estimate the systematic error associated with the assumed $\beta$ by adopting nearby values of 0.20 and 0.24, resulting in variations of 0.15 days in the inferred $t_{\rm exp}$ (see Figure~\ref{fig:siexp}, bottom panel).
The quality of the fit does not change significantly when $\beta$ is decreased to 0.20, although \chisqr\ increases to $\sim$ 33 when $\beta$ is increased to 0.24.
We adopt the conservative uncertainty of 0.16 days for the explosion epoch of MJD 59529.32 (Table~\ref{tab:param}), summing the uncertainties from fitting the $\beta$ = 0.22 power-law and varying the power-law index independently.

\subsection{Summary of Important Epochs for Early Evolution}\label{subsec:earlysum}

We summarize the important epochs of the early evolution of \uname\ from our analyses above as follows.

\begin{enumerate}
    \item The power-law rise of \uname\ began at $t_{\rm PL}$ = MJD 59529.85 $\pm$ 0.55, associated with the onset of emission from the main distribution of centrally-concentrated \ni56\ in the ejecta.
    While the technique of power-law fitting is often used to estimate the epoch of "first light" in \tase, it likely over-estimates the onset of emission in \uname\ (see point 2 below).
    
    \item There is evidence for excess emission beginning prior to $t_{\rm PL}$. The bulk of this excess emission peaks in the ``excess emission phase'' prior to 1.75 days since $t_{\rm PL}$ in the rest frame, during which the SN undergoes redward \bv\ color evolution, and in the optical bands.

    \item By fitting the early velocity evolution of \uname, we estimate an explosion epoch of MJD 59529.32 $\pm$ 0.04 days, $\sim$ 0.5 days prior to $t_{\rm PL}$ and 1.0 days prior to the peak of the early excess emission in rest frame.

    \item The spectra of \uname\ show both HVFs and PVFs of \siii\ $\lambda$6355~\AA\ until $-$7.8 days since $B$-band maximum in rest frame (= 8.9 days since $t_{\rm PL}$), and PVFs alone until 52.3 days. \cii\ is seen in a few epochs prior to the disappearance of the HVFs.

    \item The \siii\ HVF and PVF velocities of \uname\ both follow the power-law decline expected from homologous expansion of polytropic (n=3) \tas\ ejecta for $\sim$ 3.6 days post-explosion, after which the HVF velocity plateaus. The PVF velocity plateaus near $B$-band maximum, as expected for \tase.
\end{enumerate}

\begin{figure*}[t!]
\epsscale{\scl}
\begin{center}
\includegraphics[width=0.95\textwidth]{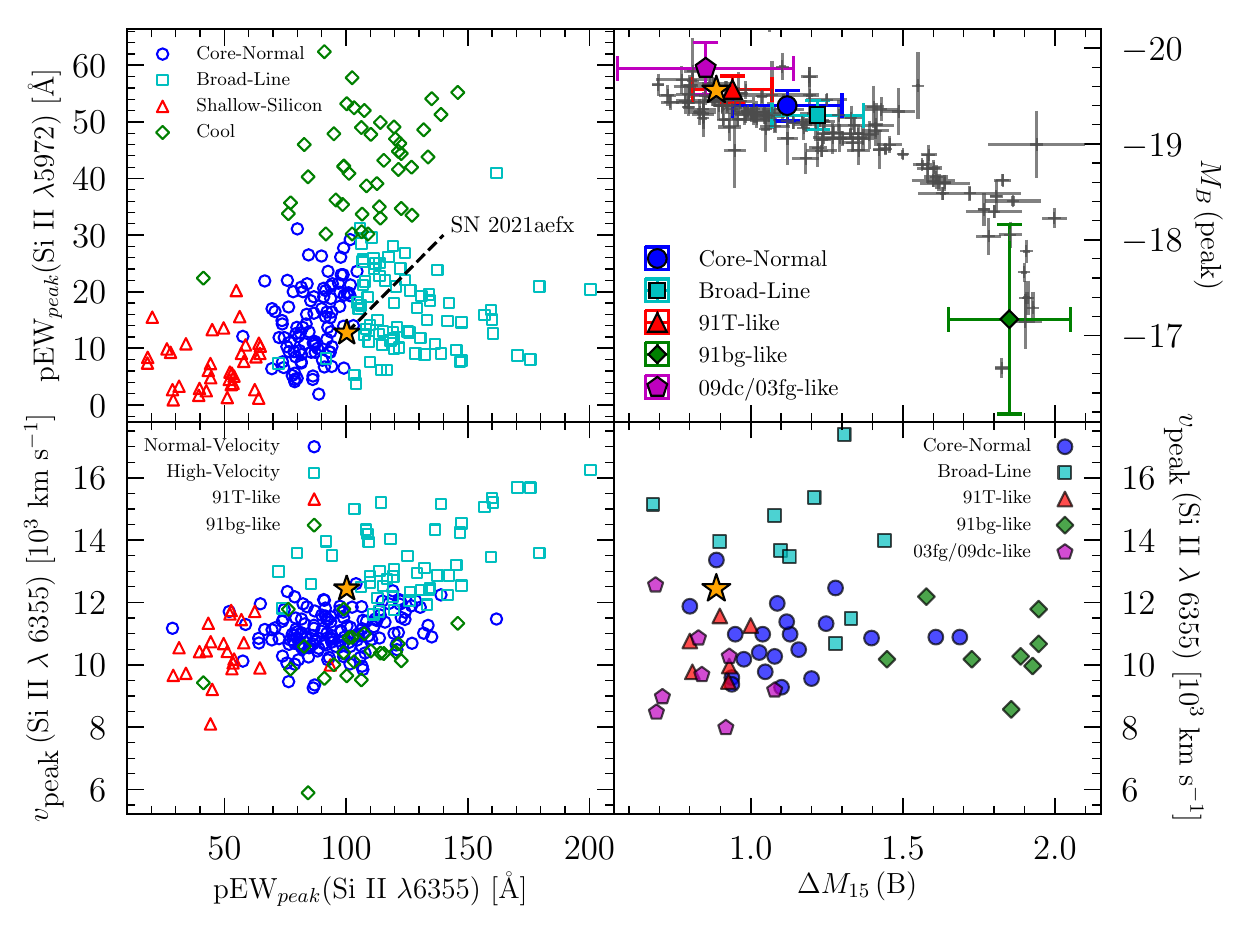}
\end{center}
\caption{(Top-Left) \citet{Branch2006pasp} classification diagram comparing the pseudo-equivalent widths of \siii\ lines of \uname\ from a spectrum near $B$-band maximum ($-$0.5 days; orange star) with those a population of \tase\ \citep{Blondin2012aj}. The colored symbols represent events from the four main subtypes of \tase: CN (blue circles), BL (cyan squares), Shallow-Silicon (red triangles), and Cool (green diamonds).
Note that CN and BL are both subsets of normal \tase, while Shallow-Silicon and Cool are considered peculiar.
(Bottom-Left) \citet{Wang2009apj} classification diagram comparing the \sii~$\lambda$6355~\AA\ equivalent width and Doppler velocity of \uname\ from near $B$-band maximum (orange star) with those of the same population of \tase\ from the top-left panel.
The NV, HV, 91T-like, and 91bg-like subtypes represented by the colored symbols approximately correspond to the CN, BL, Shallow-Silicon, and Cool subtypes respectively.
(Top-Right) Comparison of the $M_B$ (peak) and \dm15\ of \uname\ (orange star) with those of other \tase\ \citep[grey crosses;][]{Burns2018apj}.
The average values for the four main subtypes of \tase\ are represented by their correspondingly colored symbols, while the magenta pentagon represents the average for the extremely slow-declining and over-luminous events once thought be from super-Chandrasekhar-mass explosions \citep{Parrent2014apss} called ``03fg/09dc-like'' \citep{Taubenberger2019mnras, Ashall2020apjl}.
(Bottom-Right) Comparison of the \dm15\ and peak \siii\ velocity of \uname\ (orange star) with a population of other \tase\ \citep{Parrent2014apss} with colored symbols representing the same subtypes as in the top-right panel.
\label{fig:class}}
\end{figure*}

\section{Supernova Classification and Explosion Parameters}\label{sec:peak}

\subsection{Spectroscopic Classification} \label{subsec:class}

\tase\ are typically classified using their spectra near the epoch of $B$-band maximum light based on the percent equivalent widths of their \siii\ $\lambda$6355~\AA\ and $\lambda$5972~\AA\ absorption features \citep[``pEW$_{\rm peak}$'';][]{Branch2006pasp} and \siii\ velocity \citep[$\varv_{\rm peak}$;][]{Wang2009apj}.
Using our spectrum of \uname\ obtained from $-$0.5 days since the epoch of $B$-band maximum, we measure pEW$_{\rm peak}$ (\siii~$\lambda$6355~\AA) and pEW$_{\rm peak}$ (\siii~$\lambda$5972~\AA) of 100.3 $\pm$ 1.0 and 12.7 $\pm$ 0.5~\AA, respectively, using the method of \citet{Branch2006pasp}, as well as $\varv_{\rm peak}$ (\siii~$\lambda$6355~\AA) = (12.44 $\pm$ 0.02) $\times$ 10$^3$~km~s$^{-1}$ (see Table~\ref{tab:sifit}).
Figure~\ref{fig:class} (left panels) compares these peak parameters to those of a sample of \tase\ from the CN/NV (blue circles) and BL/HV (cyan squares) normal subtypes as well as the 91T-like (``or Shallow Silicon''; red triangles) and 91bg-like (or ``Cool''; green diamonds) peculiar subtypes.
\uname\ appears to be on the boundary between the CN/NV and BL/HV events, while it is clearly inconsistent with the peculiar events.
Thus, we classify \uname\ as a normal \tas\ that is intermediate between the CN/NV and BL/HV subtypes, confirming the classifications based on spectra from pre-peak \citep[$-$8 days;][]{Ashall2022apj} and post-peak \citep[$+$4 days;][]{Hosseinzadeh2022apj} epochs.

The light curves of \uname\ appear to be relatively slow-declining and over-luminous among \tase\ with normal spectroscopic classification.
Figure~\ref{fig:class} (right panels) compares the $B$-band peak absolute magnitude, \dm15, and $\varv_{\rm peak}$ (\siii~$\lambda$6355~\AA) of \uname\ with those of other \tase\ from the CN, BL, 91T-like, and 91bg-like subtypes, as well as the extremely luminous and slow-declining \tase\ from the 03fg/09dc-like subtype once thought to be from super-Chandrasekhar-mass explosions \citep{Taubenberger2019mnras, Ashall2020apjl}.
The template-fitted DM of \uname\ (Section~\ref{subsec:templ}) places its light curves
near the slow and luminous boundary of the normal events, consistent with 91T-like and 03fg/09dc-like events, while less accurate DM estimates allow even higher luminosities (Section~\ref{subsec:host}).
However, we note that the extreme 03fg/09dc-like events, in general, lack $i$-band secondary maxima and reach $i$-band primary maxima at a later phase than $B$-band maximum \citep{Ashall2021apj}, both of which are inconsistent with \uname.
A super-Chandrasekhar ejecta mass is also less likely for \uname\
based on its bolometric luminosity and UV light curves as detailed in Section~\ref{subsec:bolo} below.

\subsection{Bolometric Light Curve and SN Explosion Parameters} \label{subsec:bolo}

\begin{figure}[t!]
\epsscale{\scl}
\plotone{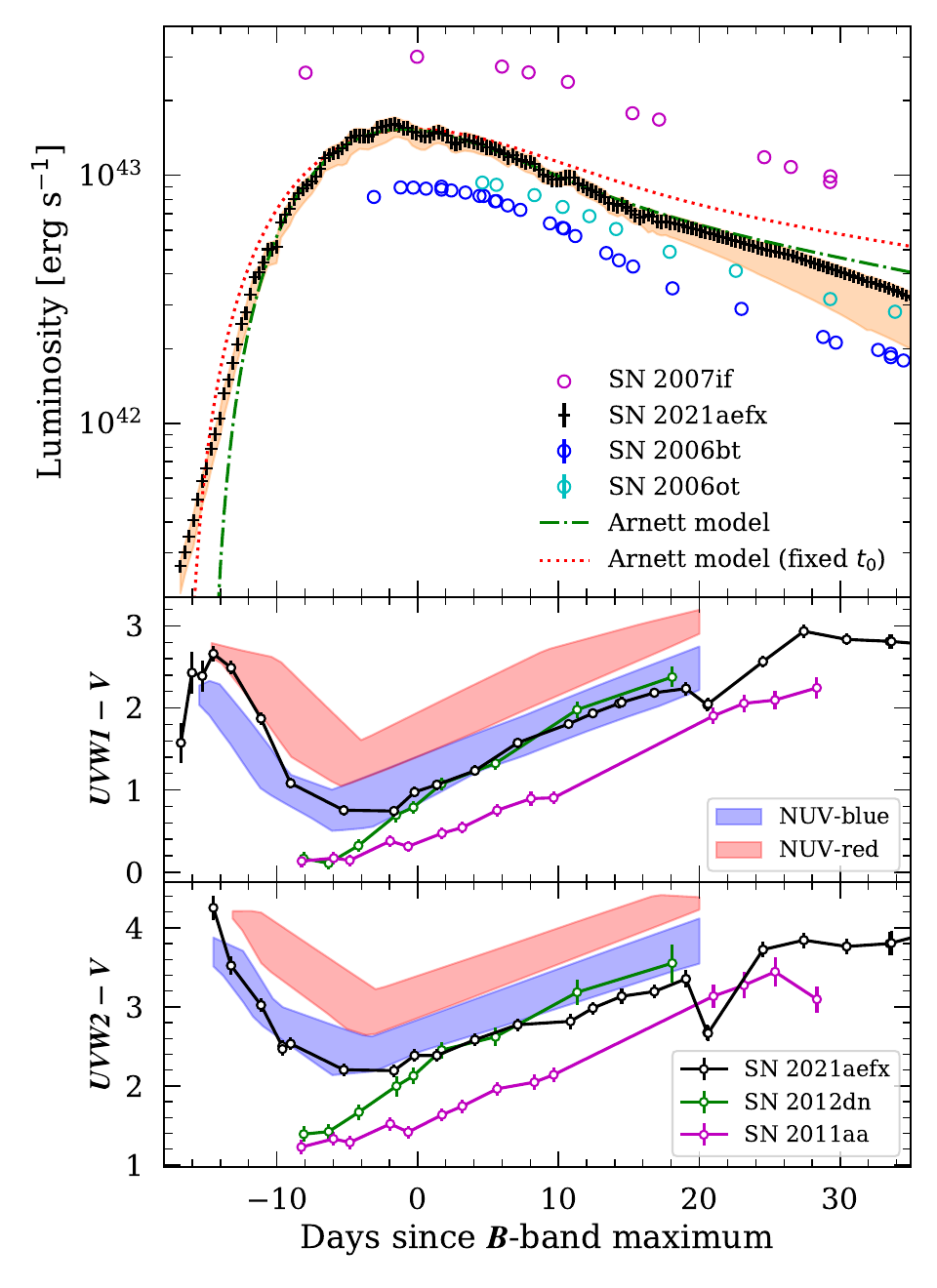}
\caption{(Top) UVOIR bolometric light curve of \uname\ (black crosses) are compared to those of three events with super-Chandrasekhar inferred ejecta masses \citep{Scalzo2019mnras} and two models: (1) Arnett model \citep{Arnett1982apj} matched to the rise time and peak luminosity of the bolometric light curve (red dotted
curve); and (2) Arnett model fitted to the bolometric light curve near peak (green dot-dashed curve).
The bounds of the pumpkin colored region corresponds to the larger of two uncertainties: (1) the 1-$\sigma$ uncertainty level of the bolometric light curve; and (2) the potential uncertainty from using the best-fit SNooPy template to complete the UVOIR waverange where observations are absent.
(Middle and Bottom) The observed (non-dereddened) UV-optical colors of \uname\ (black open circles) are compared to those of the 03fg/09dc-like events SNe~2012dn and 2011aa \citep[green and magenta circles, respectively;][]{Brown2014apj}, and the expectations for the NUV-red/blue groups of normal \tase\
\citep[red/blue colored shaded areas, respectively;][]{Milne2013apj}.
\label{fig:uvcolor}}
\end{figure}

During the photospheric phase within the first $\sim$ 30 days after
the explosion, the majority of \tas\ emission is expected to fall within the
ultraviolet–optical–infrared (UVOIR) waverange \citep{Contardo2000aap}.
Figure~\ref{fig:uvcolor} (top panel) presents the bolometric light curve of \uname\ obtained by integrating the spectral energy distribution (SED) constructed from the KSP $BVi$-band light curves and Swift UV light curves \citep{Hosseinzadeh2022apj} over the UVOIR waverange from $UVM2$ to $K$ band (2246 to 23763~\AA).
The waverange is completed where observations are lacking by extrapolating the best-fit SNooPy template (\sbv\ = 1.05; see Section~\ref{subsec:templ}) following the method of \citet{Ni2022natas}, which accounts for $<$ 20\% of the total bolometric luminosity within the photospheric phase. 

We compare the bolometric light curve of \uname\ to the \ni56-powered SN light curve model of \citet{Arnett1982apj} in order to estimate the \ni56\ mass, ejecta mass, and ejecta kinetic energy of the explosion.
In the case where the distribution of \ni56\ is strongly peaked toward the center of the ejected mass and the ejecta opacity is
constant, the model predicts the radioactively powered luminosity of a \tas\ during the photospheric phase as a self-similar solution with the luminosity axis scaled by the \ni56\ mass ($M_{\rm Ni}$) and the time axis scaled by the geometric mean of the diffusion and
expansion timescales \citep[$\tau_m$;][]{Ni2022natas, Ni2023apj}:
\begin{equation}
    \tau_m = \left(\frac{\kappa}{13.8\ c}\right)^{1/2} \left(\frac{6M_{\rm ej}^3}{5 E_{\rm ej}}\right)^{1/4}
    \label{eq:taum}
\end{equation}
\citep[Note that the opacity of \tase\
in the photospheric phase is expected to be nearly constant $\sim$ 0.1~cm$^2$~g$^{-1}$;][]{Pinto&Eastman2000apj}.
We fit the model for two cases: (1) the onset of the model is set to be equal to $t_{\rm PL}$ measured in Section~\ref{subsec:gaus}; and (2) the onset is allowed to vary.

In case (1), the model (Figure~\ref{fig:uvcolor}, red dotted curve) is matched to the peak bolometric luminosity and the rise time between $t_{\rm PL}$ and bolometric peak, where we measure the bolometric peak luminosity and epoch to be 1.535 $\pm$ 0.020 $\times$ 10$^{43}$ ergs~s$^{-1}$ and $-$0.80 $\pm$ 0.21 days since $B$-band maximum, respectively, in \uname.
However, this model fails to represent the observed bolometric light curve near peak, with slower pre-peak rise and post-peak decline compared to the observations.
The inferred explosion parameters of $M_{\rm  Ni}$ = 0.73 $\pm$ 0.01~\msol\ and $\tau_m$ = 14.47 $\pm$ 0.23 days are also less reasonable. Adopting a characteristic ejecta velocity of $v_{\rm ej}$ = (12.44 $\pm$ 0.02) $\times$ 10$^3$~km~s$^{-1}$ (= peak \siii\ velocity; Table~\ref{tab:sifit}), and $E_{\rm ej} = \frac{3}{10} M_{\rm ej} v_{\rm ej}^2$ for uniformly expanding spherical ejecta, they imply an ejecta mass of $M_{\rm ej}$ = 2.01 $\pm$ 0.06~\msol, which far exceeds the Chandrasekhar mass and is extremely large even among events with super-Chandrasekhar ejecta masses ranging in $\sim$ 1.6--2.1~\msol\ \citep{Scalzo2019mnras}.

In case (2), the model (Figure~\ref{fig:uvcolor}, green dot-dashed curve) is fitted to the near-peak bolometric light curve from $-$10 days to 10 days since the bolometric peak,
setting the onset of the model as a fit parameter $t_{\rm Ni}$ that represents the onset of the central \ni56-powered light curve.
The best-fit (\chisqr\ = 0.2) model with $M_{\rm Ni}$ = 0.631 $\pm$ 0.009~\msol, $\tau_m$ = 11.80 $\pm$ 0.31 days, and $t_{\rm Ni}$ = $-$14.91 $\pm$ 0.17 days since $B$-band maximum in rest-frame provides an excellent comparison to the observed bolometric light
curve near the peak, though it underpredicts the observed luminosity in earlier epochs.
The underprediction by the best-fit model is similar to what is seen in SN~2018aoz and KSP-OT-201509b \citep{Ni2022natas, Moon2021apj}, and likely to be caused by a shallower \ni56\ distribution and/or related to the presence of excess emission, which provides additional heating at early times.

Adopting $\tau_m$ = 11.80 $\pm$ 0.31 days for \uname\ and the approximate opacity of $\kappa$ = 0.1~cm$^2$~g$^{-1}$ expected for \tase\ dominated by \ni56\ line transitions in the photospheric phase, we find that the bolometric light curve is consistent with a near-Chandrasekhar-mass ($M_{\rm ej}$ = 1.34 $\pm$ 0.07~\msol) explosion with kinetic energy of $E_{\rm ej}$ = (1.24 $\pm$ 0.06)~$\times$~10$^{51}$ ergs.
The inferred ejecta mass is equal to the Chandrasekhar mass \citep[1.38~\msol;][]{Mazzali2007sci} when a slightly smaller opacity of 0.097~cm$^2$~g$^{-1}$ is used.
By extending the calculation to include opacities in the range of 0.08--0.10~cm$^2$~g$^{-1}$ that have been used in modelling the bolometric luminosity of \tase\ \citep{Arnett1982apj, Piro&Nakar2014apj, Li2019apj, Moon2021apj, Ni2022natas}, we obtain ejecta masses of \uname\ in the range of $\sim$ 1.3--1.7~\msol, which is consistent with nebular-phase studies favouring a near-Chandrasekhar-mass explosion (Section~\ref{sec:neb}) and also includes super-Chandrasekhar ejecta masses.

In order to examine the possibility of \uname\ having a super-Chandrasekhar ejecta mass, we examine its \sbv\ parameter, bolometric light curve shape, and UV-Optical color evolution.
First, the events with moderately super-Chandrasekhar ejecta masses ($\sim$ 1.5--1.7~\msol) have \sbv\ ranging in $\sim$ 1.1--1.3, while the \sbv\ of 1.00 for \uname\ is lower than those of all other \tase\ inferred to have super-Chandrasekhar ejecta masses based on their bolometric light curves \citep{Scalzo2019mnras}.
Second, the top panel of Figure~\ref{fig:uvcolor} compares the bolometric light curve evolution of \uname\ to those of three other events with super-Chandrsekhar ejecta masses \citep[SNe~2007if, 2006bt, and 2006ot;][]{Scalzo2019mnras}.
In this case, the luminosity of \uname\ is apparently within the range of luminosities seen in those other events, and the decline rates of the bolometric light curves in the figure also appear to be similar.
We measure the decline in luminosity after 15 days since peak to be
$\Delta M_{15, {\rm bol}}$ = 0.75 $\pm$ 0.04 mag for \uname, 
consistent with other \tase\ inferred to have normal and mildly super-Chandrasekhar ejecta masses \citep[$\sim$ 1.2--1.7 \msol;][]{Scalzo2019mnras}.
Finally, the UV-Optical colors of \uname\ are compared in the middle ($UVW1 - V$) and bottom ($UVW2 - V$) panels to those of two peculiar 03fg/09dc-like events (SNe~2012dn and 2011aa), which are thought to have super-Chandrasekhar ejecta masses \citep{Taubenberger2019mnras}, as well as the expectations for normal \tase\ from the NUV-red and NUV-blue groups \citep{Milne2013apj}.
Though the color evolution of \uname\ is extremely blue even among the NUV-blue group in some phases, its pre-peak color evolution clearly diverges from the 03fg/09dc-like events.
Overall, \uname\ is more consistent with being a normal \tas\ with relatively blue UV-Optical color and slow bolometric decline rate than with having a super-Chandrasekhar ejecta mass.

\section{Early Excess Emission Models} \label{sec:excess}

\begin{figure*}[t!]
\epsscale{\scl}
\begin{center}
\includegraphics[width=0.88\textwidth]{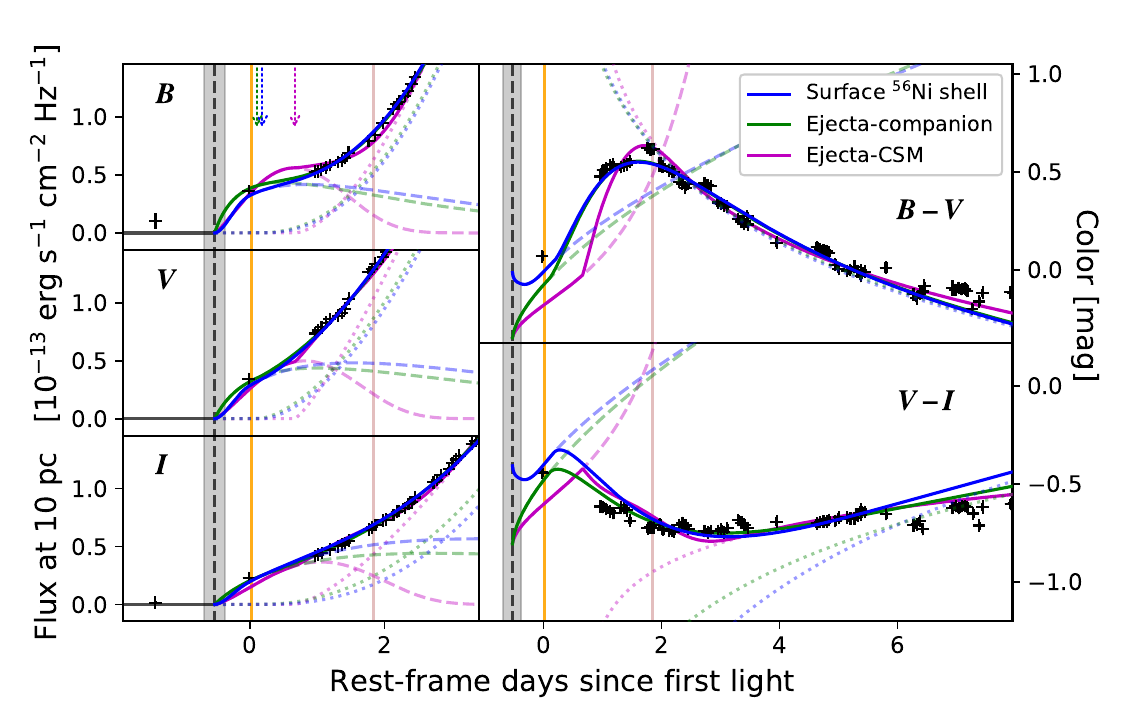}
\end{center}
\caption{(Left) The dereddened early $BVi$ (from top to bottom) light curves of \uname\ in rest frame are compared to what is expected from power-law $+$ three models of early excess emission in \tase: (1) surface \ni56\ shell (blue solid curves), (2) ejecta interaction with the companion (green solid curves); and (3) ejecta interaction with CSM (magenta solid curves).
The transparent curves show the early excess emission (dashed) and underlying power-law (dotted) components, respectively, of each correspondingly colored model, while the inverted arrows with dotted lines mark the onsets of the power-law components.
The black dashed vertical line and shaded region represents the epoch of explosion and its 1-$\sigma$ uncertainty.
The orange and brown vertical lines mark the epochs of the correspondingly colored spectra in Figure~\ref{fig:uvsup} from near the excess emission peak (see Figure~\ref{fig:gaus}) and the reddest \bv\ color (see top-right), respectively.
(Right) The dereddened \bv\ (top) and \vi\ (bottom) colors of \uname\ are compared with those predicted by the same models from the left panels.
\label{fig:excessmod}}
\end{figure*}

We examine the origin of the early excess emission in \uname\ using models combining two luminosity components, following the methods described in \citet{Ni2023apj}: (1) a power-law for the underlying SN emission from the centrally-concentrated main distribution of \ni56\ in the ejecta; and (2) excess emission.
The former is modelled as $f_{\nu} \propto (t - t_{\rm PL})^{\alpha_{\nu}}$ as in Section~\ref{subsec:gaus}.
For the latter, we model three conceivable mechanisms in early \tase\ that can produce excess emission: (1) radioactive heating by \ni56\ near the ejecta surface, (2) ejecta-companion interaction, and (3) ejecta-CSM interaction.
The onset of each model is adopted to be the epoch of explosion, $t_{\rm exp}$ = MJD 59529.32 (Section~\ref{subsec:expl}).
We briefly describe each model below.

For the \ni56\ model, we use the analytic form described in \citet[][Equation 4]{Ni2023apj} for a spherical shell of \ni56\ at the ejecta surface as well as the following equation describing the fractional mass of outer ejecta ($\Delta M/M_{\rm ej}$) that is visible via photon diffusion at any given time
\begin{equation}
        \frac{\Delta M}{M_{\rm ej}} \approx 1.3 
        \left(\frac{t-t_{\rm exp}}{\tau_m}\right)^{1.76}
    \label{eq:mdiff}
\end{equation}
adopting $\tau_m$ = 11.80 days from bolometric light curve fitting (Section~\ref{subsec:bolo}).
This model has free parameters of $M_s$, $t_s$, and $\tau_s$, representing the total mass of \ni56\ in the shell, the rest-frame epoch when $\Delta M$ reaches the inner radius of the shell, and a parameter related to the optical depth where the \ni56\ radioactive emission is thermalized, respectively.

For the case of ejecta-companion interaction, we use the model of \citep[][K10 hereafter]{Kasen2010apj}, adopting the electron scattering opacity of $\kappa$ = 0.2~cm$^2$~g$^{-1}$ for H-poor \tas\ ejecta and explosion parameters from Section~\ref{subsec:bolo} ($M_{\rm ej}$ = 1.34~\msol, $E_{\rm ej}$ = 1.24~\msol).
The model is parameterized by the binary separation distance ($a$) and observer viewing angle ($\theta$) as described in \citet{Olling2015nat}.

For ejecta-CSM interaction, we use the model of \citep{Piro2015apj} for ejecta interaction with a uniform, spherical distribution of CSM, adopting the same opacity and explosion parameters as for the ejecta-companion interaction model above. 
The model has free parameters of $M_{\rm CSM}$ and $R_{\rm CSM}$ for the mass and radius of the CSM, respectively.

The power-law $+$ excess emission models are fitted to the early $BVi$-band light curves of \uname\ before $-$10 days since $B$-band maximum, or $\lesssim$ 40\% of maximum brightness where the power-law rise is applicable in \tase\ following Section~\ref{subsec:gaus}.
(Note that UV light curves are not included in this fit because they are dominated by line effects in the excess emission phase which are not considered in the simple blackbody models; see Section~\ref{sec:uvsup} below for a detailed discussion of these effects).
Figure~\ref{fig:excessmod} compares the best-fits obtained using the surface \ni56\ heating (blue solid curves), ejecta-companion interaction (green solid curves), and ejecta-CSM interaction (magenta solid curves) models.
All three models appear to be capable of providing optical excess emission with a similar timescale and luminosity as what is observed in \uname.
However, the best-fit parameters and mechanism of the surface \ni56\ heating model appear to be the most compatible with other observations of \uname\ as detailed in Section~\ref{subsec:mod} below.

\subsection{Results and Interpretation of Modelling} \label{subsec:mod}

\begin{figure*}[!]
\epsscale{\scl}
\begin{center}
\includegraphics[width=0.92\textwidth]{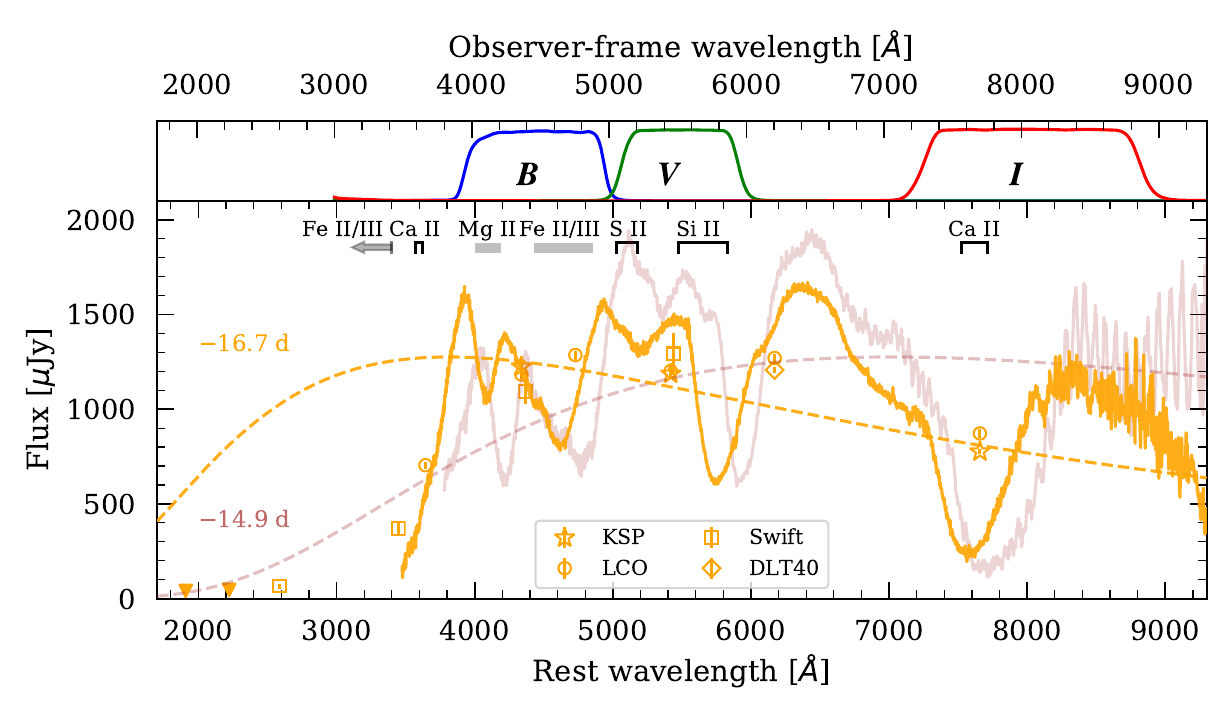}
\end{center}
\caption{(Top) The KSP $BVI$-band transmission curves used for $BVi$-band photometry. (Bottom) The spectrum (orange solid curve) and SED (orange symbols) of \uname\ from $-$16.7 rest-frame days since $B$-band maximum, near the peak epoch of the early excess emission ($-$16.2 days; Section~\ref{subsec:gaus}), is compared to the blackbody (orange dashed curve) predicted by the best-fit surface \ni56\ heating model (Section~\ref{subsec:mod}) from the same post-explosion phase (0.6 days in rest frame).
The SED consists of interpolated photometry from KSP, LCO, Swift and DLT40.
The spectrum is flux-calibrated to the KSP $BVi$-band photometry (orange stars).
The wavelengths of key absorption features in \tase\ are labelled at the top of the panel.
The spectrum from $-$14.9 days (brown), near the epoch of the reddest early \bv\ color ($-$15.2 days; Section~\ref{sec:early}), is also shown compared to the modelled blackbody for that phase (2.4 days post-explosion; brown dashed curve). Both spectrum and SED are normalized to the same KSP $BVi$-band observations (orange stars) for better comparison to the observations from the prior phase.
The model explains the observed redward color evolution in \uname\ (Figure~\ref{fig:col}) with a shift in continuum temperature between the two epochs.
\label{fig:uvsup}}
\end{figure*}

For the case of the surface \ni56\ heating, we obtained the best-fit with \chisqr\ = 6.5 and parameters $t_s$ = 0.52 days since explosion in rest frame, $M_s$ = 6.9~$\times$~10$^{-3}$~\msol, $\tau_s$ = 1.58, $t_{\rm PL}$ = 0.71 rest-frame days since explosion, and $\alpha_{B,V,i}$ = (2.23, 1.68, 2.39).
The \ni56\ model parameters correspond to a 6.9~$\times$~10$^{-3}$~\msol\ shell of excess \ni56\ in the outer 0.53\% of the ejecta mass (Equation~\ref{eq:mdiff}) along the line of sight.
This is consistent with what can be expected in double-detonation or off-center ignited delayed-detonation explosion mechanisms that predict over-densities of \ni56\ near the ejecta surface, which can contain a similar amount of excess \ni56\ \citep[$\sim$ 10$^{-2}$~\msol;][]{Ni2022natas, Maeda2010apj, Seitenzahl2013mnras}.
The presence of this slight surface \ni56\ enrichment may also be compatible with likely explanations for other observed features in \uname, including UV suppression (see Section~\ref{sec:uvsup}) and \siii\ HVFs (see Section~\ref{subsec:homo}).

For the case of ejecta-companion interaction, we obtained the best-fit with \chisqr\ = 3.4, $a$ = 4.7~$\times$~10$^{11}$~cm, $\theta \sim$ 0\degr, $t_{\rm PL}$ = 0.631 rest-frame days since explosion, and $\alpha_{B,V,i}$ = (2.23, 1.67, 2.17).
Note that the power-law components of both the best-fit surface \ni56\ shell and ejecta-companion interaction models infer dark phases of $\sim$ 0.6--0.7 days between the epochs of explosion and power-law onset---consistent with what has been found in other \tase\ \citep[$\sim$ 0.4--1.6][]{Piro&Nakar2014apj, Ni2022natas}---as well as indices that are near the \tas\ population average \citep[$\sim$ 2.0;][]{Miller2020apj}.
The binary separation distance of the K10 model fit corresponds to the expected Roche separation of a 2~\msol\ main sequence subgiant.
Due to the assumption of local thermodynamic equilibrium (LTE) between the shock-heated ejecta and its radiated emission in the K10 model and the 0\degr\ viewing angle, both of which tend to over-estimate the luminosity of the ejecta-companion interaction emission \citep{Kutsuna2015pasj, Olling2015nat}, the required companion may be even larger \citep[e.g., see \S4 discussion in][]{Ni2023apj}.
However, the presence of a large non-degenerate companions is disfavoured by nebular-phase spectroscopy of \uname\ showing an absence of emission from swept-up H-rich companion material (see Section~\ref{sec:neb}).

For the case of ejecta-CSM interaction, the best-fit is obtained with \chisqr\ = 15.4, $M_{\rm CSM}$ = 0.024~\msol, $R_{\rm CSM}$ = 3.7~$\times$~10$^{11}$~cm, $t_{\rm PL}$ = 1.19 rest-frame days since explosion, and $\alpha_{B,V,i}$ = (1.67, 1.22, 1.42), though the radius may be up to $\sim$ 2$\times$ larger for equatorially-concentrated CSM geometries \citep{Ni2023apj}. 
The fit is slightly worse than the surface \ni56\ shell and ejecta-companion interaction models because the predicted emission cools (i.e., evolves redward) slower immediately post-explosion, which also results in significantly different best-fit power-law indices and $t_{\rm PL}$ compared to the other two models (see above), though it is still capable of reproducing the observed light curves overall.
The earliest spectra of \uname\ that were taken during the excess emission phase (see Figure~\ref{fig:specevol}) place additional constraints on the ability of this model to explain the observed early excess emission.
For CSM originating from the binary mass transfer process that triggers the \tas\ explosion, narrow emission lines \citep[e.g., from ionized C, O, and He;][]{Yaron2017natph} could be produced by extended secondary components of the CSM---located $\gtrsim$ 10$^{13}$~cm from the progenitor star given the prompt phase of the earliest spectrum of \uname\ $\sim$ 0.6 days post-explosion---which are apparently absent in the earliest spectra of \uname.
This may require the SN to explode on a sufficiently rapid timescale after mass transfer begins \citep[e.g., on a viscous timescale $\lesssim$ 10$^6$~s for accretion disk originated CSM;][]{Raskin2013apj}.

\subsection{Excess Emission and UV Suppression} \label{sec:uvsup}

Figure~\ref{fig:uvsup} shows the evolution of the SED of \uname\ from near the peak of the early excess emission ($-$16.2 days; Section~\ref{subsec:gaus}) to the reddest early \bv\ color ($-$15.2 days; Section~\ref{sec:early}).
For the former, we compare the photometric SED of the KSP, LCO, Swift, and DLT40 observations from $-$16.7 days and the spectrum (orange) from the same epoch normalized to the KSP photometric wavebands with the 13380~K blackbody predicted by the best-fit surface \ni56\ shell model from the same phase post-explosion (0.6 days; orange dashed curve).
Note that ejecta-companion interaction predicts a very similar SED evolution.
For the latter, we compare the spectrum (orange) from the nearest epoch of $-$14.9 days to the same model from the corresponding phase (7260~K at 2.4 days post-explosion; brown dashed curve), where both are normalized to the KSP $BVi$-band observations.
While the observed optical continuum emission is consistent with the modelled blackbody evolution, the UV emission appears to be suppressed in the earliest phase. 
As seen in the figure, any pure blackbody SED that does not include UV emission would be unreasonably cold for the SED at $-$16.7 days in the optical bands---i.e., even colder than $\sim$ 7000~K (brown dashed curve). Thus, the suppression is likely due to line absorption rather than a cold continuum temperature.

The suppression of UV emission in \uname\ in this phase resembles the drop-off of the photometric SED of SN~2018aoz $\lesssim$ 5000~\AA\ at $\sim$ 0.9 days post-explosion \citep{Ni2022natas}, which has been attributed to the absorption features of \fex\ in bluer wavelengths.
The sharp drop-off of the photometric SED of \uname\ $\lesssim$ 3800~\AA\ at a similar post-explosion phase (0.6 days) coincides with the expected wavelengths of \caii\ and \fex\ absorption features, indicating that those elements are likely responsible for the suppression.
The required 0.6-day phase of UV suppression by the Fe-line forming region and Equation~\ref{eq:mdiff} indicates that these elements are present within some outer fraction of the ejecta mass along the line of sight $<$ 0.69\%, consistent with the location of excess surface \ni56\ that can explain the early excess emission (outer 0.53\%; Section~\ref{sec:excess}).
Although a smaller amount of Fe-peak elements from either the tail of the main distribution of centrally-concentrated \ni56\ synthesized in the explosion \citep{Mazzali2014mnras} or an enhanced-metallicity progenitor star \citep{Walker2012mnras} may also be capable of explaining the observed UV suppression,
the possibility for the excess surface \ni56\ to explain both the early excess emission and UV suppression of \uname\ calls for future exploration by radiative transfer simulations.

\section{Origin of \siii\ High-Velocity Features}\label{sec:velocity}

HVFs in \tase\ are thought to be produced by detached components of a particular ion's line-forming region at a higher velocity than the primary SN photosphere.
Several possibilities have been considered for the origin of detached components, including density enhancements in the high-velocity outer ejecta produced by the explosion mechanism \citep[e.g., pulsational or off-center ignited delayed-detonations, or double-detonations;][]{Yarbrough2023mnras, Seitenzahl2013mnras, Boos2021apj}, interaction with CSM, and enhanced populations of atomic states in the outer ejecta \citep{Mazzali2005mnras}.
These origins usually invoke three main mechanisms \citep[e.g., see][]{Mulligan2018mnras}
involving (1) two distributions of matter, one at the normal SN ejecta velocity comprising most of the ejecta mass and a secondary distribution (e.g., a shell, clump, or bullet) at a higher velocity in the outer ejecta, which we call ``split-velocity'' ejecta; (2) a shock interaction that sends a reverse shock into the ejecta; and/or (3) a line opacity enhancement.

We explore the origin of the observed \siii\ HVFs in \uname\ 
by modelling the evolution of its line-forming region as follows.
First, we show that two ejecta mass components 
with similar density distributions expanding homologously since the epoch of explosion can explain the \siii\ HVF and PVF velocity evolution. In this case,
the main distribution of ejecta mass in the SN and a low-mass secondary component moving at a higher velocity produce the PVFs and HVFs, respectively, by forming primary and secondary photospheres (Section~\ref{subsec:homo}).
Secondly, we find that a change in the ejecta density structure caused by shock interaction with material in the vicinity of the progenitor system is not capable of explaining the HVF presence (Section~\ref{subsec:shock}).
Finally, we examine line opacity changes that can be expected from enhanced \ni56\ fraction or shock interaction in the outer ejecta, finding that these are less capable of explaining the observed characteristics of the HVF velocity evolution (Section~\ref{subsec:opac}).
Overall, a split-velocity ejecta appears to be the most likely origin of the observed \siii\ HVFs in \uname.

\subsection{Homologous Expansion of Split-Velocity Ejecta} \label{subsec:homo}

\begin{figure*}[hbtp]
\epsscale{\scl}
\begin{center}
\includegraphics[width=0.9\textwidth]{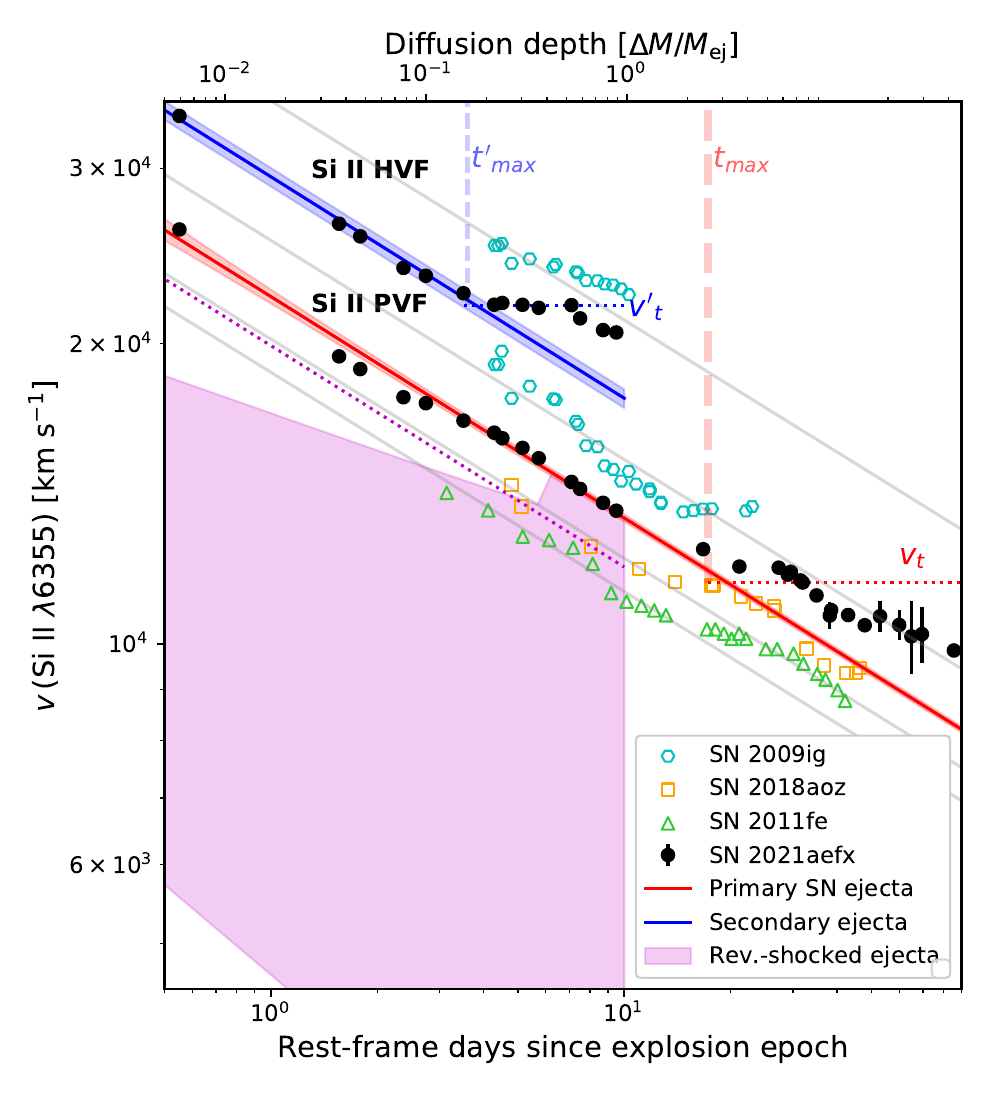}
\end{center}
\caption{The observed evolution of the \siii\ PVF and HVF velocity in \uname\ is compared to those of SNe~2009ig \citep[cyan hexagons;][]{Silverman2015mnras}, 2018aoz \citep[orange squares;][]{Ni2022natas}, and 2011fe \citep[green triangles;][]{Pereira2013aa} as well as the predictions of two models: (1) homologous expansion of the 1.34~\msol\ SN ejecta and a low-mass secondary distribution of high-velocity material (red and blue lines); and (2) reverse-shocked SN ejecta (pink shaded region).
SNe~2018aoz and 2011fe are typical \tase\ with normal \siii\ photospheric velocities, while the \siii\ features of SN~2009ig also exhibit prominent HVF and PVF components \citep{Marion2013apj} resembling those of \uname.
The light grey diagonal lines are $t^{-2/9}$ power-laws fitted to each SN.
The red vertical-dashed and horizonal-dotted lines represent the epoch of $B$-band maximum of \uname\ ($t_{\rm max}$ = 17.26 days since explosion) and the estimated transition velocity of its ejecta ($v_t$ = 0.93$v_{\rm ej}\sim$ 11,520~km~s$^{-1}$; see Section~\ref{subsec:homo}), respectively, while the blue ones represent the epoch when the HVF of \uname\ begins to plateau ($t'_{\rm max}$ = 3.6 days) and the observed plateau velocity ($v'_t$ = 21,860 $\pm$ 20~km~s$^{-1}$), respectively.
These HVF properties match the homologous expansion of $v'_{\rm ej}$ = 23,500~km~s$^{-1}$ material embedded in the outermost $\sim$ 16\% of the ejecta along the line of sight.
The magenta dotted line represents the velocity at the inner boundary of the reverse-shocked region for the case of CSM with an $s$ = 2 density profile.
\label{fig:shockhpvf}}
\end{figure*}

Photosphere evolution in SNe is often understood using the following simple model.
The ejecta density profile follows a broken power-law with a shallow-declining inner region $\rho_i \propto r^{-\delta}$ and a steep-declining outer region $\rho_0 \propto r^{-n}$ \citep{Kasen2010apj}. The two profiles join at the transition velocity $v_t = \zeta_v (E_{\rm ej}/M_{\rm ej})^{1/2}$, where $\zeta_v$ is a numerical constant. Under homologous expansion, the outer density profile is
\begin{align}
    \rho_0(r, t) = \zeta_\rho \frac{M}{v_t^3 t^3} \left(\frac{r}{v_t t} \right)^{-n}
    \label{eq:outerdens}
\end{align}
where $\zeta_\rho$ is the normalization factor. The numerical constants $\zeta_v$ and $\zeta_\rho$ follow from requiring the density profile to be continuous at the transition velocity and the density profile to integrate to the specified mass $M$ and kinetic energy $E$ of the SN.
Adopting typical values of $\delta \sim 1$ and $n \sim 10$ for \tase\ based on the simulations of \citet{Kasen2009apj} yields the values of the constants to be $\zeta_v \sim 1.69$ and $\zeta_\rho \sim 0.12$.
If $v_{\rm ej}$ = (10 $E_{\rm ej}$)/(3 $M_{\rm ej}$) is the characteristic velocity as defined in Section~\ref{subsec:bolo}---identified to be $v_{\rm peak}$ (\siii~$\lambda$6355~\AA) = 12,440~km~s$^{-1}$ (Table~\ref{tab:param})---then the transition velocity is $v_t$ = 0.93$v_{\rm ej}$ $\sim$ 11,520~km~s$^{-1}$.

The photosphere radius $r_{\rm ph}$ is the location where optical depth $\tau = \int_{r_{\rm ph}}^{\infty} \kappa \rho(r, t) dr = 1$. The nearly constant opacity of $\kappa \sim 0.1$~cm$^2$~g$^{-1}$ is expected for \ni56-dominated \tas\ ejecta \citep{Pinto&Eastman2000apj}. Integrating the outer density profile above (Equation~\ref{eq:outerdens}), we obtain 
\begin{align}
    r_{\rm ph, 0}(t) &= \left(\frac{\zeta_{\rho}\kappa M_{\rm ej} v_t^{n-3}}{n-1} \right)^{1/(n-1)} t^{(n-3)/(n-1)}\\
    &= \left(\frac{\zeta_{\rho}\kappa M_{\rm ej} v_t^7}{9} \right)^{1/9} t^{7/9}
    \label{eq:photr}
\end{align}
Under homologous expansion, the ejecta velocity at the photosphere is $r_{\rm ph, 0}/t$, which gives the photospheric velocity evolution of the SN ejecta as follows
\begin{align}
    v_{\rm ph, 0}(t) &= \left(\frac{\zeta_{\rho}\kappa M_{\rm ej} v_t^{n-3}}{n-1} \right)^{1/(n-1)} t^{-2/(n-1)}\\
    &= \left(\frac{\zeta_{\rho}\kappa M_{\rm ej} v_t^7}{9} \right)^{1/9} t^{-2/9}
    \label{eq:photv}
\end{align}

These can be written in terms of the characteristic ejecta velocity $v_{\rm ej}$ and the geometric mean of the diffusion and expansion timescales $\tau_m$ = 11.80 days (Table~\ref{tab:param}). 
\begin{multline}
    r_{\rm ph, 0}(t) \sim 9.17\times 10^{13}~{\rm cm}\\
    \left(\frac{\tau_m}{{\rm day}}\right)^{2/9} \left(\frac{v_{\rm ej}}{10^9~{\rm cm~s}^{-1}}\right)^{8/9} \left(\frac{t}{{\rm day}}\right)^{7/9}
    \label{eq:rph0}
\end{multline}
\begin{multline}
    v_{\rm ph, 0}(t) \sim 1.06\times 10^{9}~{\rm cm~s}^{-1}\\
    \left(\frac{\tau_m}{{\rm day}}\right)^{2/9} \left(\frac{v_{\rm ej}}{10^9~{\rm cm~s}^{-1}}\right)^{8/9} \left(\frac{t}{{\rm day}}\right)^{-2/9}
    \label{eq:vph0}
\end{multline}

Figure~\ref{fig:shockhpvf} compares the \siii\ HVF and PVF velocity evolution of \uname\ to those of other \tase\ and the photospheric velocity evolution predicted by Equation~\ref{eq:photv} (red curve).
The red shaded region represents the 1-$\sigma$ uncertainty associated with $\tau_m$ and $v_{\rm ej}$.
The prediction closely matches the observed PVF velocity evolution prior to approximately the epoch of $B$-band maximum ($t_{\rm max}$) when the photosphere reaches the transition velocity.
The slow-declining density distribution of the inner ejecta below the transition velocity causes the PVF velocity to plateau after $\sim t_{\rm max}$, where the observed velocity of the PVF plateau appears to be similar to $v_t$ estimated above.
The top axis of Figure~\ref{fig:shockhpvf} estimates the mass coordinate  ($\Delta M/M_{\rm ej}$ from the ejecta surface) of the diffusion depth, which applies during early times before the transition velocity is reached.
(Note that the diffusion depth reaches the transition velocity sometime prior to $t_{\rm max}$ due to it being deeper than the photosphere).

We observe that the \siii\ HVF velocity evolution exhibits key points of similarity with that of the \siii\ PVF.
The HVF follows the same $t^{-2/9}$ evolution as the PVF in the early phase, but at a $\sim$ 1.32 $\times$ higher velocity (Section~\ref{subsec:expl}), and it plateaus to a velocity of $v'_t \sim$ 21,860 $\pm$ 20~km~s$^{-1}$ after $t_{\rm max}'$ $\sim$ 3.6 days post-explosion.
This suggests that the distributions of material responsible for the PVF and HVF may be self-similar.
Relative to the bulk of the SN ejecta, the HVF velocity begins to plateau when the diffusion depth is at 0.16$\times M_{\rm ej}$, indicating that the material responsible for producing the HVF is within the outer $<$ 16\% of the ejecta mass along the line of sight.
We fit Equation~\ref{eq:photv} to the HVF velocity evolution until $t_{\rm max}' \sim$ 3.6 days, adopting the characteristic velocity of $v_{\rm ej}' = v'_t/0.93 \sim$ 23,500~km~s$^{-1}$ for the HVF material and assuming the same constant opacity of $\kappa \sim$ 0.1~cm$^2$~g$^{-1}$.
Figure~\ref{fig:shockhpvf} shows that the best-fit (blue line) provides a good fit to the observed HVF velocity evolution with $\tau_m' \sim$ = 3.2 $\pm$ 0.3 days.
Using Equation~\ref{eq:taum} and $E_{\rm ej} = \frac{3}{10} M_{\rm ej} v_{\rm ej}^2$, we estimate the mass of the HVF material to be $M_{\rm ej}'$ = 0.19 $\pm$ 0.04~\msol\ (or 14 $\pm$ 3\% of $M_{\rm ej}$, consistent with the above estimate of the diffusion depth coordinate at $t'_{\rm max}$).

The existence of a transition velocity at $v'_t \sim$ 21,860 $\pm$ 20~km~s$^{-1}$ in the ejecta mass distribution to explain the \siii\ HVF plateau timing is also evidenced by the evolution (see Figure~\ref{fig:siexp}) of the PVF component of the \caii\ NIR triplet (Appendix~\ref{sec:cafit}).
Since the PVF of the \caii\ NIR triplet traces the early \siii\ HVF prior to $\sim$ 5 days post-explosion ($\sim -$12 days since $B$-band maximum), the two features are likely formed in a common secondary photosphere (Section~\ref{subsec:hvfid}).
Between 5 and 7 days post-explosion ($\sim -$12 and $-$10 days since $B$-band maximum), which is during the \siii\ HVF plateau, the observed \caii\ NIR triplet begins to show a component (see Figure~\ref{fig:cafit}) that traces the velocity of the \siii\ PVF for the first time (together called ``co-evolving PV Ca and Si components'').
This suggests that the secondary photosphere obscures underlying \caii\ until it reaches the transition velocity $v'_t$ and the entire distribution of material responsible for the secondary photosphere has become optically thin, after which the evolution of the primary SN photosphere produces co-evolving PV Ca and Si components.

In summary, the \siii\ HVF in \uname\ is consistent with what can be produced by material that has a similar density distribution as the main distribution of ejecta mass in the SN, embedded as a low-mass secondary structure in the fast-expanding outer layers and expanding homologously post-explosion with the rest of the ejecta.
In this scenario, the photosphere of this secondary structure produces the co-evolving HV Si and Ca components.
We note that this secondary structure may also be responsible for the observed early excess emission in \uname\ if the material is slightly \ni56\ enriched, with a \ni56\ fraction of only $\sim$ 4\% being required to reproduced the observed excess (Section~\ref{sec:excess}).
It is conceivable that slightly \ni56-enriched density enhancements in the outer ejecta can be produced by explosion mechanisms such as pulsational or off-center ignited delayed-detonations \citep{Baron2012apj, Maeda2010apj, Seitenzahl2013mnras}, or double-detonations \citep[e.g.,][]{Boos2021apj}.
However, both pulsations and double-detonations tends to produce density enhancements in smooth shells,
which would result in a relatively simple photosphere geometry.
The simultaneous appearance of the \siii\ HVF with the \siii\ PVF originating from lower-velocity underlying ejecta indicates that the outer ejecta is not entirely spherically symmetric or clumpy, consistent with significant line polarization of observed HVFs \citep[$\sim$ 0.1--1.4\%;][]{Cikota2019mnras} as well as continuum polarization \citep[$\sim$ 0.2--0.3\%;][]{Wang2003apj} seen in \tase\ prior to maximum light.
Such an ejecta is much more congruous with an off-center ignited delayed-detonation, where asymmetric sub-sonic explosion processes can embed non-spherically symmetric clumps of \ni56-enriched, high-velocity material in the outer layers along the line of sight before the ejecta profile is frozen in by the subsequent detonation, resulting in a complex photosphere evolution post-explosion.
\citep[Note that, in general, SN explosion asphericities tend to increase towards the highest-velocity ejecta layers as a natural consequence of shock acceleration;][]{Matzner2013apj}.
The delayed-detonation origin for \siii\ HVFs in \uname\ is consistent with its near-Chandrasekhar ejecta mass ($\sim$ 1.34~\msol; Section~\ref{subsec:bolo}), as well as progenitor mass as inferred by some nebular-phase studies \citep[$\sim$ 1.38~\msol;][]{DerKacy2023apj}, while the observed nebular-phase line profiles of nuclear burning products in the core are consistent with an off-center ignition (Section~\ref{sec:neb}).

\subsection{Reverse-Shocked Photosphere Evolution}\label{subsec:shock}

We examine whether the \siii\ HVF can originate from a shock interaction of \uname\ with stationary material in the progenitor system, which sends a reverse shock into the ejecta that increases the density of material in the high-velocity outer regions. 
Although this process can affect the photosphere, it does not easily explain the observed properties of the HVF.
The reverse shock creates a pile-up of higher density and lower velocity material in the ejecta following the shock front, whose size in 1-D is usually a few percent of the radius of the shock front \citep{Chevalier1982apj}.
If the \siii\ line opacity $\kappa_{\rm si}$ is unaffected by the reverse shock, then the \siii\ optical depth $\tau_x = \int_{x}^{\infty} \kappa_{\rm Si} \rho\ dr$ of a fluid element at radial position $x$ is unaffected by the density pile-up. 
(Note that the forward shock pile-up, being not part of the SN ejecta, does not contain Si, and thus, it does not contribute to the \siii\ optical depth).
An optically thick reverse shock results in the \siii\ photosphere closely following the shock front, at a compressed radius and reduced velocity.
For \tas\ ejecta with an outer density profile described by $\rho\propto r^{-10}$ \citep{Kasen2009apj} interacting with CSM that can be expected to be found in \tas\ progenitor systems, with density described by $\rho\propto r^{-s}$ for $s$ as large as 3 \citep{Piro&Morozova2016apj}, the shock front radius and velocity evolve as $t^{7/(10-s)}$ and $t^{(s-3)/(10-s)}$, respectively.
The $t^{7/(10-s)}$ evolution of the reverse-shocked photosphere continues until the pile-up becomes optically thin, after which the photosphere evolution passes back into the unshocked ejecta.
This only occurs for the case of s $>$ 1, since the photosphere of free-expanding unshocked ejecta evolves as $t^{7/9}$ (see Section~\ref{subsec:homo}).

The pink shaded region in Figure~\ref{fig:shockhpvf} shows the predicted range of photospheric velocity evolution for \uname\ after the ejecta has undergone a reverse-shock, obtained by solving the fluid equations of \citet{Chevalier1982apj} for a radiation-dominated gas ($\gamma$ = 4/3) with an appropriate change in the boundary conditions.
The shaded region is bounded by two cases of circumstellar material with radial distributions characterized by $s$ = 2 (top) and $s$ = 0 (bottom), and normalized to $\rho_{\rm CSM}$ = 10$^{-4}$ $\times$ 3$M_{\rm CSM}$/(4$\pi R_{\rm CSM}^3$) $\sim$ 10$^{-8}$~g~cm$^{-3}$ at $R_{\rm CSM}$ (Section~\ref{sec:excess}).
Note that in the former case ($s>$ 1), the reverse-shocked region eventually becomes optically thin when the photosphere velocity matches the velocity at the inner boundary of the reverse-shocked region (magenta dotted line), and the velocity evolution matches the red line (unshocked ejecta) thereafter.
It is possible to achieve the same $t^{-2/9}$ power-law as the observed \siii\ HVF velocity using an $s$ = 1 CSM distribution, but the reverse-shocked photosphere would still be located at a lower velocity than what is predicted by the homologous expansion of unshocked ejecta (red line).
Thus, the pile-up of reverse-shocked ejecta alone is unable to explain the observed \siii\ HVF evolution.

\subsection{Enhancement of \siii\ opacity in the outer ejecta}\label{subsec:opac}

Next, we consider whether an enhancement of the \siii~$\lambda$6355~\AA\ line opacity in some region of the high-velocity outer ejecta along the line of sight can explain the HVF. 
Under homologous expansion, this case is accommodated by the following modification to Equation~\ref{eq:vph0}, where $\kappa'/\kappa$ is the enhancement relative to the typical photospheric opacity.
\begin{equation}
    v_{\rm ph, 1}(t) \sim \left(\frac{\kappa'}{\kappa}\right)^{1/9} v_{\rm ph, 0}(t)
    \label{eq:vph1}
\end{equation}
The observed \siii\ HVF/PVF velocity ratio of 1.32 before the HVF plateau is replicated by adopting $\kappa'/\kappa \sim$ 12 in the outer ejecta, resulting in identical HVF evolution as what is predicted by the split-velocity model (Section~\ref{subsec:homo}).
Two mechanisms come to mind that can produce a change in the line opacity.
First, excess \ni56\ in the outer 0.5\% of the ejecta can enhance the ionization and excitation level populations of species. 
Second, UV and X-ray photons produced by shock interaction could also alter the level populations.
However, it is unlikely that either of those mechanisms can explain the co-evolution of HV Si and Ca components (Section~\ref{subsec:hvfid}), as they would have to affect the line opacity of \siii~$\lambda$6355~\AA\ and the \caii\ NIR triplet in the same way.

These two mechanisms also have difficulties in explaining the observed \siii\ HVF evolution alone.
For the former, it is difficult to imagine that the level populations maintained by a clump of decaying \ni56\ can produce an opacity enhancement that remains constant during the early ($<$ 3.6 days) HVF evolution throughout some specific region of the high-velocity ejecta layers if the region itself is not kinematically distinct from the rest of the ejecta (i.e., resembling split-velocity ejecta).
For the latter, while simulations of CSM-ejecta interaction that account for changes in the level population can produce somewhat realistic \tas\ HVFs in the regime of optically thin reverse shocks \citep{Mulligan2018mnras}, this typically results in line evolution that differs substantially from the expectation for homologously expanding ejecta with constant ionization and excitation states (i.e., constant opacity).
Both cases do not readily explain the timing and velocity of the observed HVF plateau.

\section{Nebular-Phase Constraints on the Explosion and Progenitor System} \label{sec:neb}

\begin{figure}[t!]
\epsscale{\scl}
\plotone{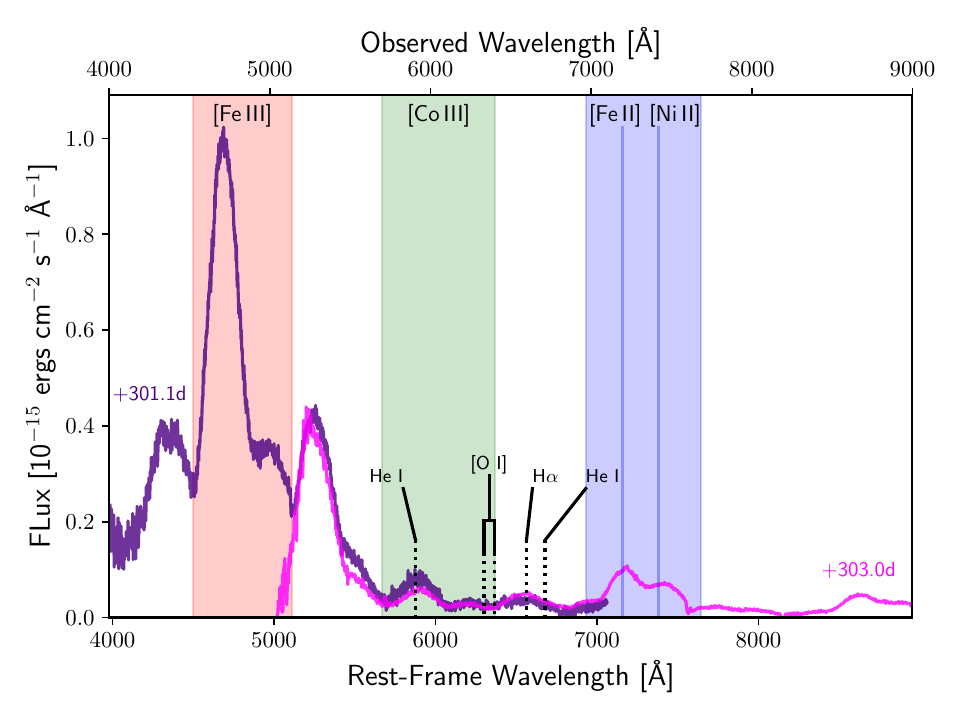}
\caption{The dereddened nebular-phase spectrum of \uname\ from $\sim$ 300 days since $B$-band maximum in rest frame obtained by GMOS with B600 and R400 filters (indigo and magenta curves, respectively). The observed emission lines of \feiii, \coiii, and the \feii\ $+$ \niii\ complex are labelled near the top with vertical shaded regions representing the wavelengths used for flux integration. 
The expected positions of non-detected H$\alpha$, \hei, and [\oi] emission lines are represented with vertical dotted lines near the bottom.
\label{fig:neb}}
\end{figure}

Figure~\ref{fig:neb} presents the nebular-phase spectrum of \uname\ from $\sim$ 300 days since $B$-band maximum obtained using GMOS through the B600 (indigo) and R400 (magenta) filters.
The B600 and R400 spectra were scaled by matching their flux to the observed $V$- and $i$-band photometry from the same epochs, respectively (Table~\ref{tab:lc}).
The spectrum shows strong emission features of \feiii~$\lambda$4658~\AA\ and \coiii~$\lambda$5888~\AA, with fluxes of (32.38 $\pm$ 0.04) and (2.48 $\pm$ 0.01) $\times$ 10$^{-14}$~ergs~s$^{-1}$~cm$^{-2}$, respectively, typically associated with \ni56\ $\rightarrow$ \co56\ $\rightarrow$ \fer56\ radioactive decay in \tase\ \citep{Kuchner1994apj}.
The noise in the spectrum is estimated by smoothing each spectrum with a second-order Savitsky–Golay filter with a width of 180~\AA, which is $\lesssim$ 1/4 of the feature widths in order to estimate the uncertainties.
There is also a double-peaked emission feature near 7290~\AA.
Sub-Chandrasekhar-mass explosion models often predict a strong emission line near 7290~\AA\ due to increased Ca production in the core.
However, the shape of the feature indicates that it is likely dominated by the \feii~$\lambda$~7155~\AA\ and \niii~$\lambda$~7378~\AA\ emission lines rather than the
[\caii]~$\lambda$7291, 7323~\AA\ doublet, since [\caii] would not be resolved as a doublet at typical \tas\ velocities \citep{Polin2021apj}.
The total flux of the feature, (4.20 $\pm$ 0.01) $\times$ 10$^{-14}$~ergs~s$^{-1}$~cm$^{-2}$, is also relatively low compared to that of \feiii\ with a ratio of only 0.130, whereas sub-Chandrasekhar-mass explosion models predict higher ratios of $\gtrsim$ 0.3 \citep{Polin2021apj}, further disfavouring a sub-Chandrasekhar-mass explosion.
Off-center ignited Chandrasekhar-mass explosion mechanisms, on the other hand, predict \feii\ and \niii\ emission features whose Doppler shifts depend on the motion of the ejecta core with respect to the rest frame of the SN \citep{Maeda2010apj, Maeda2010natur, Li2021apj}.
We observe a redshift in both the \feii\ and \niii\ features, with velocities of 1020 and 1310 km~s$^{-1}$, respectively, indicative of an asymmetric explosion where the ejecta core is receding from the observer, consistent with JWST observations of \uname\ in the nebular phase \citep{Kwok2023apj, DerKacy2023apj}.

In \tase, narrow H$\alpha$, \hei~$\lambda$5875~\AA, \hei~$\lambda$6678~\AA, [\oi]~$\lambda$6300~\AA, and [\oi]~$\lambda$6364~\AA\ lines may be produced by low-velocity material swept-up from the companion \citep[e.g.,][]{Kollmeier2019mnras} or CSM, including disrupted companion material following a violent merger \citep{Kromer2013apj, Mazzali2022mnras, Tucker2022apjl}.
By injecting synthetic emission lines of H and He with a FWHM = 1000~km~s$^{-1}$ and modelling
Doppler shifts from the rest wavelength of up to $\pm$1000~km~s$^{-1}$ into the observed nebular spectra following the methods of \citet{Sand2018apj, Sand2019apj}, we find 3$\sigma$ flux upper limits of (1.7, 4.2, and 2.6) $\times$ 10$^{-16}$~ergs~s$^{-1}$~cm$^{-2}$ for H$\alpha$, \hei~$\lambda$5875~\AA, and \hei~$\lambda$6678~\AA, respectively.
We do the same for [\oi]~$\lambda$6300~\AA\ and [\oi]~$\lambda$6364~\AA, but using FWHM = 2000~km~s$^{-1}$ and up to $\pm$2000~km~s$^{-1}$ Doppler shifts that can be expected for [\oi] \citep{Taubenberger2013apj}, obtaining 3$\sigma$ flux upper limits of (3.5 and 4.2) $\times$ 10$^{-16}$~ergs~s$^{-1}$~cm$^{-2}$, respectively.
We use the models of \citet{Botyanszki2018apj} and \citet{Dessart2020aa} to obtain the masses of H and He allowed by these upper limits, adopting the luminosity distance from template fitting (Section~\ref{subsec:templ}).
Using the former, we allow only $<$ 6.3 $\times$ 10$^{-4}$~\msol\ of H based on H$\alpha$, and $<$ (6.7 and 5.4) $\times$ 10$^{-3}$~\msol\ of He based on \hei~$\lambda$5875~\AA, and \hei~$\lambda$6678~\AA, respectively.
Using the latter, we allow $<$ 2.4 $\times$ 10$^{-3}$~\msol\ of H.
The mass of H allowed by both \citet{Botyanszki2018apj} and \citet{Dessart2020aa} models is far below what is expected for masses of stripped H-rich material from main sequence and giant companions \citep[$\sim$ 0.1; e.g.,][]{Botyanszki2018apj}, while the He constraints based on \citet{Botyanszki2018apj} would even disfavour a naked He-star companion.
These constraints are more stringent than those derived from the earlier-phase spectrum of \uname\ \citep{Hosseinzadeh2022apj}, confirming that a single-degenerate progenitor is disfavoured.

\section{Summary and Conclusion} \label{sec:conc}

The observations of \uname\ from the early ($<$ 2 day since $t_{\rm PL}$) to the nebular phase ($\gtrsim$ 200 days since $B$-band maximum) constitute one of the most extensive datasets on a \tas\ acquired to date, shedding light on the explosion processes and progenitor system of a normal \tas\ with unique early light curve and spectroscopic properties.
Here, we summarize our main results and conclusions:
\begin{itemize}
    \item Our spectrum of \uname\ from the epoch of $B$-band maximum confirms its intermediate classification between the Core-Normal and Broad-Line subtypes of normal \tase, though its near-peak light curves are slow-declining and luminous for this classification, resembling 91T-like peculiar events.
    
    \item The bolometric light curve and UV-Optical color curves in the photospheric phase ($\lesssim$ 30 days) indicate a near-Chandrasekhar ejecta mass ($\sim$ 1.34~\msol).

    \item Sub-Chandrasekhar-mass explosions are further disfavoured by the absence of \caii\ emission and the presence of stable \feii\ and \niii\ emission in the nebular phase spectrum. We observe $\sim$ 1000~km~s$^{-1}$ redshifts in both of the \feii\ and \niii\ lines which confirms the slight asymmetry of the explosion core.

    \item The early light curves up to $\lesssim$ 40\% of maximum brightness ($<$ 6.7 days since $t_{\rm PL}$) consist of an excess emission embedded in an underlying power-law that rises over the period. The onset of the power-law rise, associated with emission from the main distribution of centrally-concentrated \ni56\ in the ejecta, is estimated to be MJD 59529.85 $\pm$ 0.55 (= $t_{\rm PL}$) $\sim$ 0.5 hours after our first $BVi$ detections near the peak of the excess emission, indicating the presence of additional power sources.
    The excess emission initially peaks in the $B$-band and evolves redward thereafter, dominating the SN light curve over the ``excess emission phase'' of $\lesssim$ 1.75 days, during which it contributes $\sim$ 1.4 $\times$ 10$^{-6}$~ergs~cm$^{-2}$ of total energy along the line of sight (or 4.3 $\times$ 10$^{46}$~ergs, assuming spherically symmetric emission), mostly in the optical ($B$ to $i$) bands.
    
    \item The spectra of \uname\ show broad \siii~$\lambda$6355~\AA\ absorption features prior to $-$7.8 days since $B$-band maximum (8.9 days since $t_{\rm PL}$) that separate into two distinct Gaussian components: (1) a ``photosheric-velocity feature'' (PVF) associated with the expansion of normal \tas\ ejecta moving at 12,400~km~s$^{-1}$; and (2) a ``high-velocity feature'' (HVF) showing a (6500 $\pm$ 600)~km~s$^{-1}$ higher velocity at every epoch. 
    Both HVF and PVF components follow a $v \propto t^{-0.22}$ power-law predicted by a homologously expanding polytrope (n = 3), pointing to an explosion epoch at MJD 59529.32 (or 7:41 on November 10, 2021).

    \item The HVF and PVF features of \siii\ map to corresponding velocity components identified in the \caii\ NIR triplet that trace the \siii\ HVF and PVF at early ($\lesssim -$12 days since $B$-band maximum) and late ($\gtrsim -$10 days) times, respectively. In addition, a component of the \caii\ NIR triplet is identified showing velocities as high as (40,770 $\pm$ 100)~km~s$^{-1}$ in the earliest spectrum from $-$16.7 days days, which is the fastest-expanding ejecta material ever observed in a \tas.

    \item The evolution of the excess emission in optical bands can be fitted by three physical models predicting excess thermal emission in early \tase: (1) a $\sim$ 0.007~\msol\ shell of \ni56\ in the outer $\sim$ 0.5\% of the ejecta mass; (2) ejecta interaction with a $>$ 2~\msol\ main sequence subgiant; and (3) ejecta interaction with $\sim$ 0.02~\msol\ of CSM in a radius of $\gtrsim$ 4~$\times$~10$^{11}$~cm.

    \item Pure thermal emission is unable to fit the UV SED of \uname\ near the peak of the excess emission for any reasonable temperature, requiring line-blanket absorption in the vicinity of $\lesssim$ 3800~\AA\ by Fe and/or Ca located somewhere near the ejecta surface, above the outer 0.7\% of the ejecta mass.
    The possibility of fitting both the observed excess emission and UV suppression in \uname\ with a slight enrichment of radioactive Fe-peak elements in the outer $\sim$ 0.5\% of the ejecta along the line of sight warrants future study with 1-D radiative transfer simulations, though a smaller quantity of Fe-peak elements from either the tail of the main distribution of centrally-concentrated \ni56\ in the ejecta or a high-metallicity progenitor may also suffice for explaining the observed UV suppression.

    \item Our search for H, He, and O emission in the nebular-phase spectrum confirms there is no evidence for stripped H- or He-rich materials from the companion, nor O-rich CSM from a violent merger. This mainly confirms that a H-poor companion is more likely for the progenitor system, disfavouring the $>$ 2~\msol\ subgiant as the source of the early excess emission.

    \item If the observed early excess emission is produced by CSM from the mass transfer process that triggers the SN explosion, then the SN may need explode on a sufficiently rapid mass transfer timescale to avoid extended components of the CSM producing narrow emission lines. No such lines are seen in the earliest spectrum from only $\sim$ 0.6 days post-explosion.

    \item The \siii\ HVF velocity evolution is best-explained by a secondary component of high-velocity ($\sim$ 23,500~km~s$^{-1}$) material embedded in the outer $<$ 16\% of the ejecta mass along the line of sight.
    While \ni56\ or ejecta-CSM interaction in the outer ejecta may produce HVFs via opacity enhancement, these processes are not likely to produce co-evolving HV Ca and Si components, nor can they readily explain the \siii\ HVF power-law evolution and plateau timing/velocity.

    \item We detect weak contribution from \cii~$\lambda$6580~\AA\ to the \siii~$\lambda$6355~\AA\ feature at a few epochs prior to the disappearance of HVFs at $-$7.8 days, suggesting that the high-velocity material may be partially burnt.

    \item An off-center ignited delayed-detonation can conceivably create a low-mass, high-velocity, partially-burned plume---containing C, Si, Ca and a slight enrichment of radioactive Fe-peak elements---along the line of sight via subsonic deflagration, which is then frozen into homologous expansion by the detonation. Future modelling is needed to confirm if such an explosion can explain the observed early excess emission, UV suppression, and HVF evolution of \uname. 

    \item Double-detonations and pulsations may also create a high-velocity, possibly \ni56-enriched shell in the outer ejecta. However, double-detonations are less compatible with the near-Chandrasekhar-mass ejecta favoured by multiple lines of evidence (see above), and both mechanisms fail to produce the complex photosphere geometry required to explain the simultaneous presence of \siii\ HVFs and PVFs in the early spectra.
\end{itemize}

Only a handful of \tase\ have sufficiently early (= $\sim$ 0--2 days) multi-band/spectroscopic observations that probe the presence of excess emission, UV suppression, and HVF features at the phases seen in \uname.
The ones that have been seen so far, including normal \tase~2018aoz \citep{Ni2022natas}, 2017cbv \citep{Hosseinzadeh2017apj}, 2012fr \citep{Contreras2018apj}, 2011fe \citep{Pereira2013aa}, and 2009ig \citep{Marion2013apj}, reveal a surprising variety of features in those epochs, reflecting inhomogeneities in their origin processes.
Our analyses of \uname\ highlight how different observations, including high-cadence multi-band photometry and multi-epoch spectroscopy can be used to narrow down the origin of these early features.
Our results indicate that asymmetric and subsonic explosion processes play an important role in the explosions of normal \tase\ from near-Chandrasekhar-mass WDs, of which \uname\ is perhaps one of the best candidates to date.

\nopagebreak

\vskip 5.8mm plus 1mm minus 1mm
\vskip1sp
\section*{Acknowledgments}
\vskip4pt

This research has made use of the KMTNet system operated by the Korea Astronomy and Space Science Institute (KASI) and the data were obtained at three host sites of CTIO in Chile, SAAO in South Africa, and SSO in Australia.
Data transfer from the host site to KASI was supported by the Korea Research Environment Open NETwork (KREONET).
This research was supported by KASI under the R\&D program (Project No. 2023-1-868-03) supervised by the Ministry of Science and ICT.
This research is also based on observations obtained at the international Gemini-S Observatory, a program of NSF’s NOIRLab, which is managed by the Association of Universities for Research in Astronomy (AURA) under a cooperative agreement with the National Science Foundation on behalf of the Gemini Observatory partnership: the National Science Foundation (United States), National Research Council (Canada), Agencia Nacional de Investigaci\'{o}n y Desarrollo (Chile), Ministerio de Ciencia, Tecnolog\'{i}a e Innovaci\'{o}n (Argentina), Minist\'{e}rio da Ci\^{e}ncia, Tecnologia, Inova\c{c}\~{o}es e Comunica\c{c}\~{o}es (Brazil), and Korea Astronomy and Space Science Institute (Republic of Korea).
The Gemini-S observations were obtained under the Canadian Gemini Office (PID: GS-2021B-Q-114 and GS-2022A-Q-115) of the National Research Council and acquired through the Gemini Observatory Archive at NSF’s NOIRLab.
D.-S.M. and M.R.D. are supported by Discovery Grants from the Natural Sciences and Engineering Research Council of Canada.
D.-S.M. was supported in part by a Leading Edge Fund from the Canadian Foundation for Innovation (project No. 30951).
M.R.D. was supported in part by the Canada Research Chairs Program, the Canadian Institute for Advanced Research (CIFAR), and the Dunlap Institute at the University of Toronto.

\vspace{5mm}

\software{SNooPy \citep{Burns2011aj}, SNAP (\url{https://github.com/niyuanqi/SNAP}), IRAF}

\pagebreak

\appendix
\restartappendixnumbering

\section{Identification of \caii\ NIR Triplet High-Velocity Components}\label{sec:cafit}

\begin{figure*}[t!]
\epsscale{\scl}
\begin{center}
\includegraphics[width=0.9\textwidth]{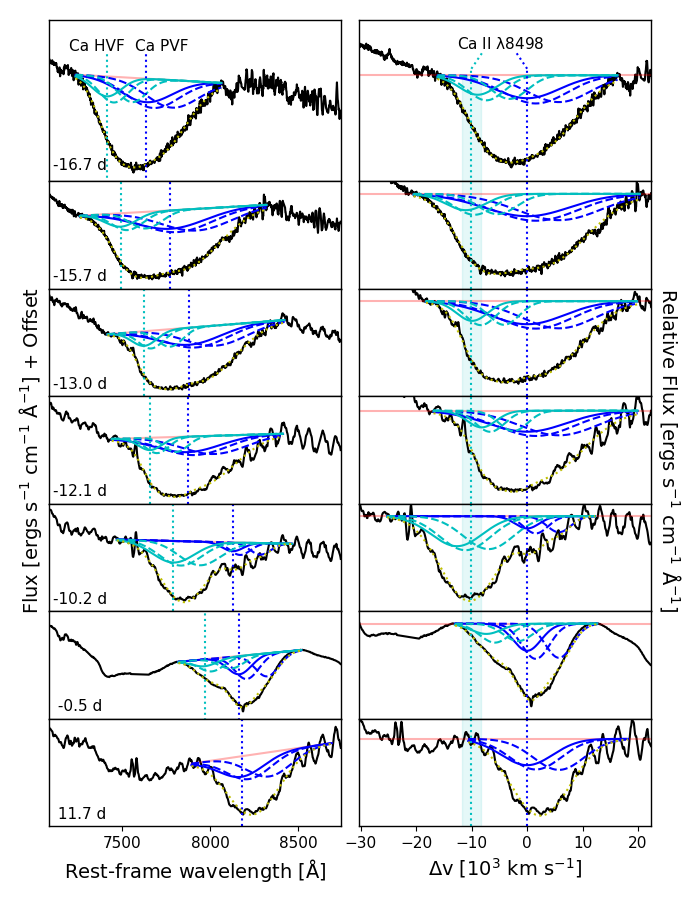}
\end{center}
\caption{Same as Figure~\ref{fig:sifit}, but showing multi-component Gaussian fits of the Ca NIR triplet with PVF (cyan curves) and HVF (blue curves) components. The solid curve in each component represent the \caii~$\lambda$8498~\AA\ line of the triplet, while the dashed curves represent the other two lines at 8542 and 8662~\AA.
Rows (1--2) are during the early \siii\ PVF and HVF decline, (3--5) span the \siii\ HVF plateau (see Figure~\ref{fig:siexp}), (6) near $B$-band maximum is the last phase when the \caii\ NIR triplet HVF is detected, and (7) is from after $B$-band maximum.
The x-axes of the right panels represent velocity in the rest frame of the fitted minima of the \caii~$\lambda$8498~\AA\ PVF at each phase (blue vertical dotted line).
The cyan vertical dotted line with shaded region represents the mean and 1-$\sigma$ velocity of the fitted minima of the \caii~$\lambda$8498~\AA\ HVF in this frame.
\label{fig:cafit}}
\end{figure*}

We apply multi-component Gaussian fitting to the \caii\ NIR triplet that often shows HVFs in \tase.
We use the method described in Section~\ref{subsec:sifit} to fit either one or two velocity components to the feature.
However, since the \caii\ NIR triplet consists of three lines ($\lambda$8498, 8542 and 8662~\AA), a triplet of symmetric Gaussians is used to fit each velocity component.
We use three fit parameters for each triplet of Gaussians based on \citet{Silverman2015mnras}: (1) the central wavelength of the first Gaussian, representing that of the \caii~$\lambda$8498~\AA\ minimum; (2) the scale height of the first Gaussian, representing the strength of \caii~$\lambda$8498~\AA; and (3) the width shared across all three Gaussians.
We fix the positions of the other two lines (8542 and 8662~\AA) according to the redshift of \caii~$\lambda$8498~\AA, and scale them according to their relative strengths using their $gf$-weights\footnote{\url{http://www.nist.gov/pml/data/asd.cfm}} in the optically-thin regime.
Similar fitting of the \caii\ NIR triplet in SN~2012fr has shown that the choice of optically-thin regime does not substantially affect the measurements, though the fit can be slightly better than the optically-thick case for early \tase\ \citep{Childress2013apj}.
For noise estimation, we apply the second-order Savitsky-Golay filter with an increased width of 200~\AA\ for the FTS spectra (Table~\ref{tab:sifit}) due to increased noise towards the red edge of the wavelength range accommodated by the increased width of the \caii\ NIR triplet feature ($>$ 500~\AA\ in all epochs).

Figure~\ref{fig:cafit} (left panels) shows the multi-component Gaussian fits (yellow dotted curves) to the \caii\ NIR triplet for a sample of the spectra of \uname.
For the eight spectra from $\leq -$0.5 days since $B$-band maximum, excellent fits are obtained using two components with \chisqr\ in the range of 1.3--8.0, while a single-component fit is inadequate.
The components closer to the red and blue edges of the feature shown with blue and cyan curves are usually called the PVF and HVF of \caii, respectively \citep{Silverman2015mnras}. 
The solid curves of each component represent the \caii~$\lambda$8498~\AA\ line while the dashed do $\lambda$8542 and 8662~\AA.

For the twelve spectra after $B$-band maximum we obtain good fits with a single velocity component (\chisqr\ in the range of 1.0--3.9) except for two spectra with particularly high S/N = 74 and 46 from $+$12.4 and $+$52.3 days, respectively.
For those two spectra, we can obtain a slightly improved \chisqr\ by fitting two velocity components.
However, the resulting \caii\ PVF velocities are $<$ 6500~km~s$^{-1}$ in both epochs---far below expectations for \tas\ photospheric velocities---while the one-component fits result in velocities that closely trace the \siii\ PVF velocity evolution in all epochs (see Figure~\ref{fig:siexp}).
Thus, we adopt the one-component fit for all twelve spectra.
The \caii\ PVF and HVF velocities measured using the fits are presented in Table~\ref{tab:sifit} and shown in Figure~\ref{fig:siexp} (bottom panel; pentagons and diamonds, respectively).

We identify three critical phases in the velocity evolution of the \caii\ NIR triplet in \uname\ as follows:
\begin{enumerate}
    \item The \caii\ PVF velocity initially traces that of the \siii\ HVF closely until $\sim$ 3.6 rest-frame days post-explosion and also the \siii\ HVF plateau until $\sim$ 5 days. The \caii\ HVF velocity averages (10.1 $\pm$ 1.7)~$\times$~10$^3$~km~s$^{-1}$ faster than that of its PVF.
    \item Sparse data from $\sim$ 5 rest-frame days post-explosion until near the end of the \siii\ HVF plateau, appear to show the velocity of the \caii\ PVF dropping to the level of the \siii\ PVF while that of the \caii\ HVF drops to near that of the \siii\ HVF at the end of its plateau.
    \item After $B$-band maximum, the \caii\ PVF velocity evolution continues to closely trace that of the \siii\ PVF, making them ``co-evolving PV Ca and Si components'', while the \caii\ HVF can no longer be identified.
\end{enumerate}

The close correlation between the velocity evolution of the PVF component of the \caii\ NIR triplet and those of the \siii\ PVF and HVF is consistent with the co-distribution of \caii\ with \siii\ as part of primary and secondary photospheres (Section~\ref{subsec:hvfid}).
The additional \caii\ HVF that appears during the first evolutionary phase indicates the presence of a third line-forming region at an even higher velocity than the secondary photosphere, though the nature of this high-velocity line-forming region remains unclear and warrants future investigation.
The observed \caii\ HVF velocity of (40,770 $\pm$ 100)~km~s$^{-1}$ in the earliest spectrum of \uname\ from $-$16.7 rest-frame days since $B$-band maximum (or 0.6 days post-explosion) is the \emph{highest expansion velocity ever observed in a \tas} \citep{Silverman2015mnras} as well as extreme for any type of SN, surpassed only by SNe associated with gamma ray bursts \citep{Izzo2019natur}.
We note that the average observed HVF/PVF velocity ratio of \caii\ during the first evolutionary phase is $\sim$ 1.4, which appears to be similar to that of \siii\ (= 1.32; Section~\ref{subsec:expl}).
If the origin of the HVF and PVF components of \caii\ in \uname\ is analogous to that of the HVF and PVF components of \siii\ (Section~\ref{sec:velocity}), then a third component of relativistically-expanding material in the outer few \% of the line of sight ejecta mass could be one possible origin of the \caii\ HVF.


\begin{thebibliography}{}
\expandafter\ifx\csname natexlab\endcsname\relax\def\natexlab#1{#1}\fi
\providecommand{\url}[1]{\href{#1}{#1}}
\providecommand{\dodoi}[1]{doi:~\href{http://doi.org/#1}{\nolinkurl{#1}}}
\providecommand{\doeprint}[1]{\href{http://ascl.net/#1}{\nolinkurl{http://ascl.net/#1}}}
\providecommand{\doarXiv}[1]{\href{https://arxiv.org/abs/#1}{\nolinkurl{https://arxiv.org/abs/#1}}}

\bibitem[Matzner et al.(2013)]{Matzner2013apj} Matzner, C.~D., Levin, Y., \& Ro, S.\ 2013, \apj, 779, 60. \dodoi{10.1088/0004-637X/779/1/60}

\bibitem[{{Afsariardchi} {et~al.}(2019){Afsariardchi}, {Moon}, {Drout},
  {Gonz{\'a}lez-Gait{\'a}n}, {Ni}, {Matzner}, {Kim}, {Lee}, {Park}, {Gal-Yam},
  {Pignata}, {Koo}, {Ryder}, {Cha}, \& {Lee}}]{Afsariardchi2019apj}
{Afsariardchi}, N., {Moon}, D.-S., {Drout}, M.~R., {et~al.} 2019, \apj, 881,
  22, \dodoi{10.3847/1538-4357/ab2be6}

\bibitem[{{Allison} {et~al.}(2014){Allison}, {Sadler}, \&
  {Meekin}}]{Allison2014mnras}
{Allison}, J.~R., {Sadler}, E.~M., \& {Meekin}, A.~M. 2014, \mnras, 440, 696,
  \dodoi{10.1093/mnras/stu289}

\bibitem[{{Arnett}(1982)}]{Arnett1982apj}
{Arnett}, W.~D. 1982, \apj, 253, 785, \dodoi{10.1086/159681}

\bibitem[{{Ashall} {et~al.}(2022){Ashall}, {Lu}, {Shappee}, {Burns}, {Hsiao},
  {Kumar}, {Morrell}, {Phillips}, {Shahbandeh}, {Baron}, {Boutsia}, {Brown},
  {DerKacy}, {Galbany}, {Hoeflich}, {Krisciunas}, {Mazzali}, {Piro},
  {Stritzinger}, \& {Suntzeff}}]{Ashall2022apj}
{Ashall}, C., {Lu}, J., {Shappee}, B.~J., {et~al.} 2022, \apjl, 932, L2,
  \dodoi{10.3847/2041-8213/ac7235}

\bibitem[{{Aznar-Sigu{\'a}n} {et~al.}(2015){Aznar-Sigu{\'a}n},
  {Garc{\'\i}a-Berro}, {Lor{\'e}n-Aguilar}, {Soker}, \&
  {Kashi}}]{Aznar2015mnras}
{Aznar-Sigu{\'a}n}, G., {Garc{\'\i}a-Berro}, E., {Lor{\'e}n-Aguilar}, P.,
  {Soker}, N., \& {Kashi}, A. 2015, \mnras, 450, 2948,
  \dodoi{10.1093/mnras/stv824}

\bibitem[{{Blondin} {et~al.}(2012){Blondin}, {Matheson}, {Kirshner}, {Mand el},
  {Berlind}, {Calkins}, {Challis}, {Garnavich}, {Jha}, {Modjaz}, {Riess}, \&
  {Schmidt}}]{Blondin2012aj}
{Blondin}, S., {Matheson}, T., {Kirshner}, R.~P., {et~al.} 2012, \aj, 143, 126,
  \dodoi{10.1088/0004-6256/143/5/126}

\bibitem[Blondin et al.(2023)]{Blondin2023arxiv} Blondin, S., Dessart, L., Hillier, D.~J., et al.\ 2023, \doarXiv{arXiv:2306.07116}

\bibitem[{{Boos} {et~al.}(2021){Boos}, {Townsley}, {Shen}, {Caldwell}, \&
  {Miles}}]{Boos2021apj}
{Boos}, S.~J., {Townsley}, D.~M., {Shen}, K.~J., {Caldwell}, S., \& {Miles},
  B.~J. 2021, \apj, 919, 126, \dodoi{10.3847/1538-4357/ac07a2}

\bibitem[Yarbrough et al.(2023)]{Yarbrough2023mnras} Yarbrough, Z., Baron, E., DerKacy, J.~M., et al.\ 2023, \mnras, 521, 3873. \dodoi{10.1093/mnras/stad758}

\bibitem[{{Bostroem} {et~al.}(2021){Bostroem}, {Jha}, {Randriamampandry},
  {Valenti}, {Sand}, {Wyatt}, {Lundquist}, {Andrews}, {Jencson}, {Dong},
  {Janzen}, {Pearson}, {Hosseinzadeh}, \& {Meza}}]{Bostroem2021tnscr}
{Bostroem}, K.~A., {Jha}, S.~W., {Randriamampandry}, S., {et~al.} 2021,
  Transient Name Server Classification Report, 2021-3888, 1

\bibitem[{{Boty{\'a}nszki} {et~al.}(2018){Boty{\'a}nszki}, {Kasen}, \&
  {Plewa}}]{Botyanszki2018apj}
{Boty{\'a}nszki}, J., {Kasen}, D., \& {Plewa}, T. 2018, \apjl, 852, L6,
  \dodoi{10.3847/2041-8213/aaa07b}

\bibitem[{{Branch} {et~al.}(1983){Branch}, {Lacy}, {McCall}, {Sutherland},
  {Uomoto}, {Wheeler}, \& {Wills}}]{Branch1983apj}
{Branch}, D., {Lacy}, C.~H., {McCall}, M.~L., {et~al.} 1983, \apj, 270, 123,
  \dodoi{10.1086/161103}

\bibitem[{{Branch} {et~al.}(2006){Branch}, {Dang}, {Hall}, {Ketchum},
  {Melakayil}, {Parrent}, {Troxel}, {Casebeer}, {Jeffery}, \&
  {Baron}}]{Branch2006pasp}
{Branch}, D., {Dang}, L.~C., {Hall}, N., {et~al.} 2006, \pasa, 118, 560,
  \dodoi{10.1086/502778}

\bibitem[{{Brown} {et~al.}(2014){Brown}, {Kuin}, {Scalzo}, {Smitka}, {de
  Pasquale}, {Holland}, {Krisciunas}, {Milne}, \& {Wang}}]{Brown2014apj}
{Brown}, P.~J., {Kuin}, P., {Scalzo}, R., {et~al.} 2014, \apj, 787, 29,
  \dodoi{10.1088/0004-637X/787/1/29}

\bibitem[{{Brown} {et~al.}(2013){Brown}, {Baliber}, {Bianco}, {Bowman},
  {Burleson}, {Conway}, {Crellin}, {Depagne}, {De Vera}, {Dilday}, {Dragomir},
  {Dubberley}, {Eastman}, {Elphick}, {Falarski}, {Foale}, {Ford}, {Fulton},
  {Garza}, {Gomez}, {Graham}, {Greene}, {Haldeman}, {Hawkins}, {Haworth},
  {Haynes}, {Hidas}, {Hjelstrom}, {Howell}, {Hygelund}, {Lister}, {Lobdill},
  {Martinez}, {Mullins}, {Norbury}, {Parrent}, {Paulson}, {Petry}, {Pickles},
  {Posner}, {Rosing}, {Ross}, {Sand}, {Saunders}, {Shobbrook}, {Shporer},
  {Street}, {Thomas}, {Tsapras}, {Tufts}, {Valenti}, {Vander Horst}, {Walker},
  {White}, \& {Willis}}]{Brown2013pasp}
{Brown}, T.~M., {Baliber}, N., {Bianco}, F.~B., {et~al.} 2013, \pasa, 125,
  1031, \dodoi{10.1086/673168}

\bibitem[{{Burns} {et~al.}(2011){Burns}, {Stritzinger}, {Phillips}, {Kattner},
  {Persson}, {Madore}, {Freedman}, {Boldt}, {Campillay}, {Contreras},
  {Folatelli}, {Gonzalez}, {Krzeminski}, {Morrell}, {Salgado}, \&
  {Suntzeff}}]{Burns2011aj}
{Burns}, C.~R., {Stritzinger}, M., {Phillips}, M.~M., {et~al.} 2011, \aj, 141,
  19, \dodoi{10.1088/0004-6256/141/1/19}

\bibitem[{{Burns} {et~al.}(2014){Burns}, {Stritzinger}, {Phillips}, {Hsiao},
  {Contreras}, {Persson}, {Folatelli}, {Boldt}, {Campillay}, {Castell{\'o}n},
  {Freedman}, {Madore}, {Morrell}, {Salgado}, \& {Suntzeff}}]{Burns2014apj}
---. 2014, \apj, 789, 32, \dodoi{10.1088/0004-637X/789/1/32}

\bibitem[{{Burns} {et~al.}(2018){Burns}, {Parent}, {Phillips}, {Stritzinger},
  {Krisciunas}, {Suntzeff}, {Hsiao}, {Contreras}, {Anais}, {Boldt}, {Busta},
  {Campillay}, {Castell{\'o}n}, {Folatelli}, {Freedman}, {Gonz{\'a}lez},
  {Hamuy}, {Heoflich}, {Krzeminski}, {Madore}, {Morrell}, {Persson}, {Roth},
  {Salgado}, {Ser{\'o}n}, \& {Torres}}]{Burns2018apj}
{Burns}, C.~R., {Parent}, E., {Phillips}, M.~M., {et~al.} 2018, \apj, 869, 56,
  \dodoi{10.3847/1538-4357/aae51c}

\bibitem[{{Chevalier}(1982)}]{Chevalier1982apj}
{Chevalier}, R.~A. 1982, \apj, 258, 790, \dodoi{10.1086/160126}

\bibitem[{{Childress} {et~al.}(2013){Childress}, {Scalzo}, {Sim}, {Tucker},
  {Yuan}, {Schmidt}, {Cenko}, {Silverman}, {Contreras}, {Hsiao}, {Phillips},
  {Morrell}, {Jha}, {McCully}, {Filippenko}, {Anderson}, {Benetti}, {Bufano},
  {de Jaeger}, {Forster}, {Gal-Yam}, {Le Guillou}, {Maguire}, {Maund},
  {Mazzali}, {Pignata}, {Smartt}, {Spyromilio}, {Sullivan}, {Taddia},
  {Valenti}, {Bayliss}, {Bessell}, {Blanc}, {Carson}, {Clubb}, {de Burgh-Day},
  {Desjardins}, {Fang}, {Fox}, {Gates}, {Ho}, {Keller}, {Kelly}, {Lidman},
  {Loaring}, {Mould}, {Owers}, {Ozbilgen}, {Pei}, {Pickering}, {Pracy}, {Rich},
  {Schaefer}, {Scott}, {Stritzinger}, {Vogt}, \& {Zhou}}]{Childress2013apj}
{Childress}, M.~J., {Scalzo}, R.~A., {Sim}, S.~A., {et~al.} 2013, \apj, 770,
  29, \dodoi{10.1088/0004-637X/770/1/29}

\bibitem[{{Contardo} {et~al.}(2000){Contardo}, {Leibundgut}, \&
  {Vacca}}]{Contardo2000aap}
{Contardo}, G., {Leibundgut}, B., \& {Vacca}, W.~D. 2000, \aap, 359, 876.
\newblock \doarXiv{astro-ph/0005507}

\bibitem[{{Contreras} {et~al.}(2018){Contreras}, {Phillips}, {Burns}, {Piro},
  {Shappee}, {Stritzinger}, {Baltay}, {Brown}, {Conseil}, {Klotz}, {Nugent},
  {Turpin}, {Parker}, {Rabinowitz}, {Hsiao}, {Morrell}, {Campillay},
  {Castell{\'o}n}, {Corco}, {Gonz{\'a}lez}, {Krisciunas}, {Ser{\'o}n},
  {Tucker}, {Walker}, {Baron}, {Cain}, {Childress}, {Folatelli}, {Freedman},
  {Hamuy}, {Hoeflich}, {Persson}, {Scalzo}, {Schmidt}, \&
  {Suntzeff}}]{Contreras2018apj}
{Contreras}, C., {Phillips}, M.~M., {Burns}, C.~R., {et~al.} 2018, \apj, 859,
  24, \dodoi{10.3847/1538-4357/aabaf8}

\bibitem[{{Deckers} {et~al.}(2022){Deckers}, {Maguire}, {Magee}, {Dimitriadis},
  {Smith}, {Sainz de Murieta}, {Miller}, {Goobar}, {Nordin}, {Rigault},
  {Bellm}, {Coughlin}, {Laher}, {Shupe}, {Graham}, {Kasliwal}, \&
  {Walters}}]{Deckers2022mnras}
{Deckers}, M., {Maguire}, K., {Magee}, M.~R., {et~al.} 2022, \mnras, 512, 1317,
  \dodoi{10.1093/mnras/stac558}

\bibitem[{{DerKacy} {et~al.}(2023){DerKacy}, {Ashall}, {Hoeflich}, {Baron},
  {Shappee}, {Baade}, {Andrews}, {Bostroem}, {Brown}, {Burns}, {Burrow},
  {Cikota}, {de Jaeger}, {Do}, {Dong}, {Dominguez}, {Galbany}, {Hsiao},
  {Karamehmetoglu}, {Krisciunas}, {Kumar}, {Lu}, {Evans}, {Maund}, {Mazzali},
  {Medler}, {Morrell}, {Patat}, {Phillips}, {Shahbandeh}, {Stangl}, {Stevens},
  {Stritzinger}, {Suntzeff}, {Telesco}, {Tucker}, {Valenti}, {Wang}, {Yang},
  {Jha}, \& {Kwok}}]{DerKacy2023apj}
{DerKacy}, J.~M., {Ashall}, C., {Hoeflich}, P., {et~al.} 2023, \apjl, 945, L2,
  \dodoi{10.3847/2041-8213/acb8a8}

\bibitem[Elagali et al.(2019)]{Elagali2019mnras} Elagali, A., Staveley-Smith, L., Rhee, J., et al.\ 2019, \mnras, 487, 2797. \dodoi{10.1093/mnras/stz1448}

\bibitem[{{Dessart} {et~al.}(2020){Dessart}, {Leonard}, \&
  {Prieto}}]{Dessart2020aa}
{Dessart}, L., {Leonard}, D.~C., \& {Prieto}, J.~L. 2020, \aap, 638, A80,
  \dodoi{10.1051/0004-6361/202037854}

\bibitem[{{Dimitriadis} {et~al.}(2019){Dimitriadis}, {Foley}, {Rest}, {Kasen},
  {Piro}, {Polin}, {Jones}, {Villar}, {Narayan}, \&
  {Coulter}}]{Dimitriadis2019apj}
{Dimitriadis}, G., {Foley}, R.~J., {Rest}, A., {et~al.} 2019, \apjl, 870, L1,
  \dodoi{10.3847/2041-8213/aaedb0}

\bibitem[{{Dimitriadis} {et~al.}(2023){Dimitriadis}, {Maguire}, {Karambelkar},
  {Lebron}, {Liu}, {Kozyreva}, {Miller}, {Ridden-Harper}, {Anderson}, {Chen},
  {Coughlin}, {Della Valle}, {Drake}, {Galbany}, {Gromadzki}, {Groom},
  {Guti{\'e}rrez}, {Ihanec}, {Inserra}, {Johansson}, {M{\"u}ller-Bravo},
  {Nicholl}, {Polin}, {Rusholme}, {Schulze}, {Sollerman}, {Srivastav},
  {Taggart}, {Wang}, {Yang}, \& {Young}}]{Dimitriadis2023mnras}
{Dimitriadis}, G., {Maguire}, K., {Karambelkar}, V.~R., {et~al.} 2023, \mnras,
  \dodoi{10.1093/mnras/stad536}

\bibitem[{{Fitzpatrick}(1999)}]{Fitzpatrick1999pasp}
{Fitzpatrick}, E.~L. 1999, \pasp, 111, 63, \dodoi{10.1086/316293}

\bibitem[{{Foley} {et~al.}(2012){Foley}, {Challis}, {Filippenko},
  {Ganeshalingam}, {Landsman}, {Li}, {Marion}, {Silverman}, {Beaton},
  {Bennert}, {Cenko}, {Childress}, {Guhathakurta}, {Jiang}, {Kalirai},
  {Kirshner}, {Stockton}, {Tollerud}, {Vink{\'o}}, {Wheeler}, \&
  {Woo}}]{Foley2012apj}
{Foley}, R.~J., {Challis}, P.~J., {Filippenko}, A.~V., {et~al.} 2012, \apj,
  744, 38, \dodoi{10.1088/0004-637X/744/1/38}

\bibitem[{{Gerardy} {et~al.}(2004){Gerardy}, {H{\"o}flich}, {Fesen}, {Marion},
  {Nomoto}, {Quimby}, {Schaefer}, {Wang}, \& {Wheeler}}]{Gerardy2004apj}
{Gerardy}, C.~L., {H{\"o}flich}, P., {Fesen}, R.~A., {et~al.} 2004, \apj, 607,
  391, \dodoi{10.1086/383488}

\bibitem[{{Guillochon} {et~al.}(2010){Guillochon}, {Dan}, {Ramirez-Ruiz}, \&
  {Rosswog}}]{Guillochon2010apj}
{Guillochon}, J., {Dan}, M., {Ramirez-Ruiz}, E., \& {Rosswog}, S. 2010, \apjl,
  709, L64, \dodoi{10.1088/2041-8205/709/1/L64}

\bibitem[{{Guillochon} {et~al.}(2017){Guillochon}, {Parrent}, {Kelley}, \&
  {Margutti}}]{Guillochon2017apj}
{Guillochon}, J., {Parrent}, J., {Kelley}, L.~Z., \& {Margutti}, R. 2017, \apj,
  835, 64, \dodoi{10.3847/1538-4357/835/1/64}

\bibitem[{{Hoeflich} {et~al.}(1995){Hoeflich}, {Khokhlov}, \&
  {Wheeler}}]{Hoeflich1995apj}
{Hoeflich}, P., {Khokhlov}, A.~M., \& {Wheeler}, J.~C. 1995, \apj, 444, 831,
  \dodoi{10.1086/175656}

\bibitem[{{Hook} {et~al.}(2004){Hook}, {J{\o}rgensen}, {Allington-Smith},
  {Davies}, {Metcalfe}, {Murowinski}, \& {Crampton}}]{Hook2004}
{Hook}, I.~M., {J{\o}rgensen}, I., {Allington-Smith}, J.~R., {et~al.} 2004,
  \pasp, 116, 425, \dodoi{10.1086/383624}

\bibitem[{{Hosseinzadeh} {et~al.}(2017){Hosseinzadeh}, {Sand}, {Valenti},
  {Brown}, {Howell}, {McCully}, {Kasen}, {Arcavi}, {Bostroem}, {Tartaglia},
  {Hsiao}, {Davis}, {Shahbandeh}, \& {Stritzinger}}]{Hosseinzadeh2017apj}
{Hosseinzadeh}, G., {Sand}, D.~J., {Valenti}, S., {et~al.} 2017, \apj, 845,
  L11, \dodoi{10.3847/2041-8213/aa8402}

\bibitem[{{Hosseinzadeh} {et~al.}(2022){Hosseinzadeh}, {Sand}, {Lundqvist},
  {Andrews}, {Bostroem}, {Dong}, {Janzen}, {Jencson}, {Lundquist}, {Meza
  Retamal}, {Pearson}, {Valenti}, {Wyatt}, {Burke}, {Howell}, {McCully},
  {Newsome}, {Gonzalez}, {Pellegrino}, {Terreran}, {Kwok}, {Jha}, {Strader},
  {Kundu}, {Ryder}, {Haislip}, {Kouprianov}, \&
  {Reichart}}]{Hosseinzadeh2022apj}
{Hosseinzadeh}, G., {Sand}, D.~J., {Lundqvist}, P., {et~al.} 2022, \apjl, 933,
  L45, \dodoi{10.3847/2041-8213/ac7cef}

\bibitem[{{Hsiao} {et~al.}(2007){Hsiao}, {Conley}, {Howell}, {Sullivan},
  {Pritchet}, {Carlberg}, {Nugent}, \& {Phillips}}]{Hsiao2007apj}
{Hsiao}, E.~Y., {Conley}, A., {Howell}, D.~A., {et~al.} 2007, \apj, 663, 1187,
  \dodoi{10.1086/518232}

\bibitem[{{Iben} \& {Tutukov}(1984)}]{Iben&Tutukov1984apjs}
{Iben}, I., J., \& {Tutukov}, A.~V. 1984, \apjs, 54, 335,
  \dodoi{10.1086/190932}

\bibitem[{{Izzo} {et~al.}(2019){Izzo}, {de Ugarte Postigo}, {Maeda},
  {Th{\"o}ne}, {Kann}, {Della Valle}, {Sagues Carracedo}, {Micha{\l}owski},
  {Schady}, {Schmidl}, {Selsing}, {Starling}, {Suzuki}, {Bensch}, {Bolmer},
  {Campana}, {Cano}, {Covino}, {Fynbo}, {Hartmann}, {Heintz}, {Hjorth},
  {Japelj}, {Kami{\'n}ski}, {Kaper}, {Kouveliotou}, {Kru{\.Z}y{\'n}ski},
  {Kwiatkowski}, {Leloudas}, {Levan}, {Malesani}, {Micha{\l}owski},
  {Piranomonte}, {Pugliese}, {Rossi}, {S{\'a}nchez-Ram{\'\i}rez}, {Schulze},
  {Steeghs}, {Tanvir}, {Ulaczyk}, {Vergani}, \& {Wiersema}}]{Izzo2019natur}
{Izzo}, L., {de Ugarte Postigo}, A., {Maeda}, K., {et~al.} 2019, \nat, 565,
  324, \dodoi{10.1038/s41586-018-0826-3}

\bibitem[{{Jiang} {et~al.}(2018){Jiang}, {Doi}, {Maeda}, \&
  {Shigeyama}}]{Jiang2018apj}
{Jiang}, J.-a., {Doi}, M., {Maeda}, K., \& {Shigeyama}, T. 2018, \apj, 865,
  149, \dodoi{10.3847/1538-4357/aadb9a}

\bibitem[{{Jiang} {et~al.}(2017){Jiang}, {Doi}, {Maeda}, {Shigeyama}, {Nomoto},
  {Yasuda}, {Jha}, {Tanaka}, {Morokuma}, {Tominaga}, {Ivezi{\'c}},
  {Ruiz-Lapuente}, {Stritzinger}, {Mazzali}, {Ashall}, {Mould}, {Baade},
  {Suzuki}, {Connolly}, {Patat}, {Wang}, {Yoachim}, {Jones}, {Furusawa}, \&
  {Miyazaki}}]{Jiang2017nat}
{Jiang}, J.-A., {Doi}, M., {Maeda}, K., {et~al.} 2017, \nat, 550, 80,
  \dodoi{10.1038/nature23908}

\bibitem[{{Jiang} {et~al.}(2021){Jiang}, {Maeda}, {Kawabata}, {Doi},
  {Shigeyama}, {Tanaka}, {Tominaga}, {Nomoto}, {Niino}, {Sako}, {Ohsawa},
  {Schramm}, {Yamanaka}, {Kobayashi}, {Takahashi}, {Nakaoka}, {Kawabata},
  {Isogai}, {Aoki}, {Kondo}, {Mori}, {Arimatsu}, {Kasuga}, {Okumura},
  {Urakawa}, {Reichart}, {Taguchi}, {Arima}, {Beniyama}, {Uno}, \&
  {Hamada}}]{Jiang2021apj}
{Jiang}, J.-a., {Maeda}, K., {Kawabata}, M., {et~al.} 2021, \apjl, 923, L8,
  \dodoi{10.3847/2041-8213/ac375f}

\bibitem[{{Kasen}(2010)}]{Kasen2010apj}
{Kasen}, D. 2010, \apj, 708, 1025, \dodoi{10.1088/0004-637X/708/2/1025}

\bibitem[{{Kasen} \& {Woosley}(2009)}]{Kasen2009apj}
{Kasen}, D., \& {Woosley}, S.~E. 2009, \apj, 703, 2205,
  \dodoi{10.1088/0004-637X/703/2/2205}

\bibitem[Raskin \& Kasen(2013)]{Raskin2013apj} Raskin, C. \& Kasen, D.\ 2013, \apj, 772, 1. \dodoi{10.1088/0004-637X/772/1/1}

\bibitem[{{Kim} {et~al.}(2016){Kim}, {Lee}, {Park}, {Kim}, {Cha}, {Lee}, {Han},
  {Chun}, \& {Yuk}}]{Kim2016jkas}
{Kim}, S.-L., {Lee}, C.-U., {Park}, B.-G., {et~al.} 2016, Journal of the Korean Astronomical Society, 49, 37,
  \dodoi{10.5303/JKAS.2016.49.1.037}

\bibitem[{{Kollmeier} {et~al.}(2019){Kollmeier}, {Chen}, {Dong}, {Morrell},
  {Phillips}, {Kushnir}, {Prieto}, {Piro}, \& {Simon}}]{Kollmeier2019mnras}
{Kollmeier}, J.~A., {Chen}, P., {Dong}, S., {et~al.} 2019, \mnras, 486, 3041,
  \dodoi{10.1093/mnras/stz953}

\bibitem[{{Kromer} {et~al.}(2013){Kromer}, {Pakmor}, {Taubenberger}, {Pignata},
  {Fink}, {R{\"o}pke}, {Seitenzahl}, {Sim}, \& {Hillebrandt}}]{Kromer2013apj}
{Kromer}, M., {Pakmor}, R., {Taubenberger}, S., {et~al.} 2013, \apjl, 778, L18,
  \dodoi{10.1088/2041-8205/778/1/L18}

\bibitem[{{Kuchner} {et~al.}(1994){Kuchner}, {Kirshner}, {Pinto}, \&
  {Leibundgut}}]{Kuchner1994apj}
{Kuchner}, M.~J., {Kirshner}, R.~P., {Pinto}, P.~A., \& {Leibundgut}, B. 1994,
  \apjl, 426, L89, \dodoi{10.1086/187347}

\bibitem[{{Kushnir} {et~al.}(2013){Kushnir}, {Katz}, {Dong}, {Livne}, \&
  {Fern{\'a}ndez}}]{Kushnir2013apj}
{Kushnir}, D., {Katz}, B., {Dong}, S., {Livne}, E., \& {Fern{\'a}ndez}, R.
  2013, \apjl, 778, L37, \dodoi{10.1088/2041-8205/778/2/L37}

\bibitem[{{Kutsuna} \& {Shigeyama}(2015)}]{Kutsuna2015pasj}
{Kutsuna}, M., \& {Shigeyama}, T. 2015, \pasj, 67, 54,
  \dodoi{10.1093/pasj/psv028}

\bibitem[{{Kwok} {et~al.}(2023){Kwok}, {Jha}, {Temim}, {Fox}, {Larison},
  {Camacho-Neves}, {Brenner Newman}, {Pierel}, {Foley}, {Andrews}, {Badenes},
  {Barna}, {Bostroem}, {Deckers}, {Fl{\"o}rs}, {Garnavich}, {Graham}, {Graur},
  {Hosseinzadeh}, {Howell}, {Hughes}, {Johansson}, {Kendrew}, {Kerzendorf},
  {Maeda}, {Maguire}, {McCully}, {O'Brien}, {Rest}, {Sand}, {Shahbandeh},
  {Strolger}, {Szalai}, {Ashall}, {Baron}, {Burns}, {DerKacy}, {Evans},
  {Fisher}, {Galbany}, {Hoeflich}, {Hsiao}, {de Jaeger}, {Karamehmetoglu},
  {Krisciunas}, {Kumar}, {Lu}, {Maund}, {Mazzali}, {Medler}, {Morrell},
  {Phillips}, {Shappee}, {Stritzinger}, {Suntzeff}, {Telesco}, {Tucker}, \&
  {Wang}}]{Kwok2023apj}
{Kwok}, L.~A., {Jha}, S.~W., {Temim}, T., {et~al.} 2023, \apjl, 944, L3,
  \dodoi{10.3847/2041-8213/acb4ec}

\bibitem[{{Lee} {et~al.}(2022){Lee}, {Kim}, {Moon}, {Park}, {Drout}, {Ni}, \&
  {Im}}]{Lee2022apj}
{Lee}, Y., {Kim}, S.~C., {Moon}, D.-S., {et~al.} 2022, \apjl, 925, L22,
  \dodoi{10.3847/2041-8213/ac4c41}

\bibitem[{{Li} {et~al.}(2019){Li}, {Wang}, {Vink{\'o}}, {Mo}, {Hosseinzadeh},
  {Sand}, {Zhang}, {Lin}, {PTSS/TNTS}, {Zhang}, {Wang}, {Zhang}, {Chen},
  {Xiang}, {Rui}, {Huang}, {Li}, {Zhang}, {Li}, {Baron}, {Derkacy}, {Zhao},
  {Sai}, {Zhang}, {Wang}, {LCO}, {Howell}, {McCully}, {Arcavi}, {Valenti},
  {Hiramatsu}, {Burke}, {KEGS}, {Rest}, {Garnavich}, {Tucker}, {Narayan},
  {Shaya}, {Margheim}, {Zenteno}, {Villar}, {UCSC}, {Dimitriadis}, {Foley},
  {Pan}, {Coulter}, {Fox}, {Jha}, {Jones}, {Kasen}, {Kilpatrick}, {Piro},
  {Riess}, {Rojas-Bravo}, {ASAS-SN}, {Shappee}, {Holoien}, {Stanek}, {Drout},
  {Auchettl}, {Kochanek}, {Brown}, {Bose}, {Bersier}, {Brimacombe}, {Chen},
  {Dong}, {Holmbo}, {Mu{\~n}oz}, {Mutel}, {Post}, {Prieto}, {Shields},
  {Tallon}, {Thompson}, {Vallely}, {Villanueva}, {Pan-STARRS}, {Smartt},
  {Smith}, {Chambers}, {Flewelling}, {Huber}, {Magnier}, {Waters}, {Schultz},
  {Bulger}, {Lowe}, {Willman}, {Konkoly/Texas}, {S{\'a}rneczky}, {P{\'a}l},
  {Wheeler}, {B{\'o}di}, {Bogn{\'a}r}, {Cs{\'a}k}, {Cseh}, {Cs{\"o}rnyei},
  {Hanyecz}, {Ign{\'a}cz}, {Kalup}, {K{\"o}nyves-T{\'o}th}, {Kriskovics},
  {Ordasi}, {Rajmon}, {S{\'o}dor}, {Szab{\'o}}, {Szak{\'a}ts}, {Zsidi},
  {Arizona}, {Milne}, {Andrews}, {Smith}, {Bilinski}, {Swift}, {Brown},
  {ePESSTO}, {Nordin}, {Williams}, {Galbany}, {Palmerio}, {Hook}, {Inserra},
  {Maguire}, {Cartier}, {Razza}, {Guti{\'e}rrez}, {North Carolina}, {Hermes},
  {Reding}, {Kaiser}, {ATLAS}, {Tonry}, {Heinze}, {Denneau}, {Weiland},
  {Stalder}, {K2 Mission Team}, {Barentsen}, {Dotson}, {Barclay},
  {Gully-Santiago}, {Hedges}, {Cody}, {Howell}, {Kepler Spacecraft Team},
  {Coughlin}, {Van Cleve}, {Cardoso}, {Larson}, {McCalmont-Everton},
  {Peterson}, {Ross}, {Reedy}, {Osborne}, {McGinn}, {Kohnert}, {Migliorini},
  {Wheaton}, {Spencer}, {Labonde}, {Castillo}, {Beerman}, {Steward}, {Hanley},
  {Larsen}, {Gangopadhyay}, {Kloetzel}, {Weschler}, {Nystrom}, {Moffatt},
  {Redick}, {Griest}, {Packard}, {Muszynski}, {Kampmeier}, {Bjella}, {Flynn},
  \& {Elsaesser}}]{Li2019apj}
{Li}, W., {Wang}, X., {Vink{\'o}}, J., {et~al.} 2019, \apj, 870, 12,
  \dodoi{10.3847/1538-4357/aaec74}

\bibitem[{{Li} {et~al.}(2021){Li}, {Wang}, {Bulla}, {Pan}, {Wang}, {Mo},
  {Zhang}, {Wu}, {Zhang}, {Zhang}, {Xiang}, {Lin}, {Sai}, {Zhang}, {Chen}, \&
  {Yan}}]{Li2021apj}
{Li}, W., {Wang}, X., {Bulla}, M., {et~al.} 2021, \apj, 906, 99,
  \dodoi{10.3847/1538-4357/abc9b5}

\bibitem[{{Maeda} {et~al.}(2010{\natexlab{a}}){Maeda}, {R{\"o}pke}, {Fink},
  {Hillebrandt}, {Travaglio}, \& {Thielemann}}]{Maeda2010apj}
{Maeda}, K., {R{\"o}pke}, F.~K., {Fink}, M., {et~al.} 2010{\natexlab{a}}, \apj,
  712, 624, \dodoi{10.1088/0004-637X/712/1/624}

\bibitem[{{Maeda} {et~al.}(2010{\natexlab{b}}){Maeda}, {Benetti},
  {Stritzinger}, {R{\"o}pke}, {Folatelli}, {Sollerman}, {Taubenberger},
  {Nomoto}, {Leloudas}, {Hamuy}, {Tanaka}, {Mazzali}, \&
  {Elias-Rosa}}]{Maeda2010natur}
{Maeda}, K., {Benetti}, S., {Stritzinger}, M., {et~al.} 2010{\natexlab{b}},
  \nat, 466, 82, \dodoi{10.1038/nature09122}

\bibitem[Baron et al.(2012)]{Baron2012apj} Baron, E., H{\"o}flich, P., Krisciunas, K., et al.\ 2012, \apj, 753, 105. \dodoi{10.1088/0004-637X/753/2/105}

\bibitem[{{Magee} \& {Maguire}(2020)}]{Magee2020aab}
{Magee}, M.~R., \& {Maguire}, K. 2020, \aap, 642, A189,
  \dodoi{10.1051/0004-6361/202037870}

\bibitem[{{Magee} {et~al.}(2020){Magee}, {Maguire}, {Kotak}, {Sim},
  {Gillanders}, {Prentice}, \& {Skillen}}]{Magee2020aa}
{Magee}, M.~R., {Maguire}, K., {Kotak}, R., {et~al.} 2020, \aap, 634, A37,
  \dodoi{10.1051/0004-6361/201936684}

\bibitem[{{Maoz} {et~al.}(2014){Maoz}, {Mannucci}, \&
  {Nelemans}}]{Maoz2014araa}
{Maoz}, D., {Mannucci}, F., \& {Nelemans}, G. 2014, \araa, 52, 107,
  \dodoi{10.1146/annurev-astro-082812-141031}

\bibitem[{{Marion} {et~al.}(2013){Marion}, {Vinko}, {Wheeler}, {Foley},
  {Hsiao}, {Brown}, {Challis}, {Filippenko}, {Garnavich}, {Kirshner},
  {Landsman}, {Parrent}, {Pritchard}, {Roming}, {Silverman}, \&
  {Wang}}]{Marion2013apj}
{Marion}, G.~H., {Vinko}, J., {Wheeler}, J.~C., {et~al.} 2013, \apj, 777, 40,
  \dodoi{10.1088/0004-637X/777/1/40}

\bibitem[{{Marion} {et~al.}(2016){Marion}, {Brown}, {Vink{\'o}}, {Silverman},
  {Sand}, {Challis}, {Kirshner}, {Wheeler}, {Berlind}, {Brown}, {Calkins},
  {Camacho}, {Dhungana}, {Foley}, {Friedman}, {Graham}, {Howell}, {Hsiao},
  {Irwin}, {Jha}, {Kehoe}, {Macri}, {Maeda}, {Mandel}, {McCully}, {Pandya},
  {Rines}, {Wilhelmy}, \& {Zheng}}]{Marion2016apj}
{Marion}, G.~H., {Brown}, P.~J., {Vink{\'o}}, J., {et~al.} 2016, \apj, 820, 92,
  \dodoi{10.3847/0004-637X/820/2/92}

\bibitem[{{Matteucci}(2012)}]{Matteucci2012book}
{Matteucci}, F. 2012, {Chemical Evolution of Galaxies},
  \dodoi{10.1007/978-3-642-22491-1}

\bibitem[{{Mazzali} {et~al.}(2022){Mazzali}, {Benetti}, {Stritzinger}, \&
  {Ashall}}]{Mazzali2022mnras}
{Mazzali}, P.~A., {Benetti}, S., {Stritzinger}, M., \& {Ashall}, C. 2022,
  \mnras, 511, 5560, \dodoi{10.1093/mnras/stac409}

\bibitem[Mazzali et al.(2014)]{Mazzali2014mnras} Mazzali, P.~A., Sullivan, M., Hachinger, S., et al.\ 2014, \mnras, 439, 1959. \dodoi{10.1093/mnras/stu077}

\bibitem[{{Mazzali} {et~al.}(2007){Mazzali}, {R{\"o}pke}, {Benetti}, \&
  {Hillebrandt}}]{Mazzali2007sci}
{Mazzali}, P.~A., {R{\"o}pke}, F.~K., {Benetti}, S., \& {Hillebrandt}, W. 2007,
  Science, 315, 825, \dodoi{10.1126/science.1136259}

\bibitem[Mazzali et al.(2005)]{Mazzali2005mnras} Mazzali, P.~A., Benetti, S., Stehle, M., et al.\ 2005, \mnras, 357, 200. \dodoi{10.1111/j.1365-2966.2005.08640.x}

\bibitem[Walker et al.(2012)]{Walker2012mnras} Walker, E.~S., Hachinger, S., Mazzali, P.~A., et al.\ 2012, \mnras, 427, 103. \dodoi{10.1111/j.1365-2966.2012.21928.x}

\bibitem[{{Meikle} {et~al.}(1996){Meikle}, {Cumming}, {Geballe}, {Lewis},
  {Walton}, {Balcells}, {Cimatti}, {Croom}, {Dhillon}, \&
  {Economou}}]{Meikle1996mnras}
{Meikle}, W.~P.~S., {Cumming}, R.~J., {Geballe}, T.~R., {et~al.} 1996, \mnras,
  281, 263, \dodoi{10.1093/mnras/281.1.263}

\bibitem[{{Miller} {et~al.}(2020){Miller}, {Yao}, {Bulla}, {Pankow}, {Bellm},
  {Cenko}, {Dekany}, {Fremling}, {Graham}, {Kupfer}, {Laher}, {Mahabal},
  {Masci}, {Nugent}, {Riddle}, {Rusholme}, {Smith}, {Shupe}, {van Roestel}, \&
  {Kulkarni}}]{Miller2020apj}
{Miller}, A.~A., {Yao}, Y., {Bulla}, M., {et~al.} 2020, \apj, 902, 47,
  \dodoi{10.3847/1538-4357/abb13b}

\bibitem[{{Milne} {et~al.}(2013){Milne}, {Brown}, {Roming}, {Bufano}, \&
  {Gehrels}}]{Milne2013apj}
{Milne}, P.~A., {Brown}, P.~J., {Roming}, P. W.~A., {Bufano}, F., \& {Gehrels},
  N. 2013, \apj, 779, 23, \dodoi{10.1088/0004-637X/779/1/23}

\bibitem[{{Moffat}(1969)}]{Moffat1969aap}
{Moffat}, A.~F.~J. 1969, \aap, 3, 455

\bibitem[{{Moon} {et~al.}(2016){Moon}, {Kim}, {Lee}, {Pak}, {Park}, {He},
  {Antoniadis}, {Ni}, {Lee}, {Kim}, {Park}, {Kim}, {Cha}, {Lee}, \&
  {Gonzalez}}]{Moon2016spie}
{Moon}, D.-S., {Kim}, S.~C., {Lee}, J.-J., {et~al.} 2016, in Society of
  Photo-Optical Instrumentation Engineers (SPIE) Conference Series, Vol. 9906,
  Ground-based and Airborne Telescopes VI, 99064I, \dodoi{10.1117/12.2233921}

\bibitem[{{Moon} {et~al.}(2021){Moon}, {Ni}, {Drout},
  {Gonz{\'a}lez-Gait{\'a}n}, {Afsariardchi}, {Park}, {Lee}, {Kim},
  {Antoniadis}, {Kim}, \& {Lee}}]{Moon2021apj}
{Moon}, D.-S., {Ni}, Y.~Q., {Drout}, M.~R., {et~al.} 2021, \apj, 910, 151,
  \dodoi{10.3847/1538-4357/abe466}

\bibitem[{{Mould} {et~al.}(2000){Mould}, {Huchra}, {Freedman}, {Kennicutt},
  {Ferrarese}, {Ford}, {Gibson}, {Graham}, {Hughes}, {Illingworth}, {Kelson},
  {Macri}, {Madore}, {Sakai}, {Sebo}, {Silbermann}, \&
  {Stetson}}]{Mould2000apj}
{Mould}, J.~R., {Huchra}, J.~P., {Freedman}, W.~L., {et~al.} 2000, \apj, 529,
  786, \dodoi{10.1086/308304}

\bibitem[{{Mulligan} \& {Wheeler}(2018)}]{Mulligan2018mnras}
{Mulligan}, B.~W., \& {Wheeler}, J.~C. 2018, \mnras, 476, 1299,
  \dodoi{10.1093/mnras/sty027}

\bibitem[{{Ni} {et~al.}(2022){Ni}, {Moon}, {Drout}, {Polin}, {Sand},
  {Gonz{\'a}lez-Gait{\'a}n}, {Kim}, {Lee}, {Park}, {Howell}, {Nugent}, {Piro},
  {Brown}, {Galbany}, {Burke}, {Hiramatsu}, {Hosseinzadeh}, {Valenti},
  {Afsariardchi}, {Andrews}, {Antoniadis}, {Arcavi}, {Beaton}, {Bostroem},
  {Carlberg}, {Cenko}, {Cha}, {Dong}, {Gal-Yam}, {Haislip}, {Holoien},
  {Johnson}, {Kouprianov}, {Lee}, {Matzner}, {Morrell}, {McCully}, {Pignata},
  {Reichart}, {Rich}, {Ryder}, {Smith}, {Wyatt}, \& {Yang}}]{Ni2022natas}
{Ni}, Y.~Q., {Moon}, D.-S., {Drout}, M.~R., {et~al.} 2022, Nature Astronomy, 6,
  568, \dodoi{10.1038/s41550-022-01603-4}

\bibitem[{{Ni} {et~al.}(2023){Ni}, {Moon}, {Drout}, {Polin}, {Sand},
  {Gonz{\'a}lez-Gait{\'a}n}, {Kim}, {Lee}, {Park}, {Howell}, {Nugent}, {Piro},
  {Brown}, {Galbany}, {Burke}, {Hiramatsu}, {Hosseinzadeh}, {Valenti},
  {Afsariardchi}, {Andrews}, {Antoniadis}, {Beaton}, {Bostroem}, {Carlberg},
  {Cenko}, {Cha}, {Dong}, {Gal-Yam}, {Haislip}, {Holoien}, {Johnson},
  {Kouprianov}, {Lee}, {Matzner}, {Morrell}, {McCully}, {Pignata}, {Reichart},
  {Rich}, {Ryder}, {Smith}, {Wyatt}, \& {Yang}}]{Ni2023apj}
---. 2023, \apj, 946, 7, \dodoi{10.3847/1538-4357/aca9be}

\bibitem[{{Olling} {et~al.}(2015){Olling}, {Mushotzky}, {Shaya}, {Rest},
  {Garnavich}, {Tucker}, {Kasen}, {Margheim}, \& {Filippenko}}]{Olling2015nat}
{Olling}, R.~P., {Mushotzky}, R., {Shaya}, E.~J., {et~al.} 2015, \nat, 521,
  332, \dodoi{10.1038/nature14455}

\bibitem[{{Pakmor} {et~al.}(2012){Pakmor}, {Kromer}, {Taubenberger}, {Sim},
  {R{\"o}pke}, \& {Hillebrandt}}]{Pakmor2012apj}
{Pakmor}, R., {Kromer}, M., {Taubenberger}, S., {et~al.} 2012, \apjl, 747, L10,
  \dodoi{10.1088/2041-8205/747/1/L10}

\bibitem[{{Park} {et~al.}(2017){Park}, {Moon}, {Zaritsky}, {Pak}, {Lee}, {Kim},
  {Kim}, \& {Cha}}]{Park2017apj}
{Park}, H.~S., {Moon}, D.-S., {Zaritsky}, D., {et~al.} 2017, \apj, 848, 19,
  \dodoi{10.3847/1538-4357/aa88ab}

\bibitem[{{Parrent} {et~al.}(2014){Parrent}, {Friesen}, \&
  {Parthasarathy}}]{Parrent2014apss}
{Parrent}, J., {Friesen}, B., \& {Parthasarathy}, M. 2014, \apss, 351, 1,
  \dodoi{10.1007/s10509-014-1830-1}

\bibitem[{{Parrent} {et~al.}(2012){Parrent}, {Howell}, {Friesen}, {Thomas},
  {Fesen}, {Milisavljevic}, {Bianco}, {Dilday}, {Nugent}, {Baron}, {Arcavi},
  {Ben-Ami}, {Bersier}, {Bildsten}, {Bloom}, {Cao}, {Cenko}, {Filippenko},
  {Gal-Yam}, {Kasliwal}, {Konidaris}, {Kulkarni}, {Law}, {Levitan}, {Maguire},
  {Mazzali}, {Ofek}, {Pan}, {Polishook}, {Poznanski}, {Quimby}, {Silverman},
  {Sternberg}, {Sullivan}, {Walker}, {Xu}, {Buton}, \&
  {Pereira}}]{Parrent2012apj}
{Parrent}, J.~T., {Howell}, D.~A., {Friesen}, B., {et~al.} 2012, \apjl, 752,
  L26, \dodoi{10.1088/2041-8205/752/2/L26}

\bibitem[{{Pereira} {et~al.}(2013){Pereira}, {Thomas}, {Aldering}, {Antilogus},
  {Baltay}, {Benitez-Herrera}, {Bongard}, {Buton}, {Canto}, {Cellier-Holzem},
  {Chen}, {Childress}, {Chotard}, {Copin}, {Fakhouri}, {Fink}, {Fouchez},
  {Gangler}, {Guy}, {Hillebrandt}, {Hsiao}, {Kerschhaggl}, {Kowalski},
  {Kromer}, {Nordin}, {Nugent}, {Paech}, {Pain}, {P{\'e}contal}, {Perlmutter},
  {Rabinowitz}, {Rigault}, {Runge}, {Saunders}, {Smadja}, {Tao},
  {Taubenberger}, {Tilquin}, \& {Wu}}]{Pereira2013aa}
{Pereira}, R., {Thomas}, R.~C., {Aldering}, G., {et~al.} 2013, \aap, 554, A27,
  \dodoi{10.1051/0004-6361/201221008}

\bibitem[{{Perlmutter} {et~al.}(1999){Perlmutter}, {Aldering}, {Goldhaber},
  {Knop}, {Nugent}, {Castro}, {Deustua}, {Fabbro}, {Goobar}, {Groom}, {Hook},
  {Kim}, {Kim}, {Lee}, {Nunes}, {Pain}, {Pennypacker}, {Quimby}, {Lidman},
  {Ellis}, {Irwin}, {McMahon}, {Ruiz-Lapuente}, {Walton}, {Schaefer}, {Boyle},
  {Filippenko}, {Matheson}, {Fruchter}, {Panagia}, {Newberg}, {Couch}, \&
  {Project}}]{perlmutter1999apj}
{Perlmutter}, S., {Aldering}, G., {Goldhaber}, G., {et~al.} 1999, \apj, 517,
  565, \dodoi{10.1086/307221}

\bibitem[{{Phillips} {et~al.}(1999){Phillips}, {Lira}, {Suntzeff}, {Schommer},
  {Hamuy}, \& {Maza}}]{Phillips1999aj}
{Phillips}, M.~M., {Lira}, P., {Suntzeff}, N.~B., {et~al.} 1999, \aj, 118,
  1766, \dodoi{10.1086/301032}

\bibitem[{{Pinto} \& {Eastman}(2000)}]{Pinto&Eastman2000apj}
{Pinto}, P.~A., \& {Eastman}, R.~G. 2000, \apj, 530, 757,
  \dodoi{10.1086/308380}

\bibitem[{{Piro}(2015)}]{Piro2015apj}
{Piro}, A.~L. 2015, \apj, 808, L51, \dodoi{10.1088/2041-8205/808/2/L51}

\bibitem[{{Piro} \& {Morozova}(2016)}]{Piro&Morozova2016apj}
{Piro}, A.~L., \& {Morozova}, V.~S. 2016, \apj, 826, 96,
  \dodoi{10.3847/0004-637X/826/1/96}

\bibitem[{{Piro} \& {Nakar}(2013)}]{Piro&Nakar2013apj}
{Piro}, A.~L., \& {Nakar}, E. 2013, \apj, 769, 67,
  \dodoi{10.1088/0004-637X/769/1/67}

\bibitem[{{Piro} \& {Nakar}(2014)}]{Piro&Nakar2014apj}
---. 2014, \apj, 784, 85, \dodoi{10.1088/0004-637X/784/1/85}

\bibitem[{{Polin} {et~al.}(2019){Polin}, {Nugent}, \& {Kasen}}]{Polin2019apj}
{Polin}, A., {Nugent}, P., \& {Kasen}, D. 2019, \apj, 873, 84,
  \dodoi{10.3847/1538-4357/aafb6a}

\bibitem[{{Polin} {et~al.}(2021){Polin}, {Nugent}, \& {Kasen}}]{Polin2021apj}
---. 2021, \apj, 906, 65, \dodoi{10.3847/1538-4357/abcccc}

\bibitem[Phillips et al.(2013)]{Phillips2013apj} Phillips, M.~M., Simon, J.~D., Morrell, N., et al.\ 2013, \apj, 779, 38. \dodoi{10.1088/0004-637X/779/1/38}

\bibitem[{{Poznanski} {et~al.}(2012){Poznanski}, {Prochaska}, \&
  {Bloom}}]{Poznanski2012mnras}
{Poznanski}, D., {Prochaska}, J.~X., \& {Bloom}, J.~S. 2012, \mnras, 426, 1465,
  \dodoi{10.1111/j.1365-2966.2012.21796.x}

\bibitem[{{Quimby} {et~al.}(2006){Quimby}, {H{\"o}flich}, {Kannappan},
  {Rykoff}, {Rujopakarn}, {Akerlof}, {Gerardy}, \& {Wheeler}}]{Quimby2006apj}
{Quimby}, R., {H{\"o}flich}, P., {Kannappan}, S.~J., {et~al.} 2006, \apj, 636,
  400, \dodoi{10.1086/498014}

\bibitem[{{Riess} {et~al.}(1998){Riess}, {Filippenko}, {Challis},
  {Clocchiatti}, {Diercks}, {Garnavich}, {Gilliland}, {Hogan}, {Jha},
  {Kirshner}, {Leibundgut}, {Phillips}, {Reiss}, {Schmidt}, {Schommer},
  {Smith}, {Spyromilio}, {Stubbs}, {Suntzeff}, \& {Tonry}}]{Riess1998aj}
{Riess}, A.~G., {Filippenko}, A.~V., {Challis}, P., {et~al.} 1998, \aj, 116,
  1009, \dodoi{10.1086/300499}

\bibitem[{{Riess} {et~al.}(2016){Riess}, {Macri}, {Hoffmann}, {Scolnic},
  {Casertano}, {Filippenko}, {Tucker}, {Reid}, {Jones}, {Silverman},
  {Chornock}, {Challis}, {Yuan}, {Brown}, \& {Foley}}]{Riess2016apj}
{Riess}, A.~G., {Macri}, L.~M., {Hoffmann}, S.~L., {et~al.} 2016, \apj, 826,
  56, \dodoi{10.3847/0004-637X/826/1/56}

\bibitem[{{Sai} {et~al.}(2022){Sai}, {Wang}, {Elias-Rosa}, {Yang}, {Zhang},
  {Lin}, {Mo}, {Piro}, {Zeng}, {Reguitti}, {Brown}, {Burns}, {Cai}, {Fiore},
  {Hsiao}, {Isern}, {Itagaki}, {Li}, {Li}, {Pessi}, {Phillips}, {Schuldt},
  {Shahbandeh}, {Stritzinger}, {Tomasella}, {Vogl}, {Wang}, {Wang}, {Wu},
  {Yang}, {Zhang}, {Zhang}, \& {Zhang}}]{Sai2022mnras}
{Sai}, H., {Wang}, X., {Elias-Rosa}, N., {et~al.} 2022, \mnras, 514, 3541,
  \dodoi{10.1093/mnras/stac1525}

\bibitem[{{Sand} {et~al.}(2018){Sand}, {Graham}, {Boty{\'a}nszki}, {Hiramatsu},
  {McCully}, {Valenti}, {Hosseinzadeh}, {Howell}, {Burke}, {Cartier},
  {Diamond}, {Hsiao}, {Jha}, {Kasen}, {Kumar}, {Marion}, {Suntzeff},
  {Tartaglia}, {Wheeler}, \& {Wyatt}}]{Sand2018apj}
{Sand}, D.~J., {Graham}, M.~L., {Boty{\'a}nszki}, J., {et~al.} 2018, \apj, 863,
  24, \dodoi{10.3847/1538-4357/aacde8}

\bibitem[{{Sand} {et~al.}(2019){Sand}, {Amaro}, {Moe}, {Graham}, {Andrews},
  {Burke}, {Cartier}, {Eweis}, {Galbany}, {Hiramatsu}, {Howell}, {Jha},
  {Lundquist}, {Matheson}, {McCully}, {Milne}, {Smith}, {Valenti}, \&
  {Wyatt}}]{Sand2019apj}
{Sand}, D.~J., {Amaro}, R.~C., {Moe}, M., {et~al.} 2019, \apjl, 877, L4,
  \dodoi{10.3847/2041-8213/ab1eaf}

\bibitem[{{Scalzo} {et~al.}(2019){Scalzo}, {Parent}, {Burns}, {Childress},
  {Tucker}, {Brown}, {Contreras}, {Hsiao}, {Krisciunas}, {Morrell}, {Phillips},
  {Piro}, {Stritzinger}, \& {Suntzeff}}]{Scalzo2019mnras}
{Scalzo}, R.~A., {Parent}, E., {Burns}, C., {et~al.} 2019, \mnras, 483, 628,
  \dodoi{10.1093/mnras/sty3178}

\bibitem[{{Schlafly} \& {Finkbeiner}(2011)}]{Schlafly&Finkbeiner2011apj}
{Schlafly}, E.~F., \& {Finkbeiner}, D.~P. 2011, \apj, 737, 103,
  \dodoi{10.1088/0004-637X/737/2/103}

\bibitem[{{Seitenzahl} {et~al.}(2013){Seitenzahl}, {Ciaraldi-Schoolmann},
  {R{\"o}pke}, {Fink}, {Hillebrandt}, {Kromer}, {Pakmor}, {Ruiter}, {Sim}, \&
  {Taubenberger}}]{Seitenzahl2013mnras}
{Seitenzahl}, I.~R., {Ciaraldi-Schoolmann}, F., {R{\"o}pke}, F.~K., {et~al.}
  2013, \mnras, 429, 1156, \dodoi{10.1093/mnras/sts402}

\bibitem[{{Shen} {et~al.}(2021){Shen}, {Blondin}, {Kasen}, {Dessart},
  {Townsley}, {Boos}, \& {Hillier}}]{Shen2021apjl}
{Shen}, K.~J., {Blondin}, S., {Kasen}, D., {et~al.} 2021, \apjl, 909, L18,
  \dodoi{10.3847/2041-8213/abe69b}

\bibitem[{{Silverman} {et~al.}(2015){Silverman}, {Vink{\'o}}, {Marion},
  {Wheeler}, {Barna}, {Szalai}, {Mulligan}, \&
  {Filippenko}}]{Silverman2015mnras}
{Silverman}, J.~M., {Vink{\'o}}, J., {Marion}, G.~H., {et~al.} 2015, \mnras,
  451, 1973, \dodoi{10.1093/mnras/stv1011}

\bibitem[{{Skrutskie} {et~al.}(2003){Skrutskie}, {Cutri}, {Stiening},
  {Weinberg}, {Schneider}, {Carpenter}, {Beichman}, {Capps}, {Chester},
  {Elias}, {Huchra}, {Liebert}, {Lonsdale}, {Monet}, {Price}, {Seitzer},
  {Jarrett}, {Kirkpatrick}, {Gizis}, {Howard}, {Evans}, {Fowler}, {Fullmer},
  {Hurt}, {Light}, {Kopan}, {Marsh}, {McCallon}, {Tam}, {van Dyk}, \&
  {Wheelock}}]{Skrutskie2003ycat}
{Skrutskie}, M.~F., {Cutri}, R.~M., {Stiening}, R., {et~al.} 2003, VizieR
  Online Data Catalog, VII/233

\bibitem[{{Stritzinger} {et~al.}(2002){Stritzinger}, {Hamuy}, {Suntzeff},
  {Smith}, {Phillips}, {Maza}, {Strolger}, {Antezana}, {Gonz{\'a}lez},
  {Wischnjewsky}, {Candia}, {Espinoza}, {Gonz{\'a}lez}, {Stubbs}, {Becker},
  {Rubenstein}, \& {Galaz}}]{Stritzinger2002aj}
{Stritzinger}, M., {Hamuy}, M., {Suntzeff}, N.~B., {et~al.} 2002, \aj, 124,
  2100, \dodoi{10.1086/342544}

\bibitem[{{Stritzinger} {et~al.}(2018){Stritzinger}, {Shappee}, {Piro},
  {Ashall}, {Baron}, {Hoeflich}, {Holmbo}, {Holoien}, {Phillips}, {Burns},
  {Contreras}, {Morrell}, \& {Tucker}}]{Stritzinger2018apj}
{Stritzinger}, M.~D., {Shappee}, B.~J., {Piro}, A.~L., {et~al.} 2018, \apj,
  864, L35, \dodoi{10.3847/2041-8213/aadd46}

\bibitem[{{Tanaka} {et~al.}(2008){Tanaka}, {Mazzali}, {Benetti}, {Nomoto},
  {Elias-Rosa}, {Kotak}, {Pignata}, {Stanishev}, \&
  {Hachinger}}]{tanaka2008ap}
{Tanaka}, M., {Mazzali}, P.~A., {Benetti}, S., {et~al.} 2008, \apj, 677, 448,
  \dodoi{10.1086/528703}

\bibitem[{{Taubenberger} {et~al.}(2013){Taubenberger}, {Kromer}, {Pakmor},
  {Pignata}, {Maeda}, {Hachinger}, {Leibundgut}, \& {Hillebrand
  t}}]{Taubenberger2013apj}
{Taubenberger}, S., {Kromer}, M., {Pakmor}, R., {et~al.} 2013, \apjl, 775, L43,
  \dodoi{10.1088/2041-8205/775/2/L43}

\bibitem[Taubenberger et al.(2019)]{Taubenberger2019mnras} Taubenberger, S., Floers, A., Vogl, C., et al.\ 2019, \mnras, 488, 5473. \dodoi{10.1093/mnras/stz1977}

\bibitem[Ashall et al.(2020)]{Ashall2020apjl} Ashall, C., Lu, J., Burns, C., et al.\ 2020, \apjl, 895, L3. \dodoi{10.3847/2041-8213/ab8e37}

\bibitem[Ashall et al.(2021)]{Ashall2021apj} Ashall, C., Lu, J., Hsiao, E.~Y., et al.\ 2021, \apj, 922, 205. \dodoi{10.3847/1538-4357/ac19ac}

\bibitem[{{Townsley} {et~al.}(2019){Townsley}, {Miles}, {Shen}, \&
  {Kasen}}]{Townsley2019apj}
{Townsley}, D.~M., {Miles}, B.~J., {Shen}, K.~J., \& {Kasen}, D. 2019, \apjl,
  878, L38, \dodoi{10.3847/2041-8213/ab27cd}

\bibitem[{{Trujillo} {et~al.}(2001){Trujillo}, {Aguerri}, {Cepa}, \&
  {Guti{\'e}rrez}}]{Trujillo2001mnras}
{Trujillo}, I., {Aguerri}, J.~A.~L., {Cepa}, J., \& {Guti{\'e}rrez}, C.~M.
  2001, \mnras, 328, 977, \dodoi{10.1046/j.1365-8711.2001.04937.x}

\bibitem[{{Tucker} {et~al.}(2022){Tucker}, {Ashall}, {Shappee}, {Kochanek},
  {Stanek}, \& {Garnavich}}]{Tucker2022apjl}
{Tucker}, M.~A., {Ashall}, C., {Shappee}, B.~J., {et~al.} 2022, \apjl, 926,
  L25, \dodoi{10.3847/2041-8213/ac4fbd}

\bibitem[{{Tully} \& {Fisher}(1977)}]{Tully1977aa}
{Tully}, R.~B., \& {Fisher}, J.~R. 1977, \aap, 54, 661

\bibitem[{{Valenti} {et~al.}(2021){Valenti}, {Sand}, {Wyatt}, {Lundquist},
  {Amaro}, {Andrews}, {Jencson}, {Dong}, {Davis}, \&
  {Janzen}}]{Valenti2021tnstr}
{Valenti}, S., {Sand}, D.~J., {Wyatt}, S., {et~al.} 2021, Transient Name Server
  Discovery Report, 2021-3864, 1

\bibitem[{{Wang} {et~al.}(2003){Wang}, {Baade}, {H{\"o}flich}, {Khokhlov},
  {Wheeler}, {Kasen}, {Nugent}, {Perlmutter}, {Fransson}, \&
  {Lundqvist}}]{Wang2003apj}
{Wang}, L., {Baade}, D., {H{\"o}flich}, P., {et~al.} 2003, \apj, 591, 1110,
  \dodoi{10.1086/375444}

\bibitem[Cikota et al.(2019)]{Cikota2019mnras} Cikota, A., Patat, F., Wang, L., et al.\ 2019, \mnras, 490, 578. \dodoi{10.1093/mnras/stz2322}


\bibitem[{{Wang} {et~al.}(2009){Wang}, {Filippenko}, {Ganeshalingam}, {Li},
  {Silverman}, {Wang}, {Chornock}, {Foley}, {Gates}, \&
  {Macomber}}]{Wang2009apj}
{Wang}, X., {Filippenko}, A.~V., {Ganeshalingam}, M., {et~al.} 2009, \apjl,
  699, L139, \dodoi{10.1088/0004-637X/699/2/L139}

\bibitem[{{Whelan} \& {Iben}(1973)}]{Whelan&Iben1973apj}
{Whelan}, J., \& {Iben}, Icko, J. 1973, \apj, 186, 1007, \dodoi{10.1086/152565}

\bibitem[{{Yaron} {et~al.}(2017){Yaron}, {Perley}, {Gal-Yam}, {Groh}, {Horesh},
  {Ofek}, {Kulkarni}, {Sollerman}, {Fransson}, {Rubin}, {Szabo}, {Sapir},
  {Taddia}, {Cenko}, {Valenti}, {Arcavi}, {Howell}, {Kasliwal}, {Vreeswijk},
  {Khazov}, {Fox}, {Cao}, {Gnat}, {Kelly}, {Nugent}, {Filippenko}, {Laher},
  {Wozniak}, {Lee}, {Rebbapragada}, {Maguire}, {Sullivan}, \&
  {Soumagnac}}]{Yaron2017natph}
{Yaron}, O., {Perley}, D.~A., {Gal-Yam}, A., {et~al.} 2017, Nature Physics, 13,
  510, \dodoi{10.1038/nphys4025}

\bibitem[{{Zhang} {et~al.}(2021){Zhang}, {Murakami}, {Stahl}, {Patra}, \&
  {Filippenko}}]{Zhang2021mnras}
{Zhang}, K.~D., {Murakami}, Y.~S., {Stahl}, B.~E., {Patra}, K.~C., \&
  {Filippenko}, A.~V. 2021, \mnras, 503, L33, \dodoi{10.1093/mnrasl/slab020}

\end{thebibliography}
\end{document}